\documentclass[twocolumn]{aa} 
\usepackage{graphicx}
\usepackage{amsmath}
\usepackage{txfonts}
\usepackage{natbib}
\usepackage{url}

\bibpunct{(}{)}{;}{a}{}{}
\newcommand{\changed}{}
\newcommand{\changedn}{}

\newcommand{\argmax}[1]{\operatorname{arg}\,{\underset{#1}{\operatorname{max}}\;}}
\begin{document}
\title{Wavelet-based cross-correlation analysis of structure scaling in turbulent clouds}

\author{Tigran G.\ Arshakian\inst{1,2} and Volker Ossenkopf\inst{1}}

\offprints{T.G. Arshakian}

\institute{
I. Physikalisches Institut, Universit\"at zu K\"oln, Z\"ulpicher Strasse 77, D-50937 K\"oln, Germany\\
\email{[arshakian;ossk]@ph1.uni-koeln.de}
              \and
Byurakan Astrophysical Observatory, Aragatsotn prov. 378433, Armenia, and
Isaac Newton Institute of Chile, Armenian Branch
             }

   \date{Received ...; accepted ...}


  \abstract
   {}
   {
   We propose a statistical tool to compare the scaling behaviour of turbulence in pairs 
   of molecular cloud maps. Using artificial maps with well defined spatial properties, we calibrate
   the method and test its limitations to ultimately apply it to a set of observed maps.  
}
   {
   We develop the wavelet-based weighted cross-correlation (WWCC) method to study the 
   relative contribution of structures of different sizes and their degree
   of correlation in two maps as a function of spatial scale,
   and the mutual displacement of structures in the molecular cloud 
   maps.
  }
   {
   We test the WWCC for circular structures having a single prominent scale and 
   fractal structures showing a self-similar behavior without prominent scales. 
   Observational noise and a finite map size limit the scales where the cross-correlation
   coefficients and displacement vectors can be reliably measured. For fractal maps
   containing many structures on all scales, the limitation from the observational noise
   is negligible for signal-to-noise ratios $\ga 5$. We propose an approach for 
   identification of correlated structures in the maps which allows us to localize 
   individual correlated structures and recognize their shapes and suggest a recipe for 
   the recovering of enhanced scales in self-similar structures.
   Application of the WWCC to the observed line maps of the giant molecular
   cloud G\,333 allows to add specific scale information to the results
   obtained earlier using the principle component analysis (PCA). It confirms 
   the chemical and excitation similarity of $^{13}$CO and C$^{18}$O on all scales,
   but shows a deviation of HCN at scales of up to {\changed 7~pc}. This can be interpreted
   as a chemical transition scale. The largest structures also show a
   systematic offset along the filament, probably due to a large-scale
   density gradient.
   }
   { 
   The WWCC can compare correlated structures in different maps of molecular clouds
   identifying scales that represent structural changes such as chemical and phase transitions and
   prominent or enhanced dimensions.
   }


\titlerunning{WWCC analysis of structural scaling in turbulent clouds}
\authorrunning{T.G. Arshakian and V. Ossenkopf}
\maketitle

%

\section{Introduction}

The interstellar medium has a complex dynamic structure on all scales (from 
sub-parsecs to at least tens of parsecs) as a result of various physical 
processes occurring in the multi-scale turbulent cascade of molecular and 
atomic gas. High-resolution 
and high dynamic range of observations of emission line transitions and 
continuum emission of interstellar clouds provide evidence for clumpy 
structures on all scales \citep[see e.g.][]{StutzkiGuesten1990,RomanDuval2011} 
and anisotropic clouds such as shells and filaments 
\citep[e.g.][]{Menshchikov2010,Deharveng2010}. Observations of line transitions, 
total and polarized continuum emission provide 
valuable information about physical conditions (density, temperature, 
magnetic field) and kinematics of a multi-phase gas distributed on different scales.

The correlation between {\changed different structures measured in interstellar
clouds, e.g. contours of different chemical tracers or the density, temperature
and velocity peaks, as a function of their size can be used to quantify 
commonalities and differences in the formation of these structures.
Understanding the  commonalities and differences seen in different tracers,
different velocity components or different excitation conditions as a function 
of scale length, will help to} infer the underlying physical processes {\changed in
the turbulent cascade} in interstellar clouds.
As many processes have characteristic scales, but show their signatures
only in particular tracers, it is essential to compare the scaling behaviour
of {\changed various tracers in a same region} to identify those
scales. {\changed The characteristic scales} could be, e.g., chemical transition scales 
\citep[see, e.g.,][]{Glover2010}, when comparing molecular line maps of 
different species, dynamical scales for the formation of coherent
structures \citep{Goodman1998}, IR penetration
scales showing up in dust continuum maps at different wavelengths \citep[e.g.][]{Abergel1996},
ambipolar diffusion scales {\changed for the dynamical coupling between 
ionized and neutral particles, seen when comparing maps of ionized and neutral
species \citep{mckee10,li12}}, and dissipation scales when comparing channel maps of 
individual atomic or molecular lines \citep{falgarone98}.

To address these issues, we use two different starting points. The  
wavelet analysis has been proven to be a powerful tool for detecting structures
on different spatial scales in time-series and in 2-dimensional maps.
It was used e.g. in the $\Delta$-variance analysis measuring the amount 
of structure in a molecular cloud map as a function of scale 
and to determine the slope of the power spectrum of the cloud scaling
\citep{stutzki98}.
A combination of a wavelet filtering with the cross-correlation function 
was first proposed by \cite{nesme95} to study the solar activity.
\cite{frick01} improved the method introducing the wavelet cross-correlation
function to study the correlation between galactic images as a function of 
scale. This method and its modifications were successfully applied to study 
solar physics, ionosphere fluctuations, images of astronomical objects and 
all-sky surveys \citep[e.g., ][]{vielva06,liu06,taba10,roux12,taba13}.

These approaches, however, {\changedn are not designed for recovering} 
the displacement of structures in the data sets on scale-by-scale basis and 
{\changedn accounting for a variable noise distribution 
across the maps and irregular boundaries.} 
\citet{patrikeev06} used an anisotropic wavelet transform
to isolate spiral features in images of M\,51 for different tracers 
and analysed their location and pitch angle as a function of radius
and azimuth but they did not cross-correlate the features seen in different tracers.
%
%
{\changedn The wavelet-based cross-correlation method can be improved} 
 by inheriting the corresponding
formalism from the wavelet-based $\Delta$-variance analysis
and combining it with the cross-correlation analysis. 
To account for an uneven distribution of noise in the maps, \cite{ossenkopf08a} 
(hereafter O08) 
implemented a weighting function in the improved $\Delta$-variance method to
analyse arbitrary data sets of molecular clouds. 
The weighting function corrects for the contribution of data points with a varying
signal-to-noise ratio as well as allows to perform a proper treatment of
map edges, independent of their shape, and
calculations in Fourier space and, hence, speeds up the computation by making 
use of the fast Fourier transform. 
Moreover, O08 suggested an optimal shape of the wavelet filter for analysing the
turbulent structures. 

In this paper, we make use of the advantages of weighting functions and    
develop a weighted wavelet cross-correlation (WWCC) method to {\changedn recover the 
correlation and displacement between structures of molecular clouds as a function of scale, in which no assumption about the noise or boundaries are made.}

The paper is organised as follows. Section \ref{sec:general} introduces 
the $\Delta$-variance and cross-correlation methods. The WWCC method is 
presented in Sect.~\ref{sec:wwcc}. Application of the WWCC to simulated circular 
structures and fractal structures is described in Sect.~\ref{sec:test_wwcc}.
Application of the WWCC to observed emission line maps of the giant molecular 
cloud G\,333 is presented in Sect.~\ref{sec:application}. 
Discussion and conclusions are presented in Sect.~\ref{sec:discussion}.

\section{Basic theory}
\label{sec:general}

We consider two maps, $f(\vec{x})$ and $g(\vec{x})$, $(\vec{x}=(x,y))$, 
weighted at each pixel by a significance function $w_f(\vec{x})$ and 
$w_g$(\vec{x}). The significance functions $w_f(\vec{x})$ and $w_g(\vec{x})$ 
characterize e.g. a variable signal-to-noise
ratio across the maps or can be used to treat irregular map boundaries.

\subsection{Wavelet analysis to measure the scaling in turbulent structures}
\label{sec:delta_var}

The $\Delta$-variance measures the amount of structure in an individual map 
as a function of scale, thereby identifying dominant structure sizes.
{\changed In this context ``structure'' describes the spatial variation of 
measured properties, i.e. a deviation from a flat or zero measurement.
The amount of structures quantifies the total variation in a given
map as characterized by the variance. With the $\Delta$-variance, this
is evaluated as a function of the size of the structures equivalent to 
the power spectrum} \citep[]{stutzki98}. This
method represents a 2-dimensional generalisation of the Allan-variance method 
\citep{allan96}. The $\Delta$-variance is evaluated from the map $f$ convolved 
with the wavelet $\psi$
\begin{equation}
    F(\vec{x},l) =  f(\vec{x}) \ast \psi({\vec{x}},l) = \mathop{\int \!\!\! \int}_{x,y}
      f({\vec{x}'}) \psi({\vec{x}'}-{\vec{x}},l) d{\vec{x}'},
    \label{eq:wtransform}
\end{equation}
by computing its variance
\begin{equation}
    \label{eq:delta_variance}
    \sigma^{2}_{\Delta} (l) = \mathop{\int \!\!\! \int}_{x,y} [F(\vec{x},l) - 
    \overline{F}(l)]^2  d\vec{x},
\end{equation}
where the $\ast$ symbol represents the convolution and
$\overline{F}(l)$ is the average intensity of the map.
{\changed For uniform, zero average data without boundaries, this is 
equivalent to the wavelet power spectrum.} 

The wavelet is composed of a positive core, $\psi_{\rm c}$,
and a negative annulus, $\psi_{\rm a}:
$\begin{equation}
 \psi({\vec{x}},l)= \psi_c({\vec{x}},l) +\psi_a({\vec{x}},l).
 \label{eq:wavelet}
\end{equation}
Core and annulus are normalized to an integrated weight of unity,
so that the integral over the whole wavelet cancels to zero.

For a fast computation of the convolution in Eq.~(\ref{eq:wtransform})
\citet{stutzki98} used the multiplication
in Fourier space, but \citet{bensch01} showed that this can lead to 
considerable errors from edge effects due to the incompatibility of the
implicit assumption of periodic maps in the Fourier transform and 
the boundaries of real observed maps. They suggested a 
filter that changes its shape closer to the map boundaries by truncating
it beyond the map edges and changing the amplitude of core and annulus
to retain the wavelet normalization condition. While this improves the
edge treatment its weakness is that the convolution of the map cannot
be performed in Fourier space any more. 

\citet{ossenkopf08a} suggested a modification 
that overcomes the edge treatment problems, simultaneously deals with the
effect of observational uncertainties and irregular map boundaries, and 
allows for a computation in the Fourier domain. To fix the filter function they increase 
the map size when the filter extends beyond the map edges and apply a 
zero-padding to the extended map area. 
The re-normalization of the filter is accomplished by introducing the
complementary weighting function $w(\vec{x})>0$ inside the valid map
and $w(\vec{x})=0$ in the zero-padded region. When using a generalized
weighting function $0 \le w(\vec{x}) \le 1$ this can characterize the
significance of every individual data point, including the effects of
noise and observational uncertainties.
Then the map $f_{\rm p}$ and the weights $w(\vec{x})$ have to be convolved
separately with the positive and negative filter parts
$f_{\rm c}(\vec{x},l) = f_{\rm p} \ast \psi_c({\vec{x}},l)$, 
$f_{\rm a}(\vec{x},l) = f_{\rm p} \ast \psi_a({\vec{x}},l) $, 
$w_{\rm c}(\vec{x},l) = w(\vec{x}) \ast \psi_c({\vec{x}},l)$, and
$w_{\rm a}(\vec{x},l) = w(\vec{x}) \ast \psi_a({\vec{x}},l)$
and the re-normalization is performed when computing the
map filtered on scale $l$
\begin{equation}
 F(\vec{x},l) =\frac{f_{\rm c}(\vec{x},l)}{w_{\rm c}(\vec{x},l)} 
 - \frac{f_{\rm a}(\vec{x},l)}{w_{\rm a}(\vec{x},l)}.
 \label{eq:wf_wmap}
\end{equation} 
When finally computing the $\Delta$-variance from the convolved map,
the weights are taken into account in the sum of the variations (O08):
\begin{equation}
    \label{eq:delta_var}
    \sigma^{2}_{\Delta} (l) = \frac{\mathop{\int \!\!\! \int}_{x,y} 
    [F(\vec{x},l) - \overline{F}_w(l)]^2  w_F(\vec{x},l) d\vec{x}} 
    {\mathop{\int \!\!\! \int}_{x,y} w_F(\vec{x},l) \, d\vec{x}},
\end{equation}
with
\begin{equation}
 w_F=w_c(\vec{x},l) \, w_a(\vec{x},l), 
 \label{eq:combined_weight}
\end{equation} 
and $\overline{F}_w(l)$ as the weighted average of the map
\begin{equation}
  \label{eq:meanmap}
  \overline{F}_w(l) = \frac{\mathop{\int \!\!\! \int} w_F(\vec{x},l)F(\vec{x},l)\, 
  d\vec{x}}{\mathop{\int \!\!\! \int} w_F(\vec{x},l)\, d\vec{x}}.
\end{equation}

The choice of the wavelet filter and
its optimisation is of great importance for the scaling analysis.
For isotropic wavelets, the ratio of the core 
diameter to annulus diameter is the critical parameter.
O08 found that both, French-hat and 
Mexican-hat filters with a high ratio between the diameter of the annulus and the
core of the filter, are preferred for measurements of the general power spectral slope,
while the Mexican-hat filter with low diameter ratios is more suited to sensitively
detect individual prominent scales. The Mexican-hat filter with an
annulus/core diameter ratio of $v=1.5$ was found to provide the best
compromise, revealing the correct power spectrum slope and all spectral features.
Therefore, we will use this optimal wavelet filter (Eq.~(11) in O08)
throughout the rest of the paper:
\begin{eqnarray}
  \psi_c({\vec{x}}) & = & \frac{4}{\pi l^2} \exp \left( \frac{-{\vec{x}}^2}{(l/2)^2} \right) \nonumber \\
  \psi_a({\vec{x}}) & = & -\frac{4}{\pi l^2 (v^2-1)} \left[ \exp \left( \frac{-{\vec{x}}^2}{(vl/2)^2} 
  \right) - \exp \left( \frac{-{\vec{x}}^2}{(l/2)^2} \right)  \right],
  \label{eq:wavelet_filters}
\end{eqnarray}
where $l$ is the scale of interest. The choice of the
Gaussian smoothed filter is justified by the simultaneous confinement of the
filter in ordinary and Fourier space. The usable range of scales $l$ falls between
2 pixels, needed to guarantee a reasonably isotropic filter shape on a
rectangular pixel grid, and about half the map size as discussed in Sect.~\ref{sect:dvar_spectrum}.
The method was proven to be a powerful 
tool to characterise the power spectrum of interstellar clouds \citep[][]{stutzki98,ossenkopf08b}.

\subsection{Cross-correlation}
\label{sec:cc}

{\changed We can compute the cross-correlation coefficient between two images or maps 
$f(\vec{x})$ and $g(\vec{x})$ as
\begin{equation}
  r  = \mathop{\int \!\!\! \int}_{x,y} f(\vec{x}) g(\vec{x}) \, d\vec{x}.
  \label{eq:ccc1}
\end{equation}
describing the degree of concordance between the maps. If all structures are
shifted between the two maps, the cross-correlation coefficient drops, but
the characterization of the similarity of the two maps can be recovered when
introducing a reverse offset vector $t$ while computing the cross-correlation.
This leads to the definition of the two-dimensional cross-correlation function
as a function of the offset vector, $\vec{t}$, as}
\begin{equation}
  C(\vec{t}) = \mathop{\int \!\!\! \int}_{x,y} f(\vec{x}) g(\vec{x}+\vec{t}) \, d\vec{x}.
  \label{eq:ccf}
\end{equation}
To account for different absolute scales in the maps one normalizes the cross-correlation function 
\begin{equation}
  C(\vec{t}) = \frac{\mathop{\int \!\!\! \int}_{x,y} ( f(\vec{x})-\overline{f} ) 
  ( g(\vec{x}+\vec{t})-\overline{g} ) \, d\vec{x},}{\sigma_f \,\sigma_g},
  \label{eq:nccf}
\end{equation}
where $\overline{f}$ and $\overline{g}$ are the means of signals in two maps, respectively, 
$\sigma_f$ and $\sigma_g$ are the standard deviations computed as,
\begin{equation}
\sigma_f = \left( \mathop{\int \!\!\! \int}_{x, y} (f(\vec{x})-\overline{f})^2    \,d\vec{x} \right )^{1/2},
  \label{eq:sigf}
\end{equation}
for $\sigma_f$ and equivalently for $\sigma_g$. The normalized cross correlation function has values 
between $-1$ and $+1$.

For every given offset vector $\vec{t}$, the cross-correlation function measures
the similarity (or correlation) between structures in the two maps, when the second 
map is shifted by $\vec{t}$. The correlation function for {\changed a zero shift, $\vec{t} =0$, 
recoveres the cross-correlation coefficient, 
\begin{equation}
  r = C(\vec{t}=0)
  \label{eq:cc}
\end{equation}
as the center of the cross-correlation plane.

The position of the maximum of the cross-correlation function in the $\vec{t}$ plane
indicates the recovered displacement vector between the two maps, i.e. the applied 
offset for which the two structures match best.} We
compute the displacement vector as
\begin{equation}
  \vec{\tau} = \argmax{x,y} {C(\vec{t})}.  
  \label{eq:tau}
\end{equation}
The cross-correlation function between two images always integrates 
over all scales involved in the map, not distinguishing the correlation or
mutual displacement of structures at particular scales. This is addressed
by the wavelet cross-correlation.

\section{Weighted wavelet cross-correlation}
\label{sec:wwcc}

\subsection{Theory}
\label{sec:theory}
To study the dependence of the correlation coefficient on the spatial scale, the two data
sets are filtered on scale-by-scale basis by means of wavelets and then 
cross correlated at each scale \citep[see e.g.,][]{frick01}.

Here, we introduce the weighted wavelet cross-correlation method to analyse the correlation between 
maps, $f$ and $g$, weighted at each pixel by $w_f$ and $w_g$, as a function of scale. 
For the filtering, the same formalism is exploited as for the $\Delta$-variance, i.e.
Eq.~(\ref{eq:wavelet_filters}) gives the wavelet to filter the maps on scale $l$ through
Eq.~(\ref{eq:wf_wmap}) to obtain the filtered maps
$F(\vec{x},l)$ and $G(\vec{x},l)$, and the convolved total weights at each
pixel, $w_F(\vec{x},l)$ and $w_G(\vec{x},l)$, are computed through Eq.~(\ref{eq:combined_weight}).

To study the cross correlation of the structure in two
maps, $f$ and $g$, as a function of scale, we introduce the \emph{weighted wavelet
cross-correlation function},
\begin{equation}
  C_{w}(\vec{t},l) = \frac{C_{F-G}(\vec{t},l)}{\sigma_{w,F}(l) \,\sigma_{w,G}(l)},
  \label{eq:wwccf}
\end{equation}
where $C_{F-G}$ is the weighted covariance
\begin{eqnarray}
  C_{F-G}(\vec{t}, l) & = & 
  \mathop{\int \!\!\! \int}_{x,y} \left( w_F(\vec{x},l) \, F_w(\vec{x},l) \right) \times \nonumber \\
   && w_G(\vec{x}+\vec{t},l) \, G_w(\vec{x}+\vec{t},l) \, d\vec{x},
  \label{eq:cow}
\end{eqnarray}
with 
\begin{equation}
  F_w(\vec{x},l) = F(\vec{x},l) - \overline{F}_w(l),
\end{equation}
and equivalently for $G_w(\vec{x},l)$ where $\overline{F}_w(l)$ and $\overline{G}_w(l)$
are computed from Eq.~(\ref{eq:meanmap}).
The normalization is given by the standard deviation
\begin{equation}
\sigma_{w,F}(l) = \left( \mathop{\int \!\!\! \int}_{x, y} w_F^2(\vec{x},l) \, F_w^2(\vec{x},l) \,d\vec{x} \right )^{1/2},
  \label{eq:wsigf}
\end{equation}
of the $F_w$ map. $\sigma_{w,G}(l)$ is equivalently computed for $G_w$.
The normalization WWCC function $C_{w}(\vec{t},l)$ provides values between $-1$ and $+1$.

{\changed The WWCC is therefore a three-dimensional function depending on the
offset vector $\vec{t}$ and the filter size $l$.
In the center of the offset plane, $\vec{t}=0$, the WWCC function} provides the
\emph{wavelet cross-correlation coefficient} $r(l)$, 
i.e. the degree of correlation of the two data sets on scale $l$,
\begin{equation}
  r(l) = C_w(\vec{t}=0, l).
  \label{eq:cc-l}
\end{equation}
The position of the {\changed maximum of the WWCC in the offset plane gives} the optimum offset vector 
(for scale $l$) at which the maps are best aligned. We introduce the \emph{wavelet displacement vector}, 
\begin{equation}
   \vec{\tau}(l) = \argmax{x,y} {C_w(\vec{t},l)}, 
   \label{eq:ov}
\end{equation}
which defines the amplitude and direction of displacement between two maps on scale $l$. These 
can be split into the \emph{spatial displacement function}, 
\begin{equation}
  \tau(l) = (\tau_x^2 + \tau_y^2)^{1/2},
  \label{eq:tau}
\end{equation}
where $\tau_x$ and $\tau_y$ are offsets along $x$-axis and $y$-axis, and the \emph{angular displacement function},
\begin{equation}
  \varphi(l) = \arctan \left (\frac{\tau_y}{\tau_x}\right) .
  \label{eq:tau_phi}
\end{equation}

For an efficient computation of the WWCC, the convolution in Eq.~(\ref{eq:cow})
can be obtained through a multiplication in Fourier space when transforming
the weighted functions 
\begin{equation}
  \hat{F}_w(\vec{k},l) = \mathop{\int  \int}
w_F(\vec{x},l) \, F_w(\vec{x},l) \, e^{-i\vec{k}\vec{x}}d\vec{x},
\label{eq:fFw}
\end{equation}
\begin{equation}
  \hat{G}_w(\vec{k},l) = \mathop{\int  \int}
w_G(\vec{x},l) \, G_w(\vec{x},l) \, e^{-i\vec{k}\vec{x}}d\vec{x},
\label{eq:fGw}
\end{equation}
where $\vec{k}=(k_x,k_y)$ is the wavevector. Using the translational
properties of the Fourier transform $\mathcal{F}$
\begin{equation}
\mathcal{F}\{w_G(\vec{x}+\vec{t},l) \, G_w(\vec{x}+\vec{t},l)\} = \hat{G}_w(\vec{k},l) 
\times e^{i\vec{k}\vec{t}}
\label{eq_shift_theorem}
\end{equation}
we can compute the integral for the WWCC function in Eq. (\ref{eq:wwccf})
through the inverse Fourier transform $\mathcal{F}^{-1}\{\}$ 
\begin{equation}
C_w(\vec{t},l) = \frac{\mathcal{F}^{-1}\{\hat{F}_w(\vec{k},l)
\hat{G}_w(\vec{k},l)\times \exp{i\vec{k}\vec{t}}
\}}{\sigma_{w,F}(l) \,\sigma_{w,G}(l)}.
\label{eq:fCw}
\end{equation}

\subsection{Algorithm to compute the WWCC}

\label{sec:algorithm}
The WWCC\footnote{The IDL code of the WWCC method is publicly available at \url{http://hera.ph1.uni-koeln.de/~ossk/ftpspace/wwcc/}} is computed for two images $f$ and $g$ and their weighting functions, $w_f$ and $w_g$, through the following steps:

\begin{enumerate}
 \item Generate the weighted filtered maps for a given scale $l$, $F(\vec{x},l)$ and $G(\vec{x},l)$, using the 
 convolution of $f$ and $g$ maps with the filter (Eqs.~(\ref{eq:wf_wmap}), (\ref{eq:wavelet}) and
 (\ref{eq:wavelet_filters})).
 
 \item Compute the combined wavelet-filtered weights, $w_F$ and $w_G$ (Eq.~(\ref{eq:combined_weight})).
 
 \item Calculate the $\Delta$-variance (Eq.~(\ref{eq:delta_var})).

 \item Compute the WWCC function (the degree of correlations for all possible displacements) on a
  given scale in Fourier domain (Eqs.~(\ref{eq:fCw},\ref{eq:fFw},\ref{eq:fGw})). 
  
 \item Determine the wavelet cross-correlation coefficient and 
       wavelet displacement vector (Eqs.~(\ref{eq:cc-l}) and 
       (\ref{eq:ov}--\ref{eq:tau_phi}), respectively).

 \item Repeat the steps 1--5 for every scale, $l$.       
\end{enumerate}

\section{Testing the analytic power and limitations of the weighted wavelet cross-correlation}
\label{sec:test_wwcc}

In this section we test the WWCC method by applying it to various sets of well-defined artificial 
maps to answer three questions:
\begin{itemize}
\item{}Can we determine individual characteristic scales in the two maps?
\item{}Can we reproduce a given displacement of structures at particular scales?
\item{}Can we trace a systematic change of the scale sizes between the two maps?
\end{itemize}

To address these questions, we use two different basic sets of simulated noisy
maps. In one case, we assume Gaussian intensity profiles of a given 
size, representing individual smooth clumps. Then we have only one well-known
prominent scale in the two maps to be compared and we can test the detection
of a possible scale difference and a displacement of the structures.
The other set of data is given by self-similar \emph{fractional Brownian motion} 
structures \citep[\emph{fBm, }][]{peitgen88}. They do not have any dominant
scale, but are characterized by a power-law distribution of scales. They
can be described by a power law power spectrum of fractal structures and 
random phases in Fourier space. Here, we can test whether any scale
dependent modifications (smoothing, displacement, \dots) are recovered
by the WWCC. The two test cases (circular and fBm structures) should eventually represent
limiting {\changed cases for the actual observations of molecular clouds. In
one extreme view, maps are described as a collection of ``spherical blobs'',
in the other extreme they are described as scale-free, fully self-similar fractals.
In practice, they typically contain both aspects, i.e. we find both self-similar
structures} and individual characteristic resolved structures.

For these test cases, we investigate the {\changed CC coefficient and displacement 
vector when changing the properties of the artificial clouds, such as the
size of the individual structures, the spectral index of the self-similar
scaling law,} and the assumed noise level of
the observations {\changed to test their sensitivity to those properties}. 
The  map size of all test data sets is $128 \times 128$ pixels 
(abbreviated as pix throughout the rest of the paper). 

\subsection{Structures having Gaussian intensity profiles}
\label{sec:test_wwcc_cs}

We test the WWCC method for  circular structures having Gaussian intensity
profiles and some spatial displacement to study the effect of the size of circular
structures and the noise level on the correlation coefficient and the measured
displacement vector as a function of scale.

The Gaussian intensity profile is generated with an amplitude equal to unity and
width given by standard deviation $\sigma$. A weight (significance value) of unity is 
assigned to each data point (pixel) in the map.

\begin{figure}
   \centering
   (a)\includegraphics[width=3.1cm, angle=90]{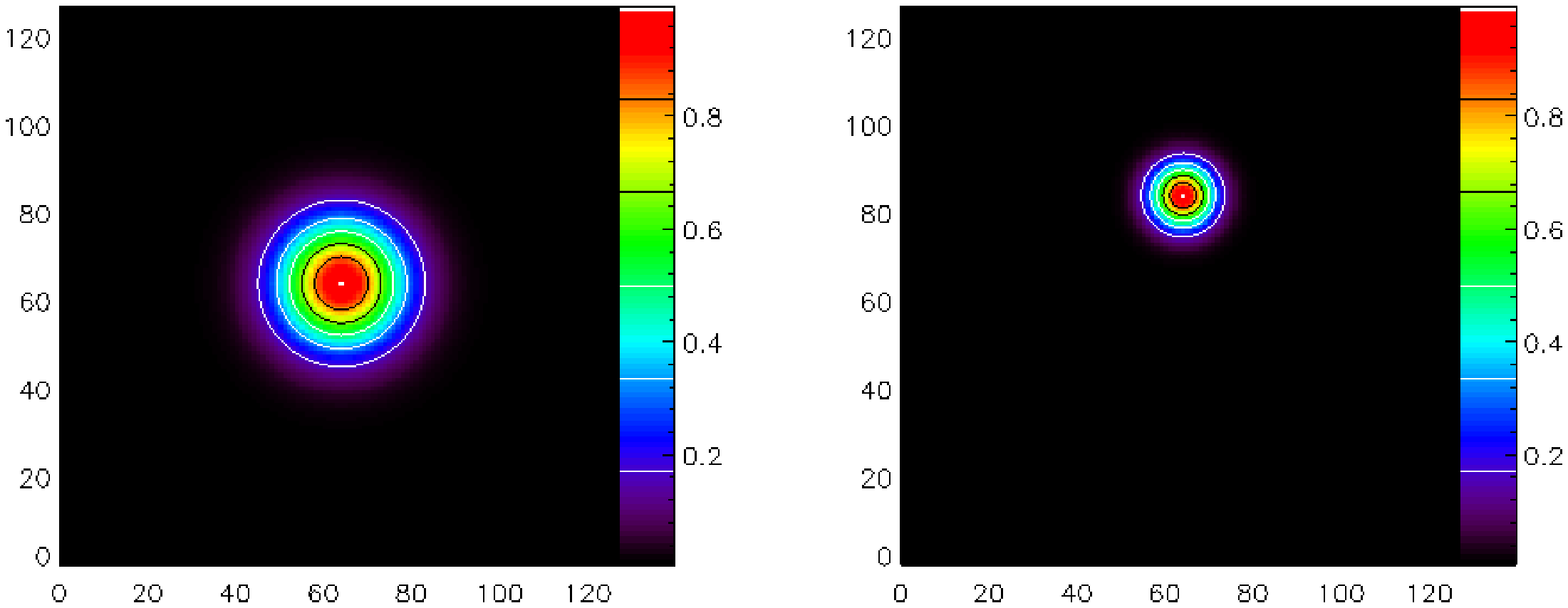}\\
   (b)\includegraphics[width=3.1cm, angle=90]{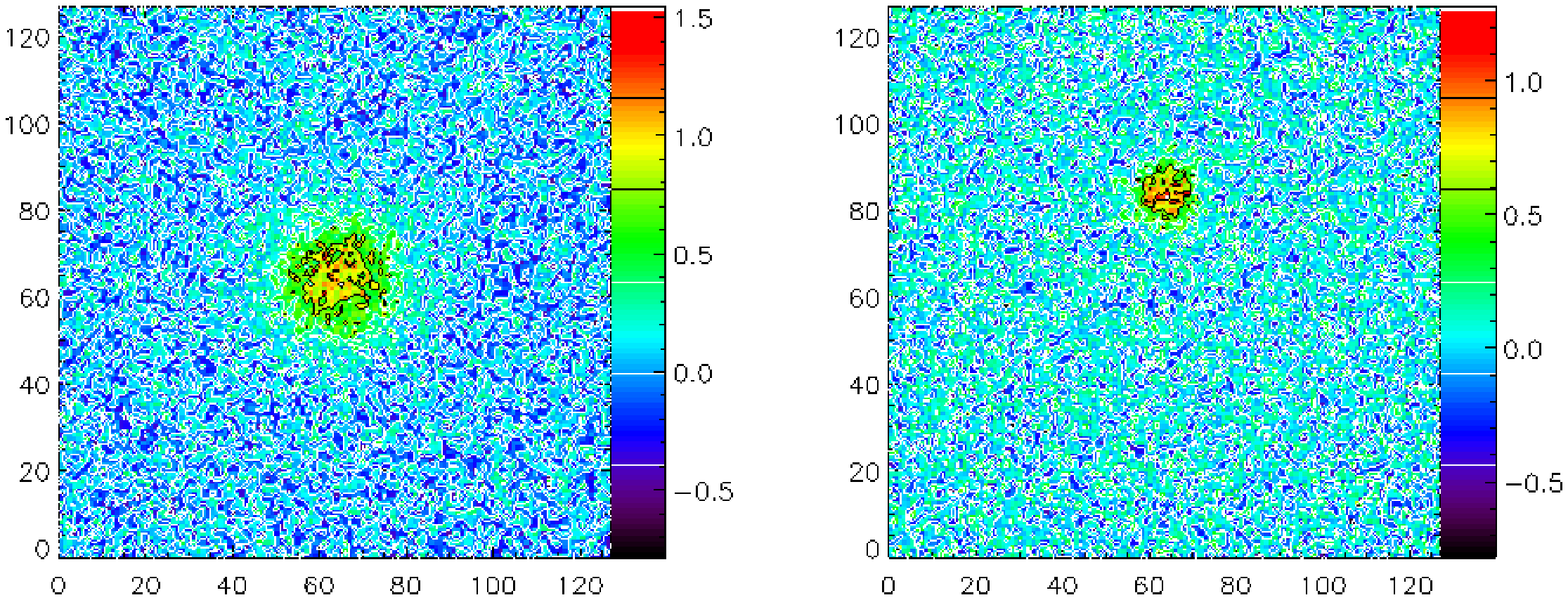}
     \caption{ (a) Maps of two circular Gaussian structures ($\sigma_1=10$ pix 
      and $\sigma_2=5$ pix) shifted by 20 pix along $y$-axis. (b) Observed with a simulated  
      $\mathrm{S/N}=5$.}
         \label{fig:gm_maps}
\end{figure}
 
In Fig.~\ref{fig:gm_maps} we show two pairs of test maps given by circular Gaussian 
structures (with $\sigma_1=10$ pix and $\sigma_2=5$ pix and a displacement of the peak
by $\vec{\tau}_i=20$ pix along $y$-axis {\changed in the second map). As the 
cross-correlation (Eqs.~(\ref{eq:ccf}) and (\ref{eq:wwccf})) only considers the relative
offset of structures between two maps, $\vec{t}$, the result is invariant with respect to an exchange of
negative offsets in map 1 by positive offsets in map 2. The WWCC always characterizes the
mutual displacement of structures within the two considered data sets. Here, we always
displace the structure in the second map relative to the first one.}
The upper panels show the original structure
representing a $\mathrm{S/N}=\infty$, the lower panels are superimposed by white noise simulating
typical observations with $\mathrm{S/N}=5$, i.e. $\sigma_{\rm noise}=0.2$. 

\begin{figure}
   \centering
   (a)\includegraphics[width=3.1cm,angle=90]{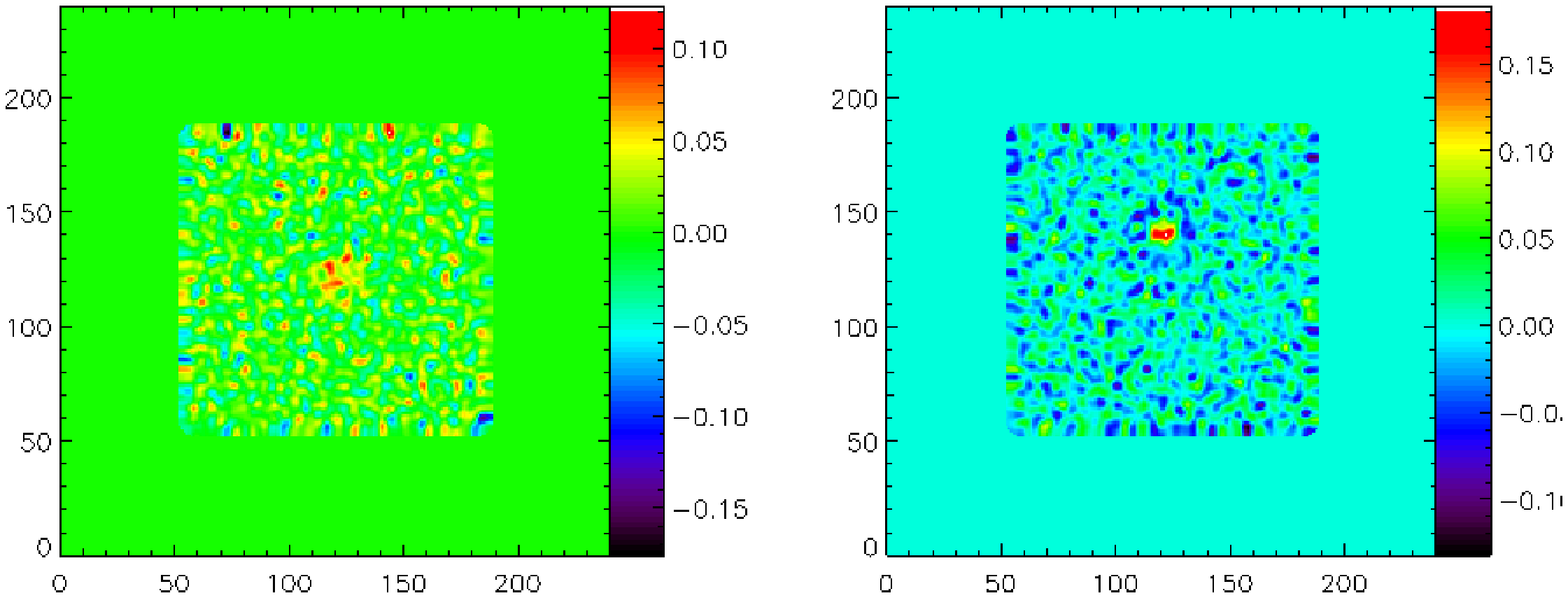}
   (b)\includegraphics[width=3.1cm,angle=90]{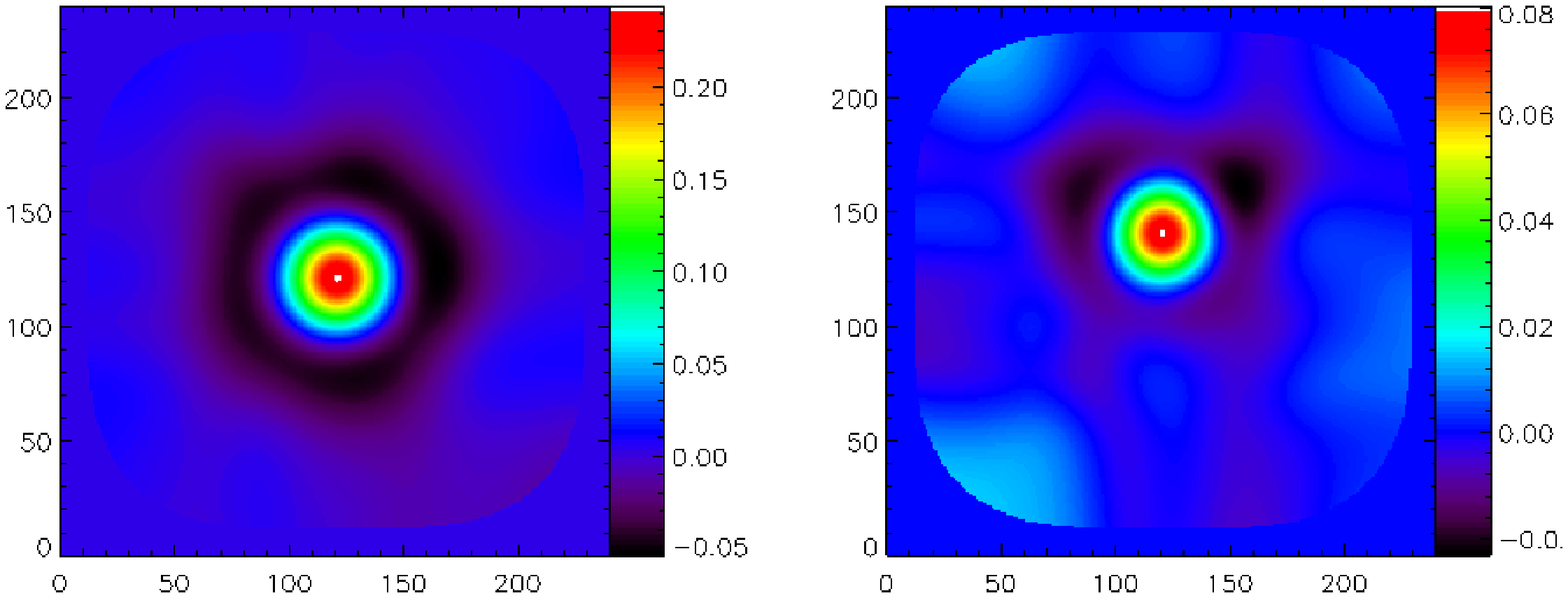}     
   \caption{Maps of the two Gaussians with $\mathrm{S/N}=5$ from Fig.~\ref{fig:gm_maps}b filtered on scale of 5 pix (a) 
    and of 45 pix (b). To allow for a filtering in Fourier
    space, {\changed the maps are extended beyond the original boundaries before filtering.} 
    The extension is treated with $w_F=0$ in the computation.
    }
   \label{fig:filtered}
\end{figure}

\subsubsection{Map filtering}

 The first step in the analysis is the wavelet convolution 
(Eq.~(\ref{eq:wtransform}))
filtering the maps on a given scale. {\changed The usable range of the 
filter sizes is inherited from the $\Delta$-variance analysis discussed 
in Sect.~\ref{sec:delta_var}. It starts at a resolution of two pixels.} 
Fig.~\ref{fig:filtered} shows the result,
$F(\vec{x},l)$ and $G(\vec{x},l)$ for the $\mathrm{S/N}=5$ case for two selected wavelet filter sizes
$l=5$ pix and $l=45$ pix. The filtered maps are larger than the original ones to 
allow for a convolution with non-truncated wavelets (see O08). The extension is treated
with zero weighting in the further computation. We see that the maps filtered on
small scales (5 pix) are dominated by noise, while the structures filtered on large scales (45 pix) 
are basically free of noise.  
Statistically, the noise will dominate the cross correlation for all scales $l$
below a critical size. This size depends on signal to noise level and will be discussed
later in this section.

\begin{figure}
   \centering
   \includegraphics[width=6.cm, angle=90]{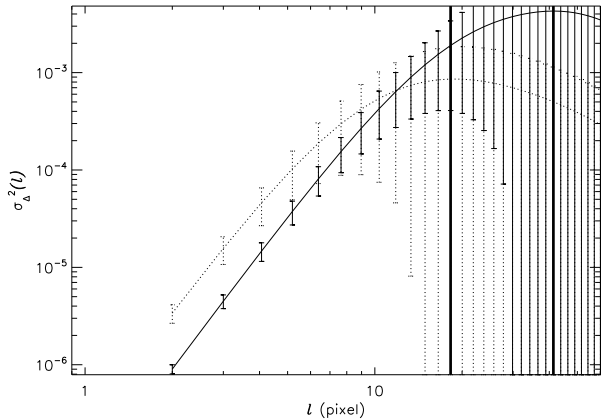}
      \caption{ $\Delta$-variance spectra of the two Gaussians with $\mathrm{S/N}=\infty$ from 
      Fig.~\ref{fig:gm_maps}\,(a). The $\Delta$-variance spectra of the large (left) 
      and small (right) circular structures (Fig.~\ref{fig:gm_maps}\,(a)) are denoted 
      by full and dotted lines respectively. Vertical thick lines are used 
      to denote the measured prominent scales of the large and small circular 
      structures, respectively.} 
         \label{fig:gm_delta1}
\end{figure}

\subsubsection{The $\Delta$-variance spectrum}
\label{sect:dvar_spectrum}

\begin{figure}
   \centering
   \includegraphics[width=6.cm, angle=90]{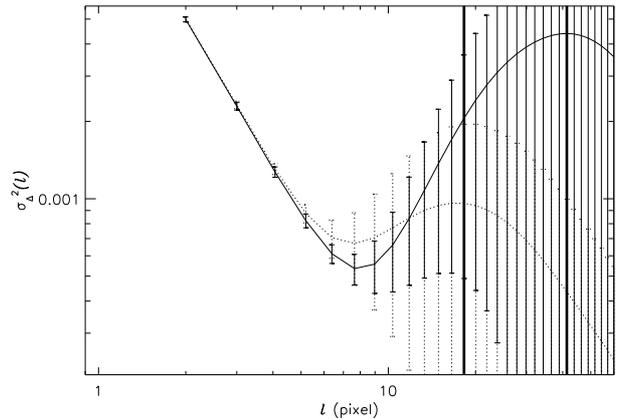} 
      \caption{ $\Delta$-variance spectra of the two Gaussians with $\mathrm{
         S/N}=5$ from Fig.~\ref{fig:gm_maps}\,(b). The $\Delta$-variance spectra of the
         large (left) and small (right) circular structures (Fig.~\ref{fig:gm_maps}\,(b)) 
         are denoted by full and dotted lines respectively. Vertical thick lines denote 
         the measured prominent scales of the large and small circular structures, respectively. }
         \label{fig:gm_delta2}
\end{figure}

Computing the variance of the wavelet-filtered maps for all filter sizes leads to the 
$\Delta$-variance spectra ($\sigma^2_{\Delta}(l)$; Eq.~(\ref{eq:delta_var})). Fig.~\ref{fig:gm_delta1} 
shows the resulting spectra for the original maps without noise and Fig.~\ref{fig:gm_delta2}
the spectra for the case with $\mathrm{S/N}=5$.  
The $\Delta$-variance characterises the amount of structure distributed on different spatial scales. 
The peaks of $\Delta$-variance spectra indicate the dominant structure size.
For structures without additional small scale contributions, the 
$\Delta$-variance falls off like $l^4$ below the dominant size. This is well seen
in Fig.~\ref{fig:gm_delta1} for scales below about 8~pix.
The increasing statistical uncertainty of the $\Delta$-variance
for larger scales due to the lower number of statistically independent entities of large 
scales in the maps practically limits the use of the $\Delta$-variance analysis to prominent sizes of less than
half of the map size (O08). The huge formal error bars in Fig.~\ref{fig:gm_delta1}
following O08 are irrelevant here, as they characterize the counting statistics of
individual structures, being very bad for a single structure in an
otherwise empty map, while every observed map typically contains much
more structure.
 
The $\Delta$-variance detects the prominent scales of the circular structures at $l_{\rm p1}=41$ 
pix and $l_{\rm p2}=18$ pix (vertical lines in Fig.~\ref{fig:gm_delta1}). 
This corresponds approximately to the diameter of the {\changed circles visible in colors in 
Fig.~\ref{fig:gm_maps}\,(a), i.e. at a level above 10\,\% of the peak intensity. For isotropic
structures, the prominent scale approximately identifies the visible size of the
structure \citep{MLO}.} We find a fixed relation between the prominent scale and the
standard deviation of a circular structure $l_{\rm p} \approx 4\sigma$.

When we compute the $\Delta$-variance spectra of the same structures simulated with 
$\mathrm{S/N}=5$ (Fig.~\ref{fig:gm_delta2}) we find a superimposed structure on 
scales less than 8 pix which is due completely to noise. The noise contribution decreases with 
$\sigma^2_{\Delta}(l) \propto l^{-2}$ \citep[see][]{bensch01} up to $\approx 8$ pix 
and becomes negligible towards larger
scales as discussed for the map filtering above, leading to a $\Delta$-variance minimum at a scale of 
about 8 pix here. At scales above the minimum the spectra behave identical to Fig.~\ref{fig:gm_delta1} 
and reach the maxima at scales of 41 pix and 18 pix.

\begin{figure}
   \centering
   (a)\includegraphics[width=6.5cm,angle=90]{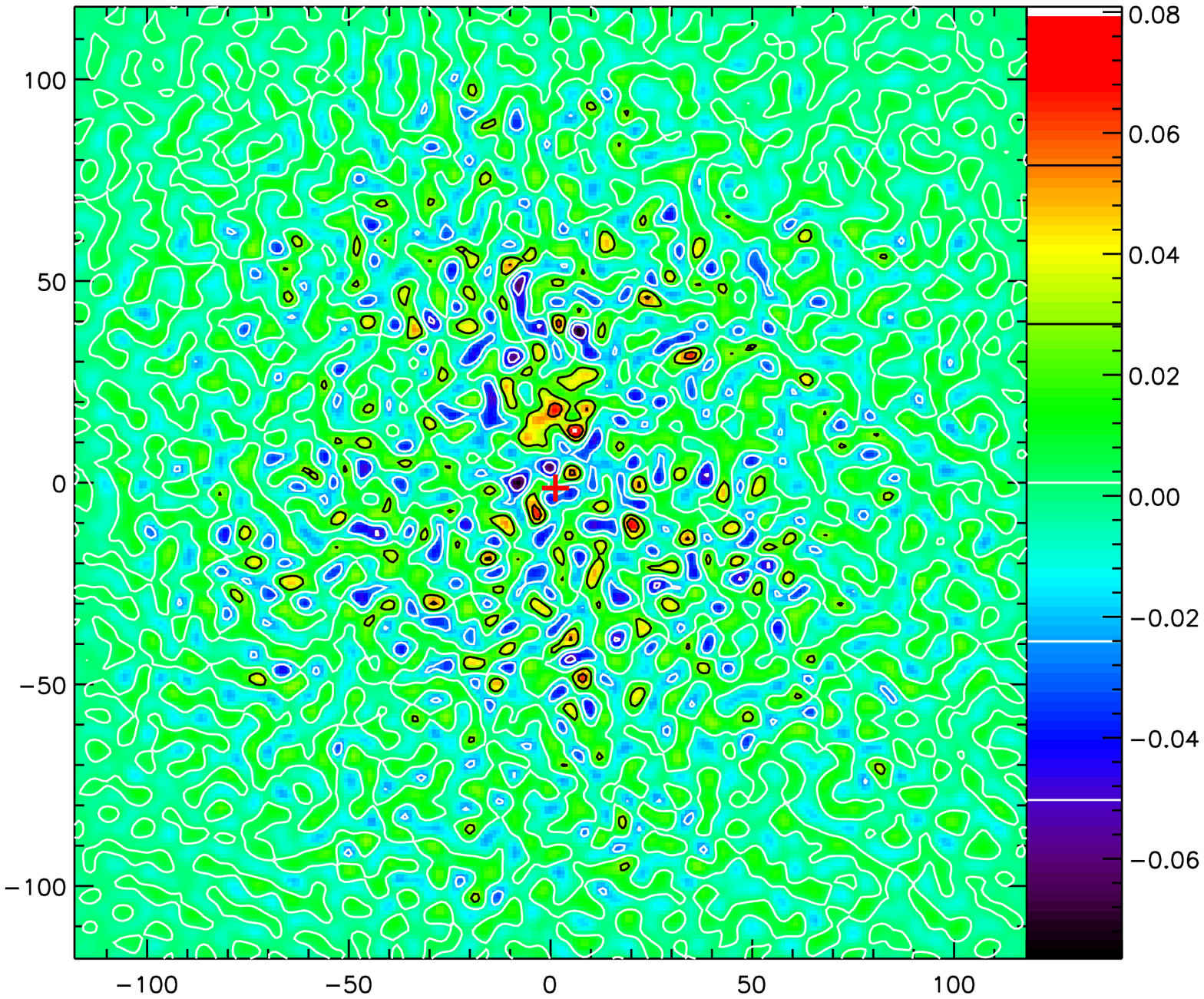}
   (b)\includegraphics[width=6.5cm,angle=90]{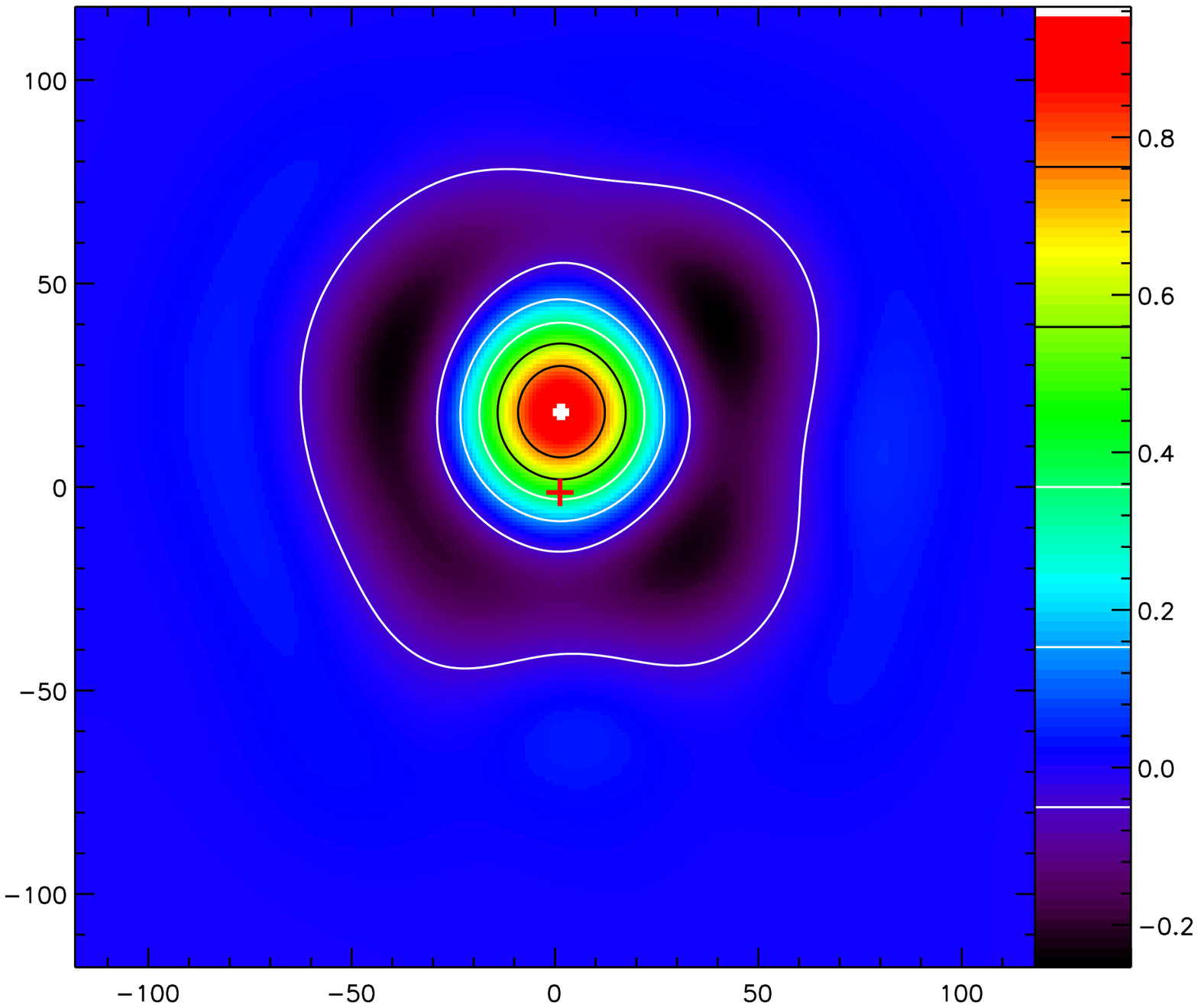} 
   \caption{Cross-correlation functions for the pairs of wavelet-filtered maps from Fig.~\ref{fig:filtered}.
   The top panel (a) shows $C_w({\vec t},5$pix), the bottom panel (b) $C_w({\vec t},45$pix).
   The red crosses denote the {\changed center of the offset plane $\vec t=0$, where one can read the
   wavelet cross-correlation coefficient $r(l)=C_w(\vec t=0,l)$. The peak of the cross-correlation
   functions is displaced from the center by $\tau(l)=\tau_i$ as the WWCC recovers the input
   displacement of the structure in the second map.} }
   \label{fig:wwccf}
\end{figure}

\subsubsection{The WWCC function}
\label{sect_wwcc_gauss} 

The WWCC map is finally computed through Eq.~(\ref{eq:fCw}) 
providing the correlation for all possible displacements on a given scale.
It is then used to compute the CC coefficient and displacement 
vector as a function of scale. 
The center of the WWCC map represents the correlation of the two structures not correcting for any displacement.
It gives the cross-correlation coefficient for the given scale (Eq.~(\ref{eq:cc-l})). 
The dominant offset between the two structures in the maps is measured as the
displacement vector between the center and the peak position of the WWCC
function (Eq.~(\ref{eq:ov})). 

This is demonstrated in Fig.~\ref{fig:wwccf} for the pairs of maps
wavelet-filtered on scales of 5 pix and 45 pix from Fig.~\ref{fig:filtered}.
The black crosses indicate the map center, i.e. a zero displacement vector. The offset of 20 pixels in $y$-direction 
entered in the simulations is easily recovered as the location of the peak in
the case of the 45-pix filter. For the maps filtered on a 5 pixel scale,
the uncorrelated noise structures are strong so that one cannot unambiguously
distinguish the maximum of the WWCC at the correct offset from other maxima caused by 
the noise. As a consequence, the position of the 
correlation peak can not be reliably determined in this case. This prevents the determination 
of the displacement vector on noise-dominated scales less of than about 8 pixels. 
\begin{figure}
   \centering
   (a)\includegraphics[width=6.cm, angle=90]{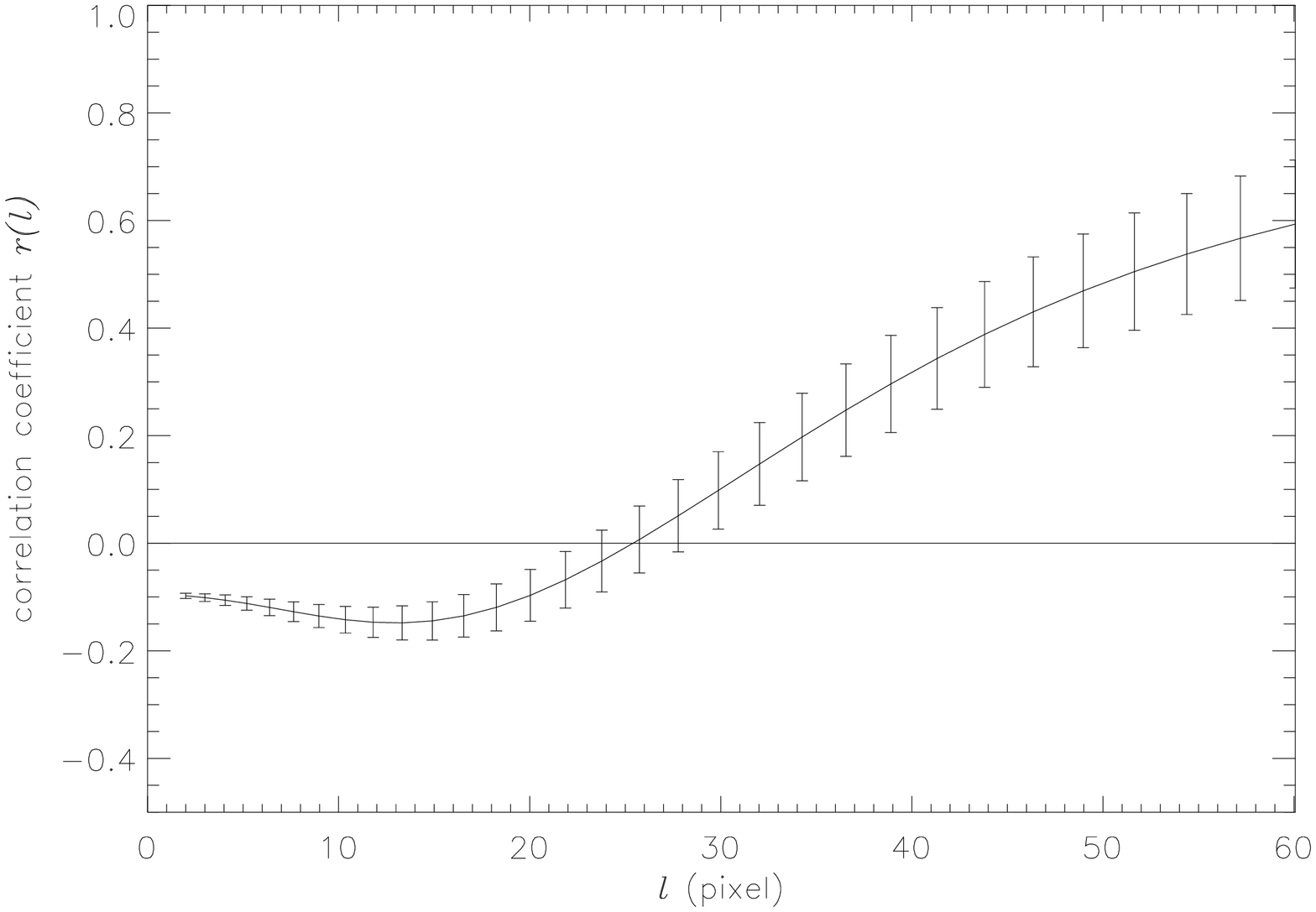} \\ 
   (b)\includegraphics[width=6.cm, angle=90]{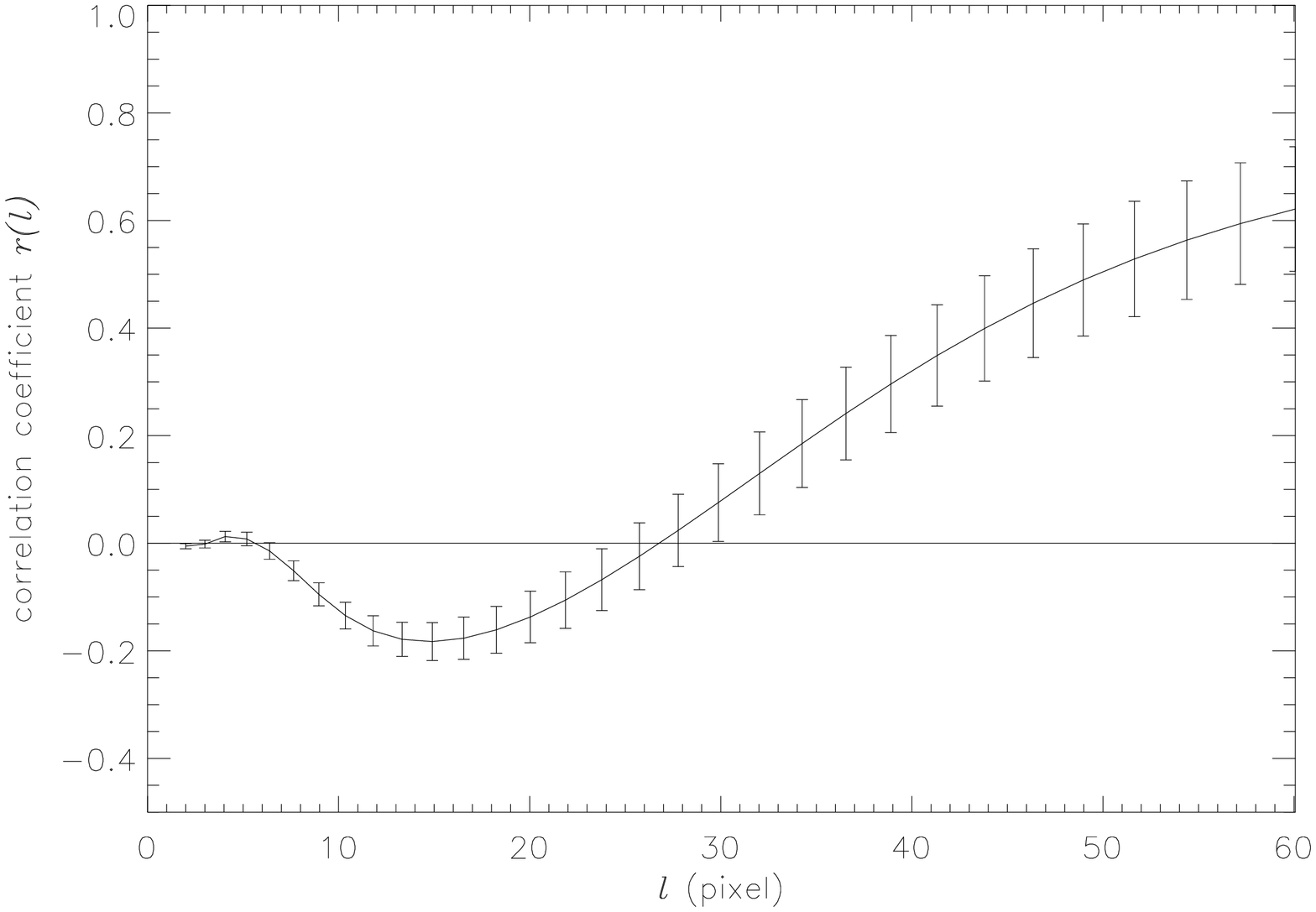}

      \caption{Correlation coefficient $r=C({\vec \tau}=0)$ as a function of scale
      for the maps from Fig.~\ref{fig:gm_maps}. (a) represents the pure Gaussians,
      (b) the maps with noise resulting in $\mathrm{S/N}=5$.
      }
      \label{fig:gm_cc0}
\end{figure}

\paragraph{Cross-correlation spectrum} ~\\

The CC coefficients are extracted from the central pixel of the WWCC function 
(Eq.~(\ref{eq:cc-l}), crosses in Fig.~\ref{fig:wwccf}) for all possible wavelet-filter sizes, $l$. 
In Fig.~\ref{fig:gm_cc0} (a,b) we show the dependence of the 
CC coefficient $r(l)$ on the scale for the circular structures with $\mathrm{S/N}=\infty$ and $\mathrm{S/N}=5$.
As the two circles in the example are clearly adjacent, the cross correlation is negative at 
small scales, showing the expected anticorrelation below the mutual shift of 20 pixels
and up to $\sim 23$ pix.
The correlation curve has a shallow minimum at $\approx 15$ pix
and monotonically increases up to 0.6 at large scales where both structures
are strongly blurred through the wavelet filter, so that they turn statistically similar.
As the offset is comparable to the diameter of the smaller circle $l_{\rm p}$, the 
coefficient remains well below unity even at large scales.
In case of $\mathrm{S/N}=\infty$, the anti-correlation is detected at all small scales, in the
case of $\mathrm{S/N}=5$ (Fig.~\ref{fig:gm_cc0}\,(b)) the correlation vanishes at 
scales $l \la 5$ pix due to the uncorrelated small-scale noise in 
the two maps. 

\begin{figure}
   \centering
   \includegraphics[width=6.2cm, angle=90]{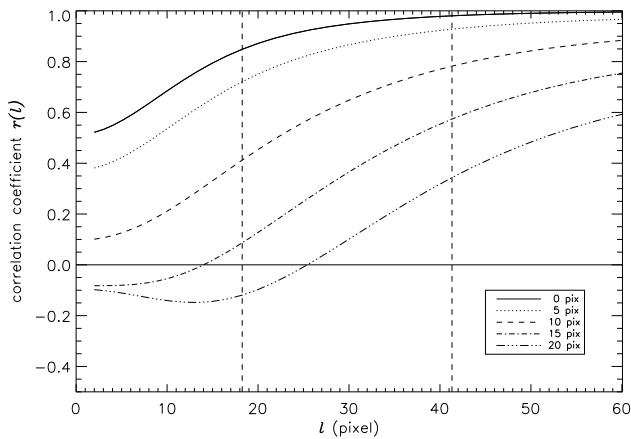}
      \caption{Correlation coefficient as a function of scale for two circular structures ($\sigma_1=10$ pix 
      and $\sigma_2=5$ pix) shifted by 0, 5, 10, 15, and 20 pix (full line, dotted, dashed, dash-dotted, 
      and dash-dot-dot-doted lines, respectively). The dashed vertical 
      lines denote the prominent scale of the two circular structures. 
      }
         \label{fig:cc-l_gauss}
\end{figure} 

There are two main parameters shaping $r(l)$, 
one is the displacement between the structures and one is the ratio of prominent 
scales of these structures, $l_{\rm p2}/l_{\rm p1}$.
To show the effect of the displacement on the correlation coefficient $r(l)$, 
we compute the WWCC for pairs of circular structures ($\sigma_1=10$ pix and $\sigma_1=5$ pix)
mutually displaced by offsets of 0, 5, 10, 15, and 20 pix (Fig.~\ref{fig:cc-l_gauss}). 

The correlation obviously becomes stronger on all scales 
if the displacement between the structures is smaller. 
An anti-correlation occurs whenever if the 
mutual shift of the two structures is larger than 1/3 of the size of the
bigger one, almost independent of the size of the smaller Gaussian. For smaller
displacements, the correlation remains positive at small scales.
Without anticorrelation we always find a monotonic increase to large scales. 
When the offset between circular structures increases the CC 
coefficient is reduced on all scales and turns to negative on small scales.

\begin{figure}
   \centering
   \includegraphics[width=6.2cm, angle=90]{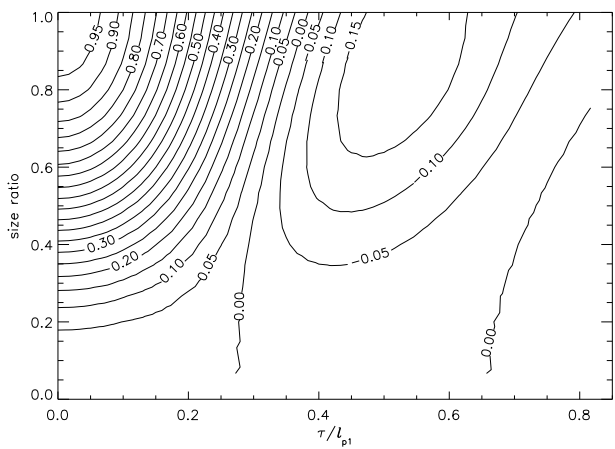}
      \caption{Contour diagram of the small-scale CC coefficient, $r(2\,{\rm pix})$, as a 
      function of prominent size ratio of circular structures ($l_{\rm p2}/l_{\rm p1}$) and 
      displacement normalised to the largest prominent scale ($\tau_i/l_{\rm p1}$). Here we 
      used $\mathrm{S/N}=\infty$ and $l_{\rm p1}=60$ pix.
      }
      \label{fig:r0_contour}
\end{figure}

To quantify the behaviour of the correlation on small scales we examine the changes of the  
CC coefficient at the smallest wavelet-filter size, $r(2\,\mathrm{pix})$, as a function
of the size ratio of the circles and their normalised displacement in Fig.~\ref{fig:r0_contour}.
The structures correlate strongly if their prominent sizes are comparable 
and the offset between them is small. The correlation weakens with decreasing the size ratio or 
increasing the offset. The correlation is about zero along the line 
$l_{\rm p2}/l_{\rm p1} \approx 0.48\tau_i/l_{\rm p1} - 0.21$ and it turns to 
negative for larger displacements reaching a minimum at 
$\tau_i/l_{\rm p1} \approx 0.5$ and $l_{\rm p2}/l_{\rm p1} \approx 1$.

\begin{figure}
   \centering
      \includegraphics[width=6.2cm, angle=90]{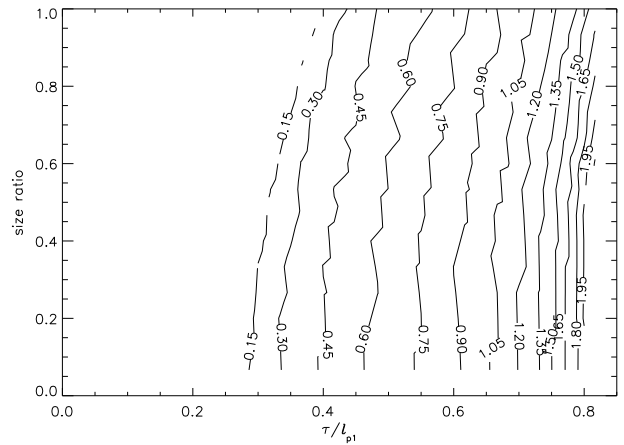}
      \caption{Contour diagram of the normalised cross correlation root $l_{\rm CC_0}/l_{\rm p1}$ as a 
      function of prominent size ratio of circular structures ($l_{\rm p2}/l_{\rm p1}$) and 
      displacement normalised to the largest prominent scale ($\tau_i/l_{\rm p1}$).
      $\mathrm{S/N}=\infty$ and $l_{\rm p1}=60$ pix.
      }
      \label{fig:l0_contour}
\end{figure}

We can also characterize the anticorrelation due to displacement through
the parameter $l_{\rm CC_0}$, defined as the scale at which the correlation
coefficient $r(l)$ turns from negative to positive, i.e.  
$r(l_{\rm CC_0})=0$. In Fig.~\ref{fig:l0_contour} we show the dependence 
of $l_{\rm CC_0}$ on the relative displacement $\tau_i/l_{\rm p1}$ and the size ratio of
the Gaussians $l_{\rm p2}/l_{\rm p1}$.
The scale of the cross-correlation root depends strongly on the offset, but hardly 
on the size ratio of the structures. For offsets below $\tau_i/l_{\rm p1}\approx 0.3$
no root is found as the cross-correlation function is positive even at small 
filter sizes. For offsets larger than $\approx 0.8 l_{\rm p1}$, i.e.
displacements comparable to the prominent scale of the largest structure, the
cross-correlation function remains negative on all scales.
At $\tau_i \approx 0.7 l_{\rm p1}$ we find $l_{\rm CC_0} \approx l_{\rm p1}$.

\begin{figure}
   \centering
   (a)\includegraphics[bb=27 119 288 730, width=3.4cm, angle=90]{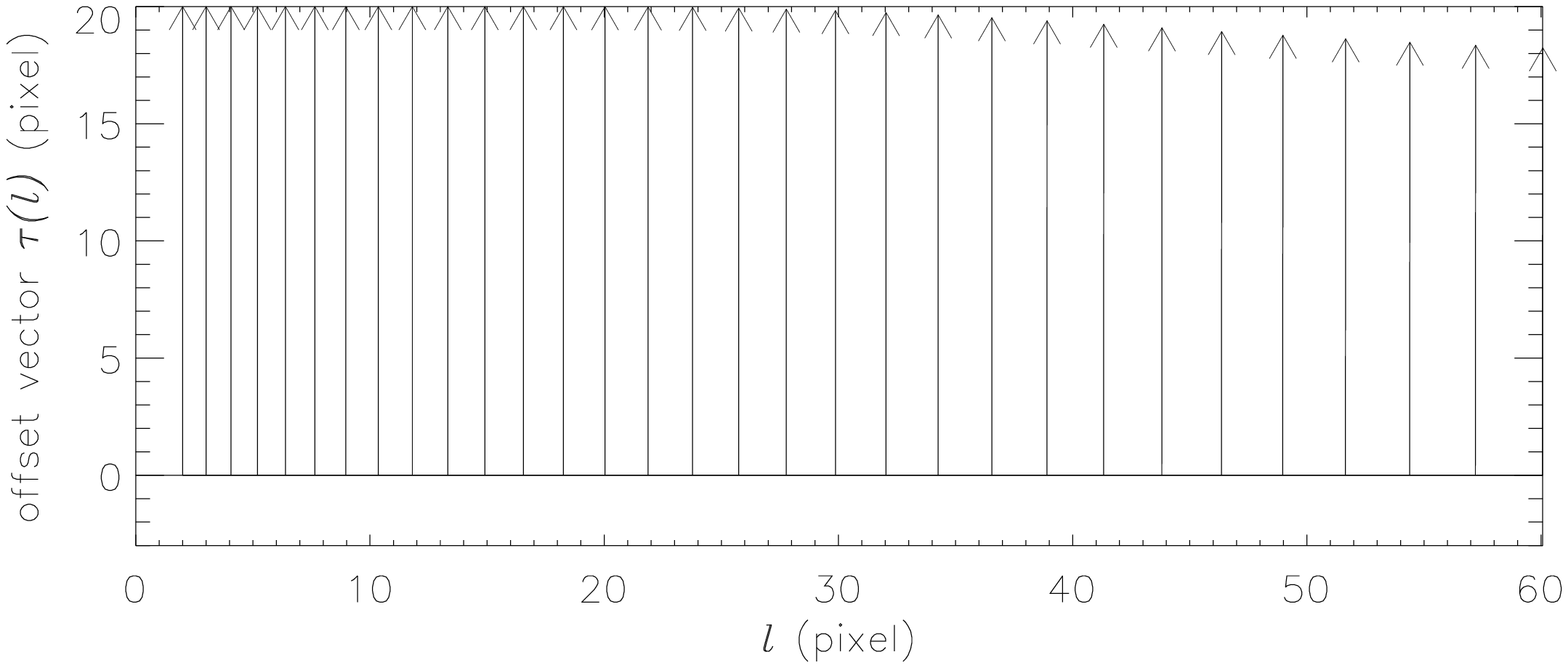}\\
   (b)\includegraphics[bb=27 119 288 730, width=3.4cm, angle=90]{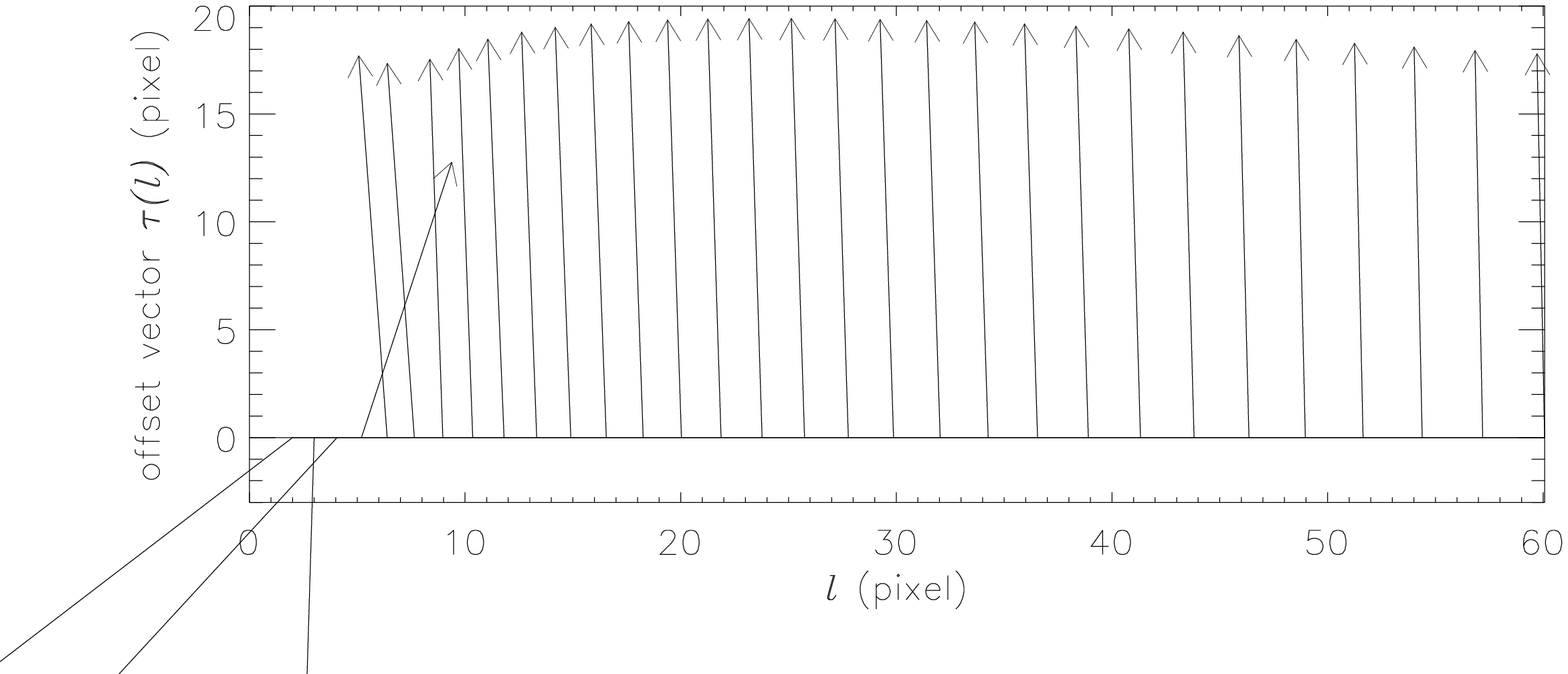}    
      \caption{Displacement vector as a function of scale for the pairs of Gaussian maps from
      Fig.~\ref{fig:gm_maps}. (a) represents the results for pure Gaussians, (b) for the maps
      with $\mathrm{S/N}=5$. The vectors are shown in the $x-y$-plane for individual discrete lags
      indicated on the $x$-axis at the origin of the plotted vectors. 
      }
         \label{fig:gm_dis}
\end{figure}

\paragraph{The spectrum of displacement vectors} ~\\

Finally, we examine the spectrum of displacement vectors, $\vec\tau(l)$, measured for all 
wavelet-filter sizes. The displacement vector gives the coordinate
of the peak of the WWCC relative to the central pixel (Eq.~(\ref{eq:ov})).
Fig.~\ref{fig:gm_dis} shows the recovered displacement vectors between circular Gaussian 
structures for $\mathrm{S/N}=\infty$ and $\mathrm{S/N}=5$. In case of infinite signal-to-noise 
(Fig.~\ref{fig:gm_dis}\,(a)), the vectors are accurately recovered at scales less 
than the $l_{\rm p1}\la 40$ pix.
At larger scales the displacement vector is slightly underestimated by up to two pixels. 
This is due to the finite map size truncating parts of the filtered circles that fall beyond 
the map boundaries. More and more of the filtered structure shows up beyond the 
map edges when going to larger wavelet-filter sizes (see Fig.~\ref{fig:filtered}) so 
that for the remaining part within the map boundaries the mutual shift 
appears smaller than the applied offset. The coinciding finite map boundaries of 
both maps impose a correlation for larger wavelet-filter sizes. This mimics a
somewhat lower displacement of the structures and limits the applicability of the
WWCC at large scales (see below) thereby affecting maps with low spatial dynamic range.

The spectrum of the displacement vectors for noisy maps ($\mathrm{S/N}=5$; Fig.~\ref{fig:gm_dis}\,(b)) shows 
completely random results for the noise dominated scales below 6 pixels where the correlation 
coefficient is close to zero. Up to the minimum of the $\Delta$-variance spectrum at 8 pixels, 
the displacement vectors remain unreliable. At scales between 8 pix and 12 pix, where the noise 
is still strong, but not dominant any more, we can recover the displacement vector with an 
accuracy $\lesssim 15$~\%. Up to 12~pixels, the cross correlation
coefficient is also dominated by noise, rather than actual structures in the map
(see Fig.~\ref{fig:gm_cc0}\,(b)). For all scales above 12~pixels, the displacement vector is 
recovered with an accuracy better than 10\,\%. 

While the finite maps size poses an upper 
limit to the usable scales, the observational noise thus poses a lower limit to the scales 
for which we can determine the correct structure offset.

\subsubsection{Limiting scales}
\label{sect:limiting-cases}

\begin{figure}
   \centering
   \includegraphics[width=7.2cm, angle=-90]{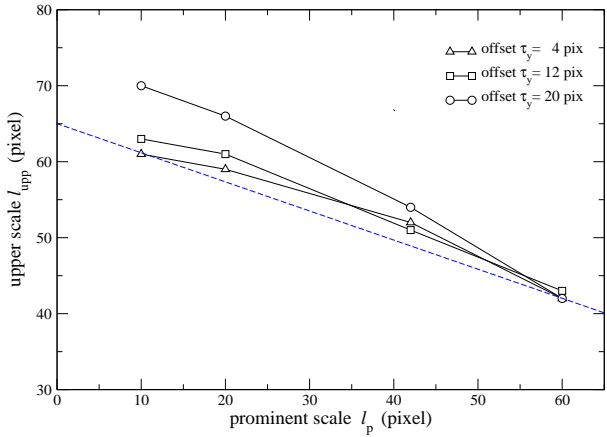}
      \caption{Upper limit for the displacement scale vs. prominent scale estimated for seven pairs of      
      equal size circular structures (having $\sigma$ = 2.5, 5, 10, 15, 20, 25 and 30 pix) offset
       by $\tau_y$ = 4, 12, and 20 pix.
      }
         \label{fig:l_P-l_D}
\end{figure} 

To systematically quantify the limits for the accurate determination
of the spectrum of displacement vectors we study the minimum 
$l_{\rm low}$ and maximum $l_{\rm upp}$ scales for which the
displacement vector can be recovered with an accuracy better than 10\,\%.

\paragraph{The upper limit for the scale, $l_{\rm upp}$} ~\\

Here, we consider the edge effects for different sizes of circular 
structures and different positions near the edge of the map.
We use seven pairs of equal size Gaussian clouds simulated with 
$\mathrm{S/N}=\infty$ and $\sigma$ = 2.5, 5, 10, 15, 20, 25, and 30 pix 
($l_{\rm p}$ = 8, 19, 40, 60, 80, 100, and 120 pix).  We
displace each pair of maps in $y$ direction, 
by $\tau_i = 4, 12,$ and 20 pix thus moving the structures closer to the map edges. 
For every given offset $\tau_i$, we compute the spectra of displacement
vectors for the seven pairs and determine
the upper limit $l_{\rm upp}$ where the measured spatial displacement $\vec{\tau}$ 
is recovered with an accuracy better than 10\,\%. This is performed by evaluating
a circle of radius $\delta\tau=0.1 \times |\vec{\tau}_i|$ around the input displacement
vector and testing whether the recovered vector falls into that circle.

The maximum scale $l_{\rm upp}$ at which the displacement vector can be recovered 
with 10\,\% accuracy depends on the truncation of the wavelet-filtered circles at the
map edges. In Fig.~\ref{fig:l_P-l_D} we show the measured limiting scale $l_{\rm upp}$ 
for a number of offsets and prominent scale sizes. We find a general decrease of 
the limiting scale $l_{\rm upp}$ with increasing prominent structure size.
For structures with prominent scales $\lesssim 1/6$ of the map size ($l_{\rm p} \lesssim 20$ pix)
the displacement vector can be accurately measured up to scale $l_{\rm upp}$ of about half 
the map size. For structures with larger prominent scales ($\gtrsim 50$ pix) a reliable
measurement of the displacement is only possible on smaller scales.
The boundary effects strongly limit the usable scales. 
The truncation of the structure itself by the finite size of the map
will result in an underestimation of the displacement at large scales.
Circular structures with sizes comparable to the map size 
($> 100$ pix) already fall beyond the map boundaries when offset by 12-20 pix,
i.e. the offset can only be measured in a small range of scales 
$l_{\rm upp} \la 10$ pix. However, these are scales that may be dominated
by noise already. 

To guarantee a reliable determination of the displacement, we fit the
lower limit of the upper scale $l_{\rm upp}$ as a function of the
prominent scale $l_{\rm p}$ by the linear relation
\begin{equation}
  l_{\rm upp} = -0.38 \,l_{\rm p} + 64 {\rm pix} = -0.38 \,l_{\rm p} + l_{\rm map}/2 .
   \label{eq1:lupp-lp}  
\end{equation}
shown as dashed line in Fig.~\ref{fig:l_P-l_D}.
This relation can be used to constrain the $l_{\rm upp}$ of circular structures 
whenever the displacement does not shift the prominent structures off the
map boundaries.

\paragraph{The lower limit for the scale, $l_{\rm low}$} ~\\

In Fig.~\ref{fig:gm_dis}\,(b) we have seen that for the $\mathrm{S/N}=5$ example
the CC coefficient and displacement vector can not be recovered at scales
below the scale of the minimum of the $\Delta$-variance spectrum $l_{\rm min}=8$
pix (Fig.~\ref{fig:gm_delta2}). 

\begin{figure}
   \centering
   \includegraphics[width=7.2cm, angle=-90]{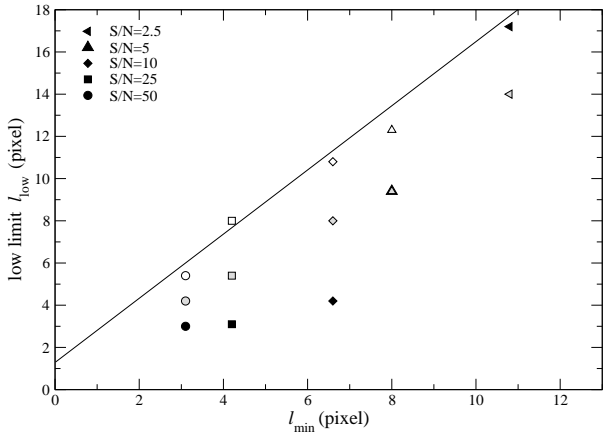}
      \caption{Lower limit for the displacement scale ($l_{\rm low}$) as a function
      of the $\Delta$-variance minimum scale ($l_{\rm min}$) for $\mathrm{S/N}=2.5, 5, 10, 25,$ and 50,
      and displacements of $\tau_i = 4, 12, 20$ pix (black, gray, and open symbols).
      An upper limit to the lower scale is described by the solid line: 
      $l_{\rm low} = 1.5\, l_{\rm min}+1.5$.   
      }
         \label{fig:llim-lmin}
\end{figure}

Figure~\ref{fig:llim-lmin} shows the lowest scale $l_{\rm low}$ 
for which the displacement vector can be retrieved with an accuracy better than 10\,\%
as a function of the  scale $l_{\rm min}$ for a number of different noise levels and displacements,
using equal size circular structures. We find a considerable scatter of 
the limit for the reliable determination of the displacement vector when
considering different offsets, but we can give a safe upper limit to 
$l_{\rm low}$ by
\begin{equation}
   l_{\rm low} = 1.5 (l_{\rm min}+1).   
   \label{eq:llow-lmin}
\end{equation} 
shown as solid line in Fig.~\ref{fig:llim-lmin}.

The range $l_{\rm min} \lesssim l \lesssim l_{\rm low}$ thus characterizes scales
where the noise does no longer hide the underlying structure -- as for
scales $l < l_{\rm min}$ --  but where it still  significantly influences the 
cross-correlation spectra and the measured displacement vectors.
 
\begin{figure}
   \centering
   \includegraphics[width=7.2cm, angle=-90]{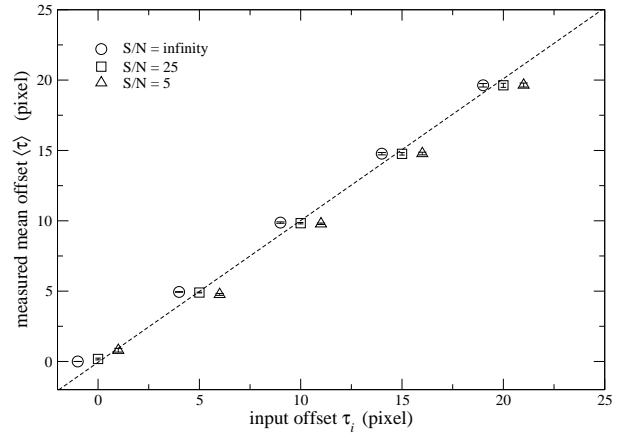}
      \caption{Measured mean displacement ($\langle{\tau}\rangle$) vs. given 
      displacement ($\tau_i$) for $\mathrm{S/N}=\infty$ (circles), $\mathrm{S/N}=25$ (squares), 
      and $\mathrm{S/N}=5$ (triangles). The circles and triangles are displaced by 1 pixel from the identity line
      (dashed line) for a better visibility. The error bars represent $1\sigma$ error of 
      the mean offset.}
      \label{fig:goff-soff}
\end{figure} 

To finally verify the reliability of the recovery of the displacement vector
within the limits given by Eqs.~(\ref{eq1:lupp-lp}) and (\ref{eq:llow-lmin}),
we restrict the spectrum of displacement vectors to that range when 
evaluating the results for 315 pairs of maps covering all combinations of 
Gaussian widths $\sigma$ = 1, 4, 7, 10, 13, and 16 pix ($l_{\rm p}$ = 4, 16, 28, 40,
52, and 64 pix), input offsets of $\tau_{i,\,y}=0, 5, 10, 15$, and 20 pix, and 
S/N values of $\infty$, 25, and 5. Fig.~\ref{fig:goff-soff} shows the 
measured  mean displacement $\langle{\tau}\rangle$ 
against the given displacement $\tau_i$ for the ensembles of 21 pairs having the same
offset and noise level. The standard deviation of the derived displacements
is very small so that the plotted error bars basically fall within the size
of the symbols so that they are hard to be seen.

Overall we find a very good agreement with a minimal scatter, proving the
method to work, but a slightly overestimated displacement at $\tau_i=0$ pix 
for high noise ($\mathrm{S/N}=5$, triangle) and a slightly underestimated displacement
for the largest offsets at all noise levels, due to the finite map size
effects. The deviations fall, however, well below one pixel size.
Hence, we conclude that the cutoff of the spectrum of the displacement vectors 
by low and upper limits allows us to recover accurate offsets, i.e.
we can reliably determine the displacement vector between two structures 
$\vec\tau(l)$ in the full range between $l_{\rm low}$ and $l_{\rm upp}$.

\subsection{Structures of fractional Brownian motion}
\label{sec:test_wwcc_fbm}

In contrast to the circular Gaussian structures that have a single
well-defined prominent scale, maps of the interstellar medium
often show a self-similar behavior without individual prominent
scales. They can be represented by the fractal structure of
\emph{fractional Brownian motion} maps. They are characterized by
a power-law distribution of scales and are periodic by generation
in Fourier space. We use a spectral index $\zeta=3$ representing a 
typical mean value observed in the cold interstellar medium \citep{Falgarone2007}. 
As for the circular Gaussian structures we evaluate how well the WWCC analysis
can detect scales enhanced in one structure compared to the other
and the displacement of structures at particular scales
within the two maps.

\subsubsection{fBm structures displaced on different scales} 
\label{sect_fbm_displaced}

In real clouds, structures can be offset with respect to each other   
on different spatial scales. This occurs, e.g., in different velocity channel maps of
line observations where the large scale structure is affected by 
systematic motions but the small scale structure by turbulent
motions, in dust emission maps at different wavelengths where
local temperature gradients create small scale displacements
independent of the possible large scale structure, or in maps
of different chemical tracers where incident UV radiation from 
one direction creates a chemical gradient on scales not mixed
by turbulent flows.
 
\begin{figure}
   \centering
   \,\,\,\includegraphics[width=3.1cm, angle=90]{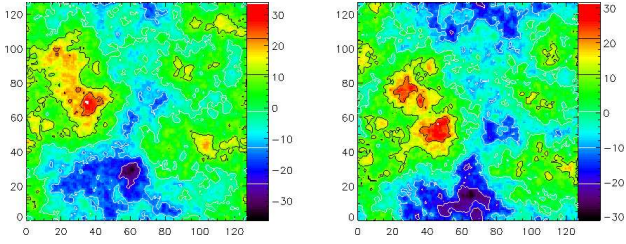}
      \caption{fBm map with $\zeta=3$ (left panel) and  the same map 
      after displacing all structures on scales $l \ge 10$ pix by $\tau_{i,x}=10$ 
      pix and $\tau_{i,y}=-15$ pix (right panel). This pair is to mimic an 
      interstellar cloud structure that could be created by the superposition 
      of small scale turbulence and a large scale velocity gradient.}
      \label{fig:fBm2m_maps}
\end{figure}

\begin{figure}
   \centering
   \includegraphics[width=6.cm, angle=90]{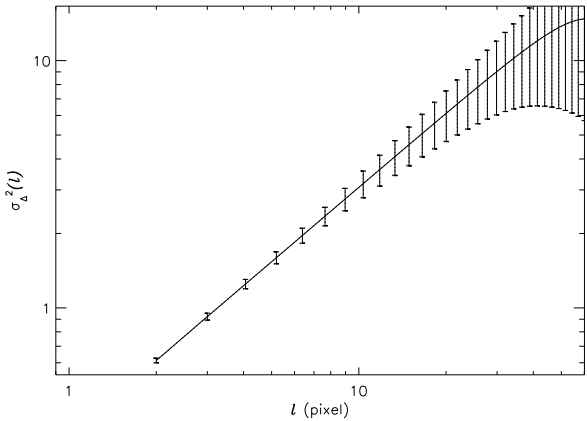}
      \caption{$\Delta$-variance spectra for the maps from
      Fig.~\ref{fig:fBm2m_maps}.}
         \label{fig:fBm2m_dvar}
\end{figure}

To test this, we start from an fBm map with a spectral index $\zeta=3$
and $\mathrm{S/N}=\infty$ (Fig.~\ref{fig:fBm2m_maps}, left panel), and
shift all {\changed structures with} scales larger than 10 pixels by $|\vec{\tau}_i|=18$ pix 
($\tau_{i,x}=10$ pix and $\tau_{i,y}=-15$ pix) using the Fourier shift
theorem (Eq.~(\ref{eq_shift_theorem})). The resulting map is shown in 
Fig.~\ref{fig:fBm2m_maps} (right panel). The spatial shift of structures
does not affect the relative contribution of structures as a function
of their size, so that both maps have an identical $\Delta$-variance spectrum,
shown in Fig.~\ref{fig:fBm2m_dvar}. It is characterized by a perfect power-law
with an exponent $\zeta - 2 = 1$, representing the fully self-similar scaling
up to the largest scales where the structures are limited by the available map size.

\begin{figure}
   \centering
   (a)\includegraphics[width=6.cm, angle=90]{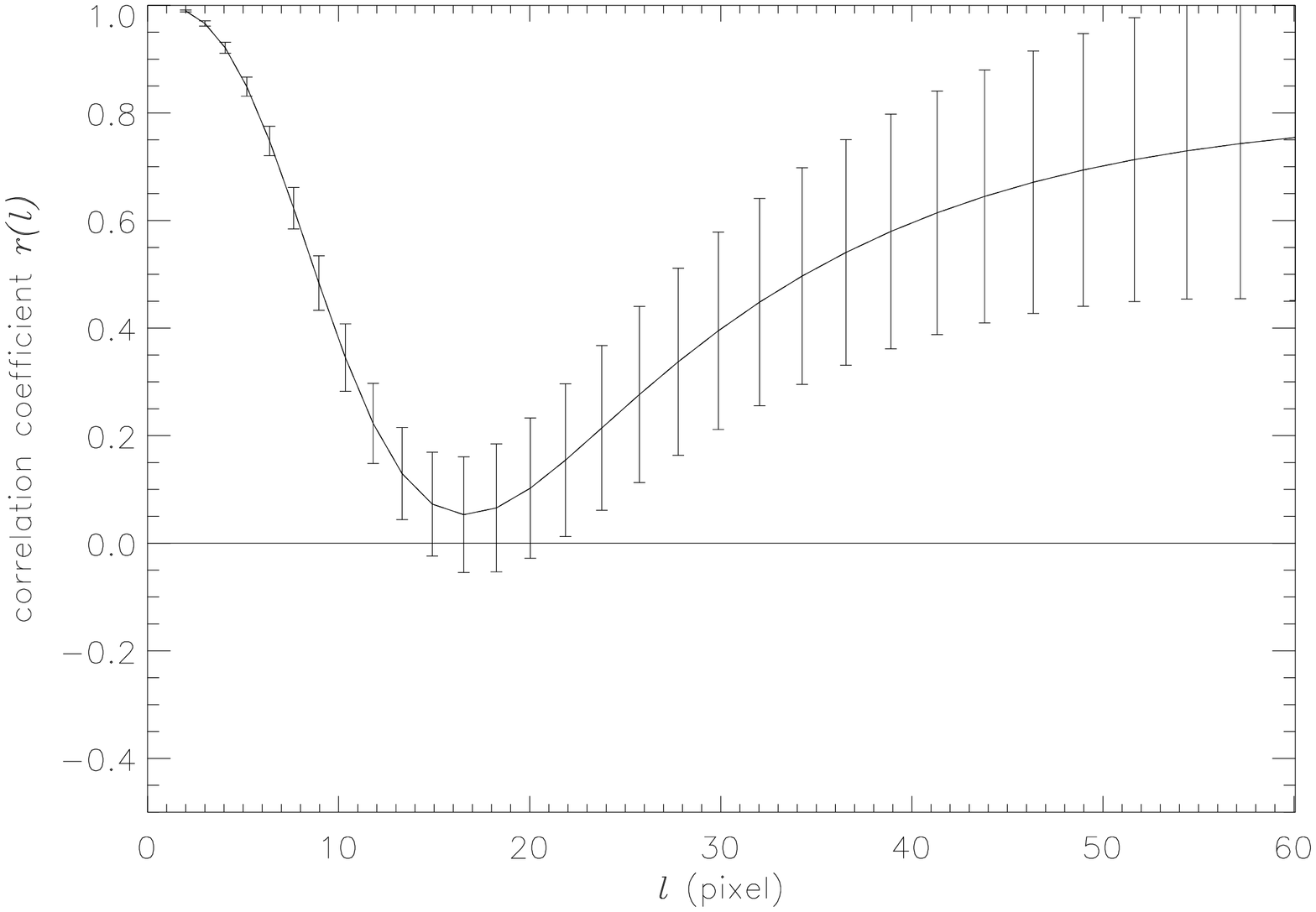}
   (b)\includegraphics[bb=27 34 288 780, width=3.24cm, angle=90]{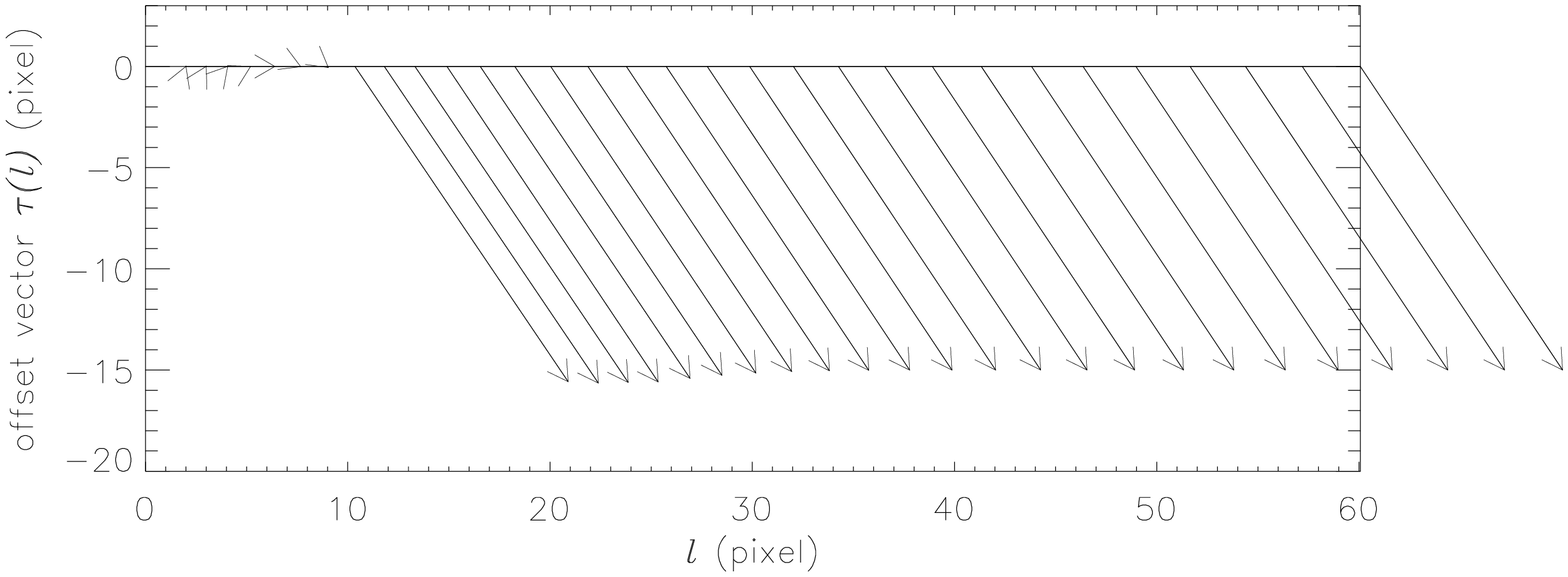} 
      \caption{Results of the WWCC for the pair of maps from Fig.~\ref{fig:fBm2m_maps}:
      (a) correlation coefficient and (d) displacement 
      vector as a function of scale. $1\sigma$ error bars are given.}
         \label{fig:fBm2m}
\end{figure}

Figure~\ref{fig:fBm2m} shows the results of the WWCC for the comparison of both
maps. The correlation
coefficient is unity at small scales (because the structures are not offset) 
and gradually decreases to a minimum at 17--18 pix due to displacement by 18 pix between
the structures (Fig.~\ref{fig:fBm2m}\,(a)). On large scales the structures are 
blurred through the wavelet filter to sizes
that exceed the displacement amplitude. This leads to monotonic
increase of the correlation there. The displacement vector is
perfectly recovered for all scales, being zero on scales $l < 10$ pix
and $\tau_x=10$ pix, $\tau_y=-15$ pix for larger scales
(Fig.~\ref{fig:fBm2m}\,(b)). 

\begin{figure}
   \centering
   (a)\,\,\,\includegraphics[width=3.1cm, angle=90]{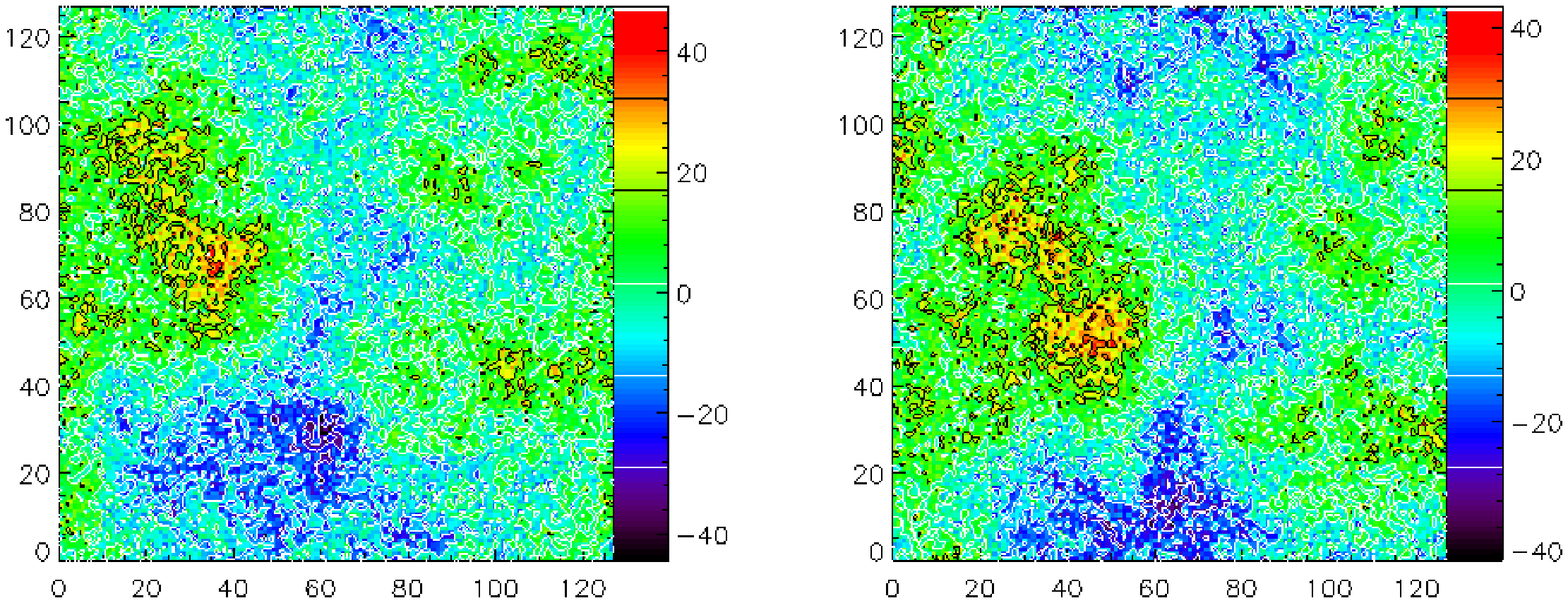}
   (b)\includegraphics[width=6.cm, angle=90]{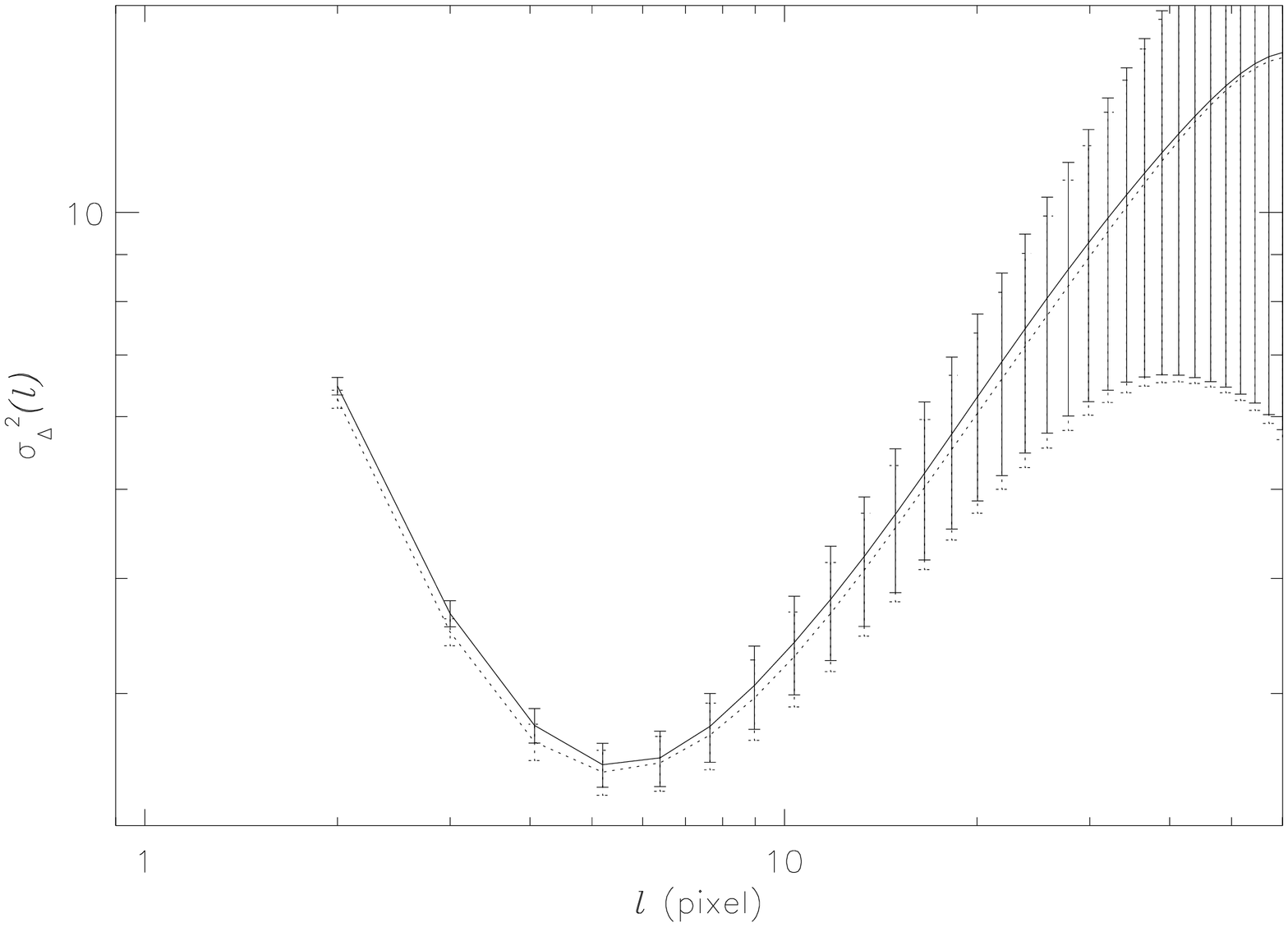}
   (c)\includegraphics[width=6.cm, angle=90]{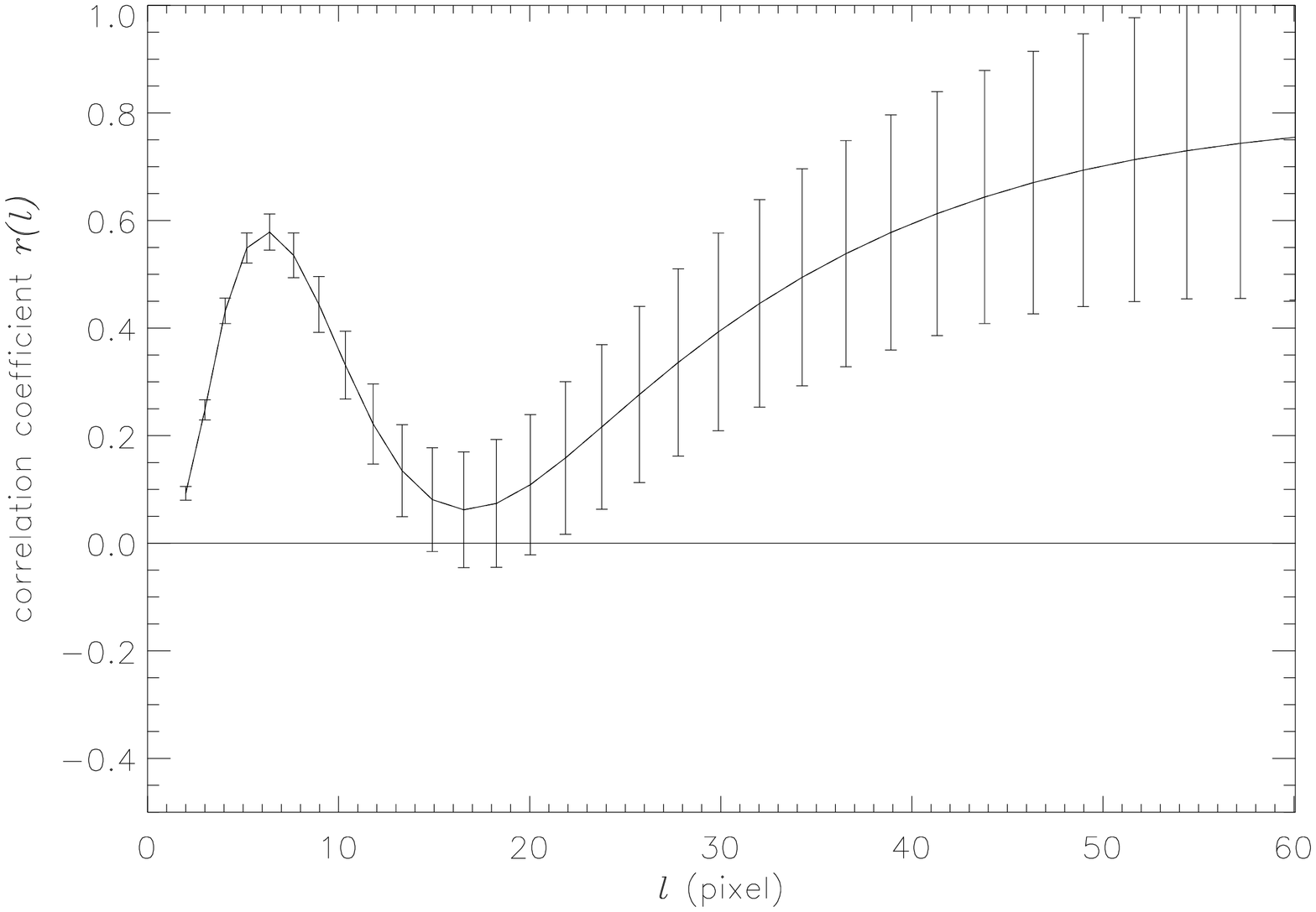}
   (d)\includegraphics[bb=27 34 288 780, width=3.24cm, angle=90]{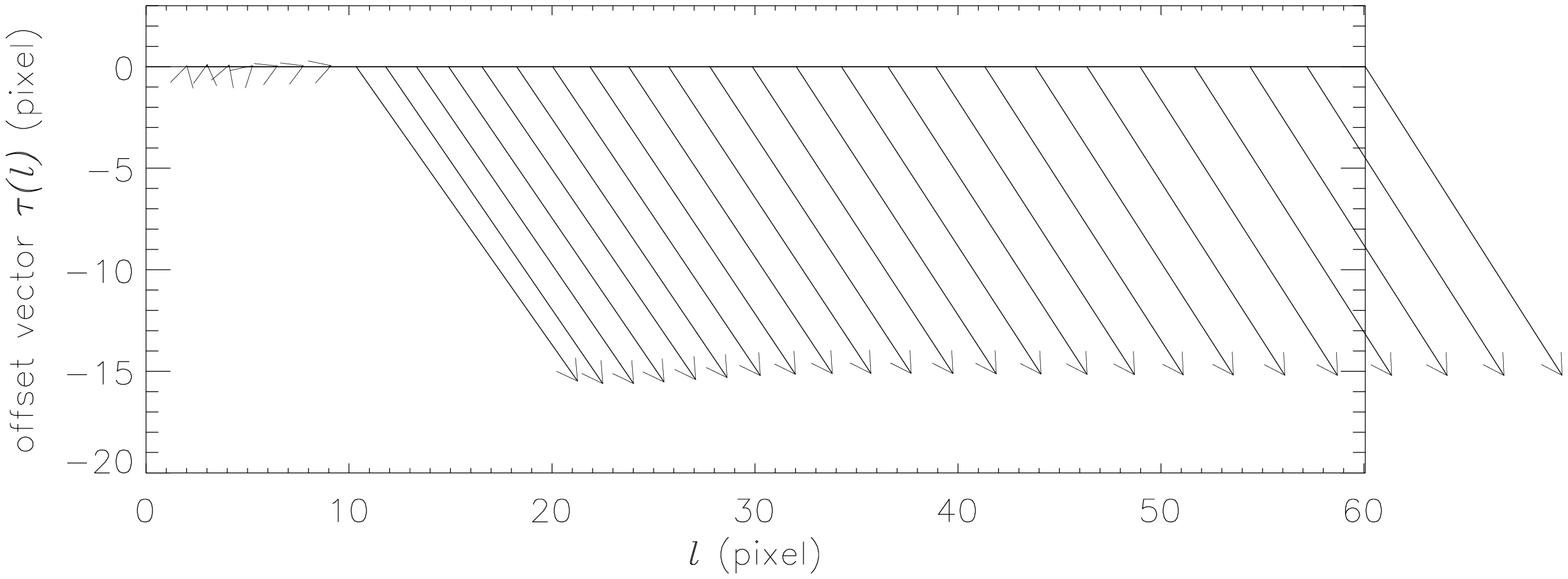} 
      \caption{(a) Original fBm map generated with $\mathrm{S/N}=5$ (left) and the same fBm map 
      displaced by $\tau_{i,x}=10$ pix and $\tau_{i,y}=-15$ pix on scales $l \ge 10$ pix (right).  
      (b) $\Delta$-variance of the original and 
      displaced fBm maps, (c) correlation coefficient and (d) displacement 
      vector as a function of scale. $1\sigma$ 
      error bars are presented.}
         \label{fig:fBm2mn}
\end{figure}

Fig.~\ref{fig:fBm2mn} shows the corresponding results when adding noise to the maps.
The noise level of $\mathrm{S/N}=5$ appears as an increased $\Delta$-variance 
at small scales ($l < 8$ pix, panel (b)). It lowers the 
CC spectrum on scales $l < 8$ pix, but has no impact on
the measurement of the displacement vector. As the map is filled with
structures on all scales, the impact of the noise is much lower here 
than for the sparsely populated maps inhabiting {\changed a single Gaussian
circular structure from Sect.~\ref{sec:test_wwcc_cs}. To ensure that
we actually find a smooth transition between the two extreme
cases of the single Gaussian circular structure and the fBm map, we also
checked the impact of the noise on recovery of the displacement spectrum
in case of multiple Gaussian structures. When we compare two maps each having three
randomly located Gaussian circular structures with the same characteristics
as in Fig.~\ref{fig:gm_maps} (b) and measure the lower limit for the displacement
scale we obtain $l_{\rm low}=5$ pix ($l_{\rm min}=7.7$ pix), i.e. a value
that is considerably smaller than that for the single Gaussian structures
($l_{\rm low}=12$ pix, see open upward triangle in Fig.~\ref{fig:llim-lmin}).
For pairs of maps inhabiting 30 Gaussian structures the impact of the noise 
further decreases. We obtain $l_{\rm low}=4$ pix ($l_{\rm min}=6.3$ pix) 
confirming that the noise dominates smaller and smaller scales if the maps
are filled with more circular structures. This is in line with a
negligible noise impact for the fBm maps filled with structures on
all scales (Figs.~\ref{fig:fBm2m}\,(b) and \ref{fig:fBm2mn}\,(d)).} 

\paragraph{Identification of correlated structures} ~\\

\begin{figure}
   \includegraphics[width=3.cm,angle=90]{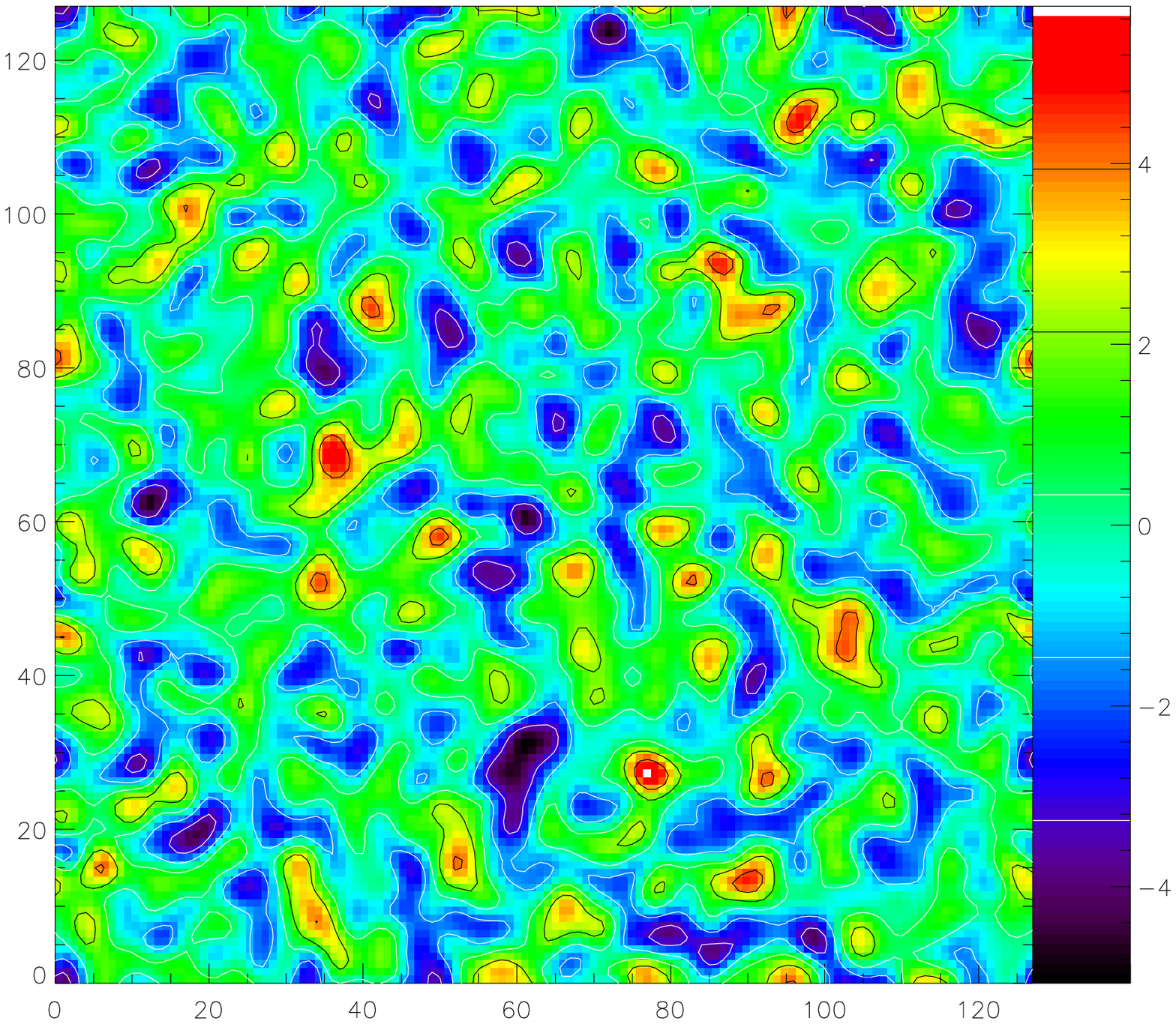}
   \includegraphics[width=3.cm,angle=90]{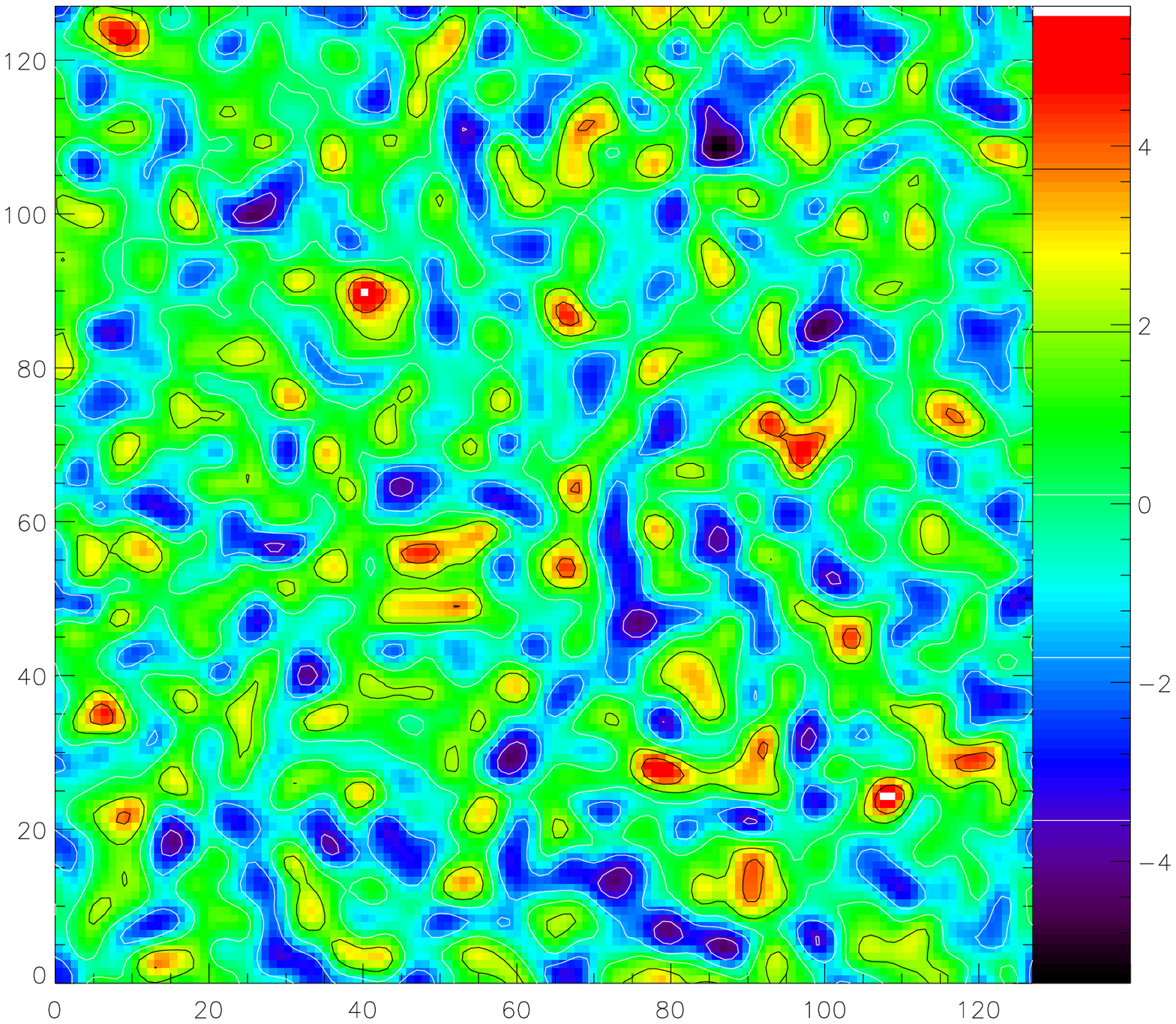}
   \includegraphics[width=3.cm,angle=90]{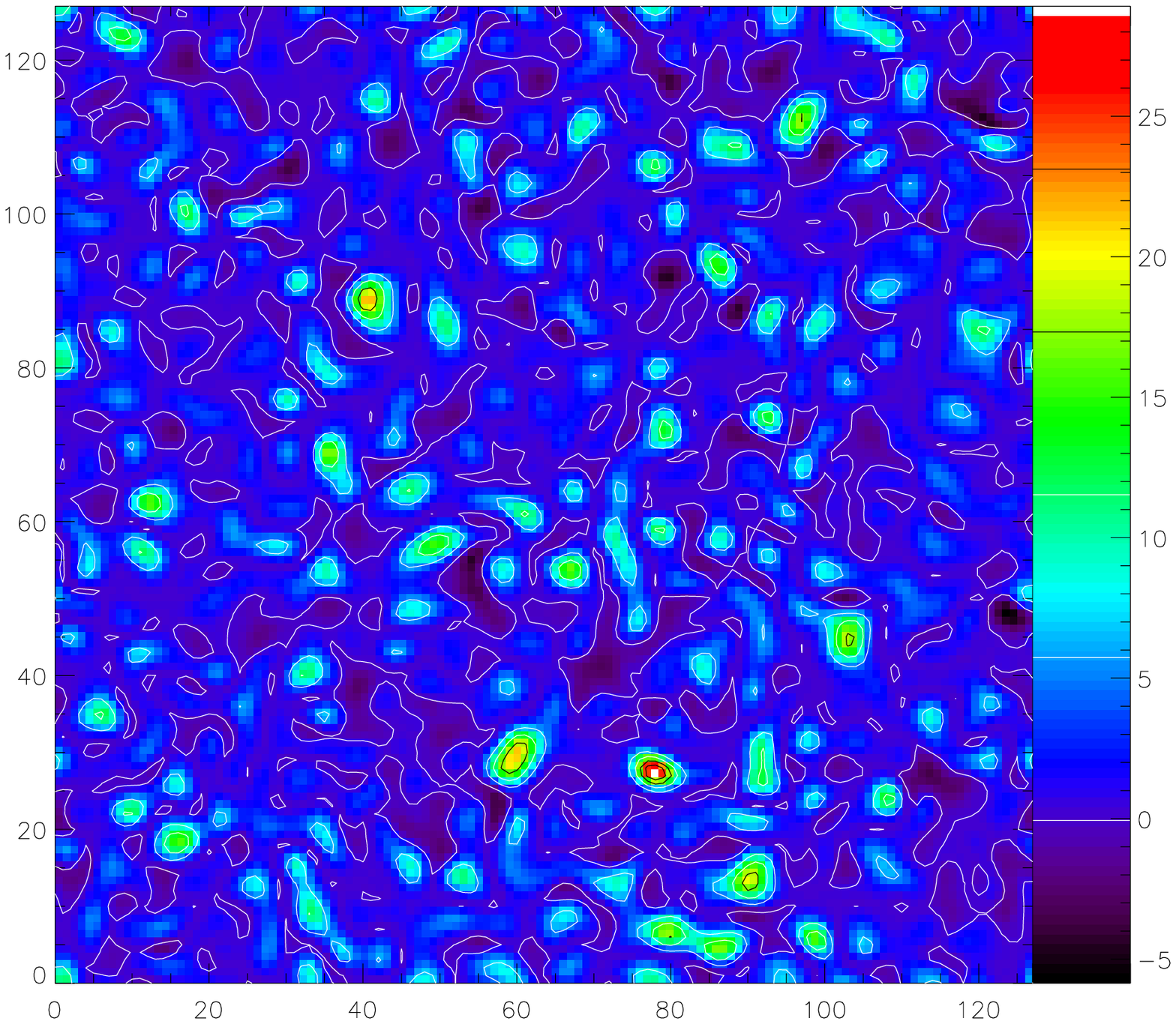}
      \caption{\emph{Top left.} The fBm map (Fig.~\ref{fig:fBm2mn}\,(a), left) filtered on scale of 6 pix. 
       \emph{Top right.} The fBm map displaced on scales $> 10$ pix (Fig.~\ref{fig:fBm2mn}\,(a), right) 
       filtered on scale of 6 pix.
       \emph{Bottom left.} The product map generated by multiplying the intensities of the top left and 
       top right fBm maps at each pixel (Eq.~(28)).
      }   
   \label{fig:fBm_maps_l=6pix}
\end{figure}
\begin{figure}%
   \centering
   \includegraphics[width=3.cm,angle=90]{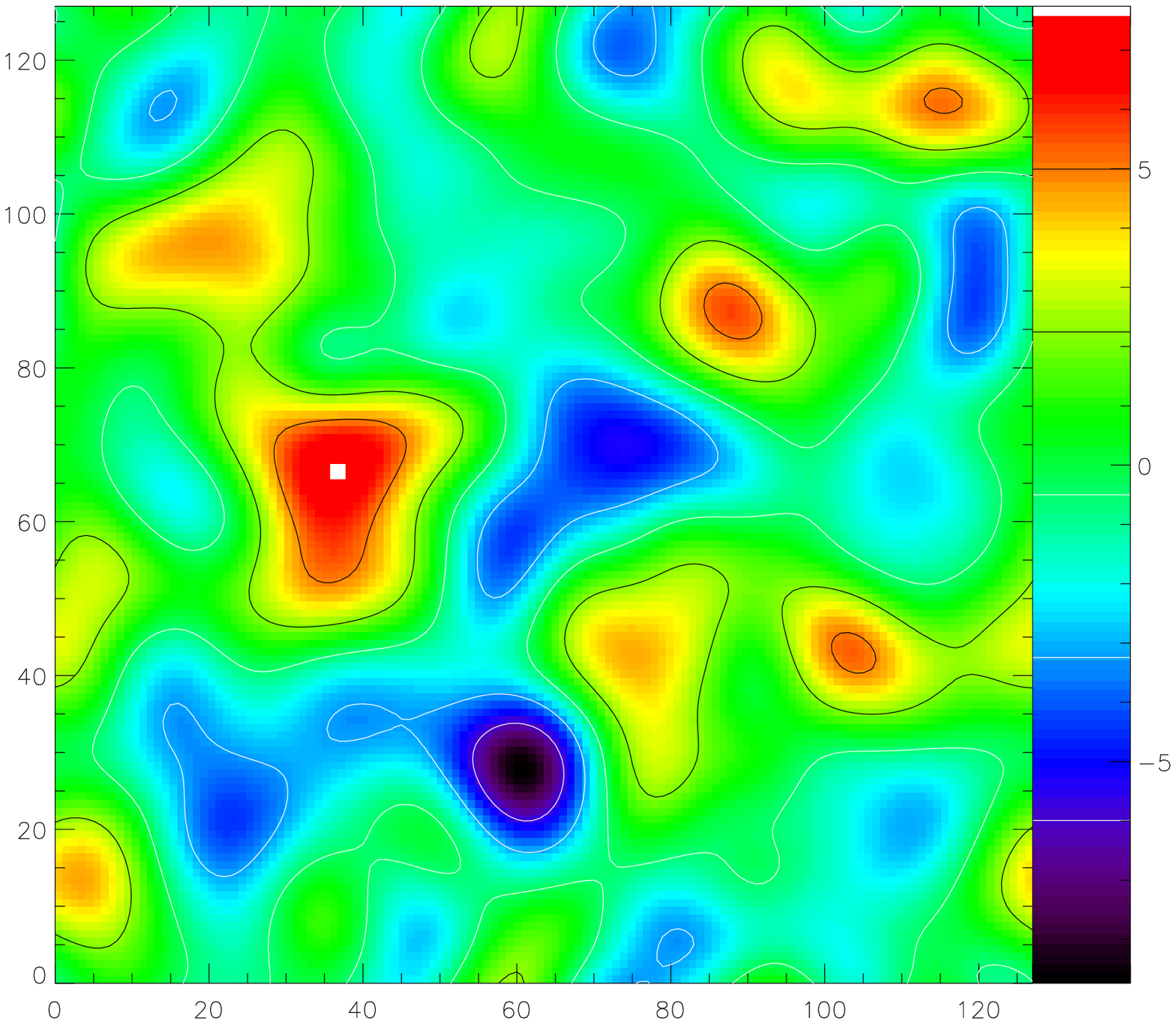}
   \includegraphics[width=3.cm,angle=90]{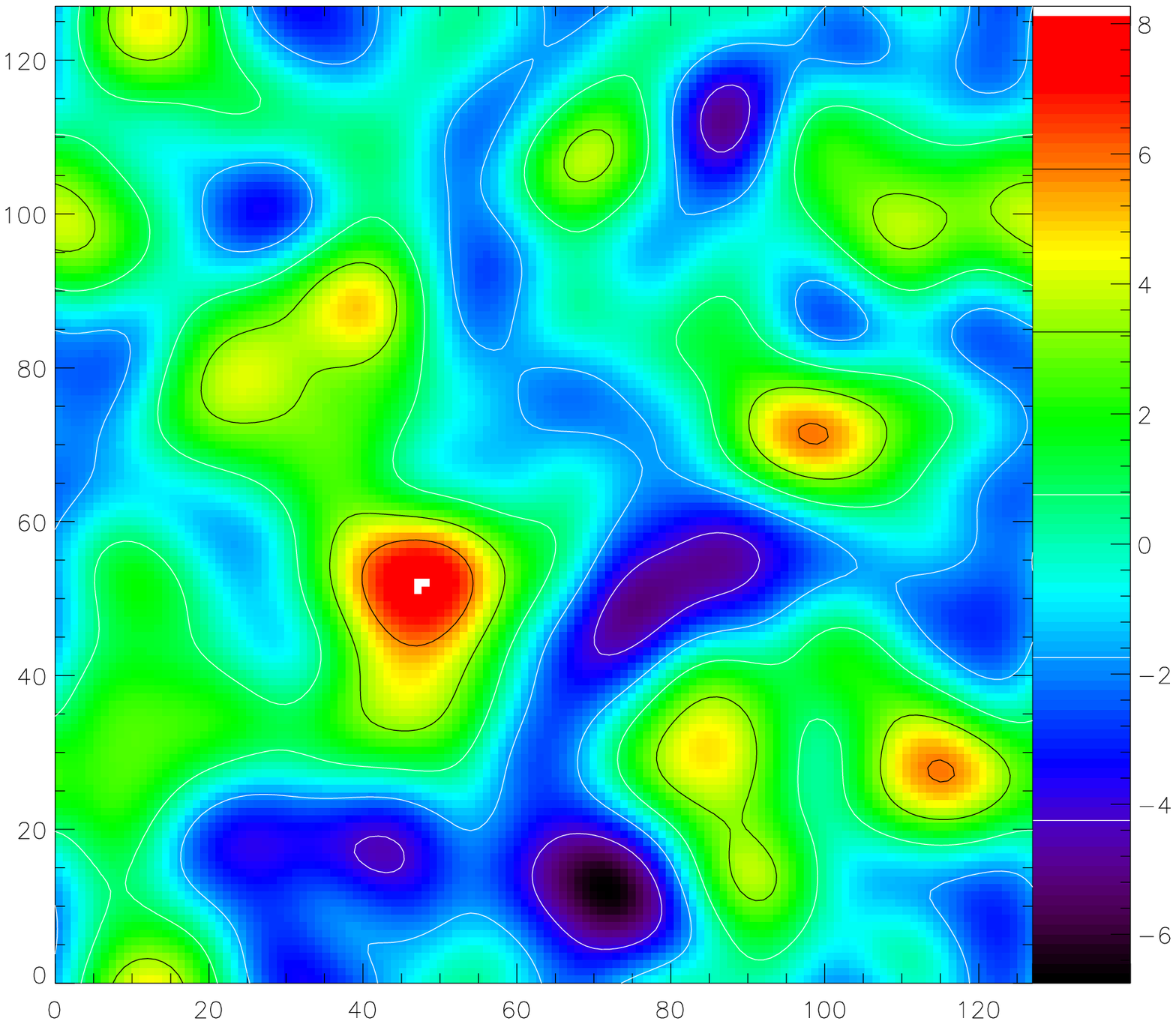}
   \includegraphics[width=3.cm,angle=90]{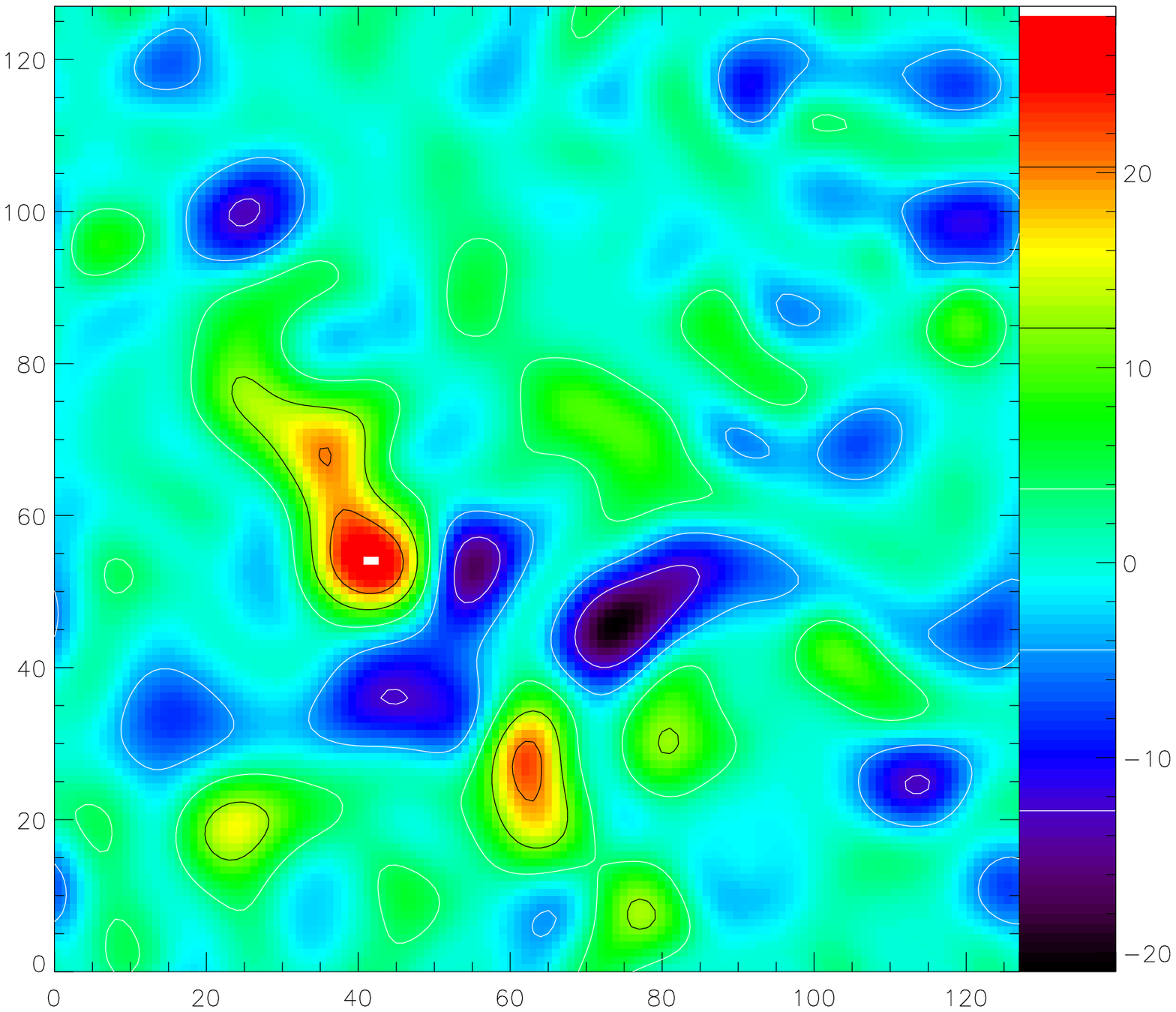}
   \includegraphics[width=3.cm,angle=90]{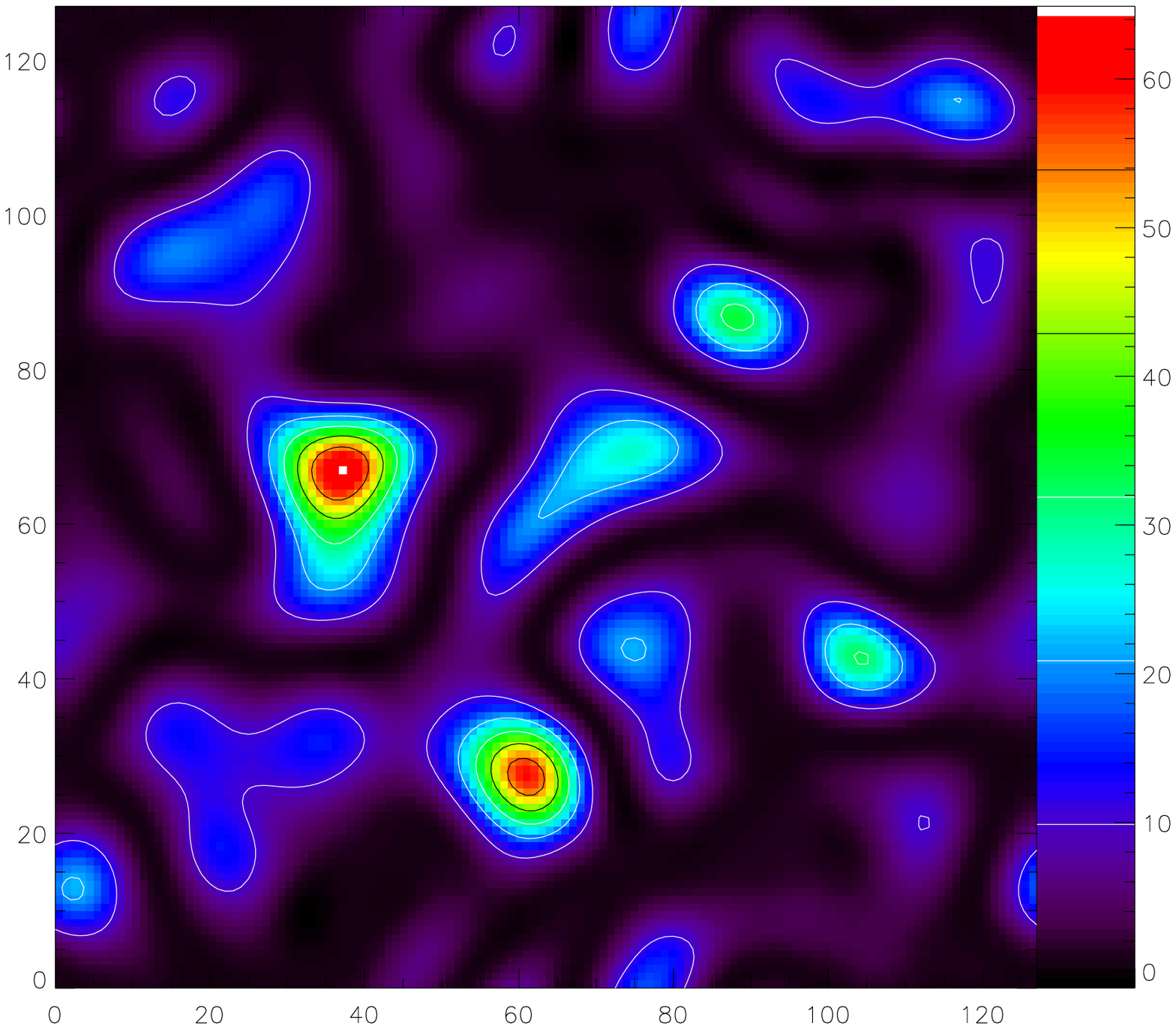}   
      \caption{\emph{Top left.} The fBm map (Fig.~\ref{fig:fBm2mn}\,(a), left) filtered on scale of 17 pix. 
       \emph{Top right.} The fBm map displaced on scales $> 10$ pix by $\tau_{i,x}=-10$ pix and $\tau_{i,y}=15$ pix
       (Fig.~\ref{fig:fBm2mn}\,(a), right) filtered on scale of 17 pix.
       \emph{Bottom left.} The product map of two fBm maps (top left and top right).
       \emph{Bottom right.} The product map of the top left and top right maps where the latter is reversely
       shifted by $\tau_x=-10$ pix and $\tau_y=15$ pix 
       (Eq.~(29)).    
      }   
   \label{fig:fBm_maps_l=17pix}
\end{figure}

To understand the nature of the correlations quantified by the WWCC, it is useful
to visually inspect the filtered maps for the individual filter scales $l$. 
In the original maps (Fig.~\ref{fig:fBm2mn} (a)) it is almost impossible to
locate individual structures that are correlated between the maps. In the 
corresponding maps that were wavelet-filtered on the scale of 6 pix
(Fig.~\ref{fig:fBm_maps_l=6pix} top), the scale where the CC spectrum
showed a peak indicating many correlated structures ($r(l=6 \,\,{\rm pix}) 
\approx 0.6$, Fig.~\ref{fig:fBm2mn} (c)), it may be possible to visually
identify some prominent structures, but this is extremely difficult due
to the noise contribution. An easy identification of correlated structures
is however possible if we visualize the integration kernel of the
correlation function Eq.~(\ref{eq:ccf}) for zero offsets $\vec{\tau}=0$, 
i.e. show the product of the two maps at each pixel,
\begin{equation}
  \label{ssec:fbm1.1}
  P(\vec{x},l) = F(\vec{x},l) \times G(\vec{x},l).
\end{equation}

In the product map $P(\vec{x},l)$ (Fig.~\ref{fig:fBm_maps_l=6pix}, bottom panel)
only relatively strong (positive or negative) features that agree between
both maps show up\footnote{In observed intensity maps, there are of course
no significant negative structures, so that we should only see the bright
structures there.}. One can easily recognize individual small ``clumps'' present
in both maps at the same location that can be identified by comparing the
product map with the original maps (Fig.~\ref{fig:fBm2mn} (a)).

In Fig.~\ref{fig:fBm_maps_l=17pix} we repeat the experiment for the larger 
scale of $l=17$ pix where the correlation coefficient is at minimum
(Fig.~\ref{fig:fBm2mn}\,(c)) because the structures are displaced by 
$\tau_{i,x}=10$ pix and $\tau_{i,y}=-15$ pix ($|\vec{\tau_i}=18$~pix, Fig.~\ref{fig:fBm2mn}\,(d)).
The product map of the filtered maps (Eq.~(\ref{ssec:fbm1.1}),
Fig.~\ref{fig:fBm_maps_l=17pix} bottom left panel) shows no structures that 
can be identified in the individual filtered maps or the original maps
in Fig.~\ref{fig:fBm2mn}\,(a) and the amplitude of the structures in the
product map is relatively small. When using, however, the recovered displacement
vector at $l=17$ pix to shift the second map back by the measured offset vector 
$\vec{\tau}(l)$ and take the product of the shifted map $G(\vec{x}+\vec{\tau}(l),l)$
and the first map $F(\vec{x},l)$ (top right), 
\begin{equation}
  \label{ssec:fbm1.2}
  P(\vec{x},\vec{\tau}(l),l) = F(\vec{x},l) \times G(\vec{x}+\vec{\tau}(l),l),
\end{equation} 
we find that the product map (Fig.~\ref{fig:fBm_maps_l=17pix}, bottom right)
recovers all of the relatively strong (positive or negative) structures seen
in the filtered maps. This allows us to localize individual correlated 
structures to address their shape and origin, independent of a mutual shift. 
{\changed In Sect.~\ref{sec:application} we encounter a pair of observed
maps of the G\,333 molecular cloud with properties that are very similar 
to the example considered here. In the observed maps we also
find correlated structures with matching locations at small scales, 
but a global displacements when considering large scale structures.}

\subsection{Structures of fBm clouds with enhanced scales}
\label{sec:fbm2}
\begin{figure}[b]
   \centering
   \includegraphics[width=3.cm,angle=90]{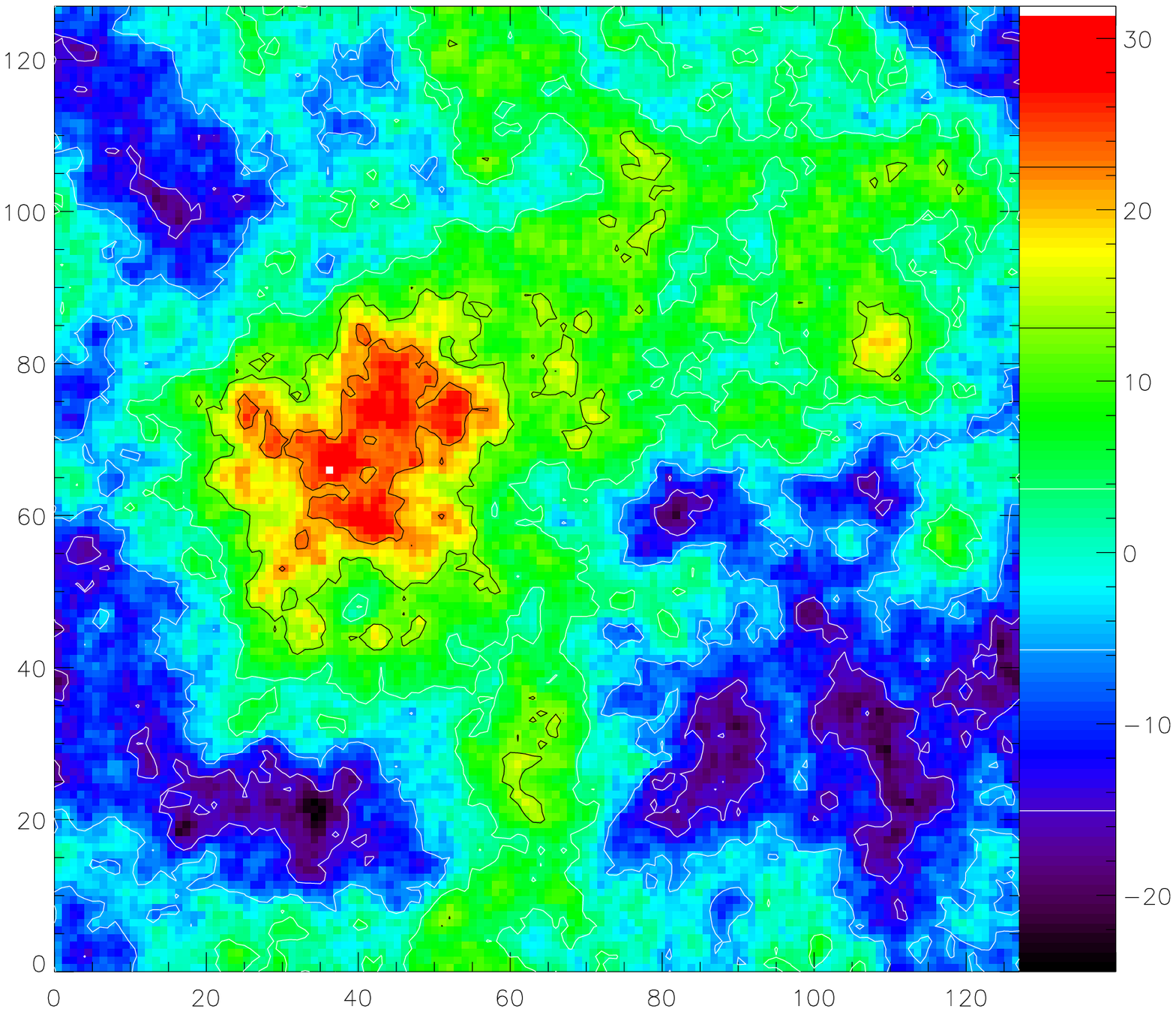}
   \includegraphics[width=3.cm,angle=90]{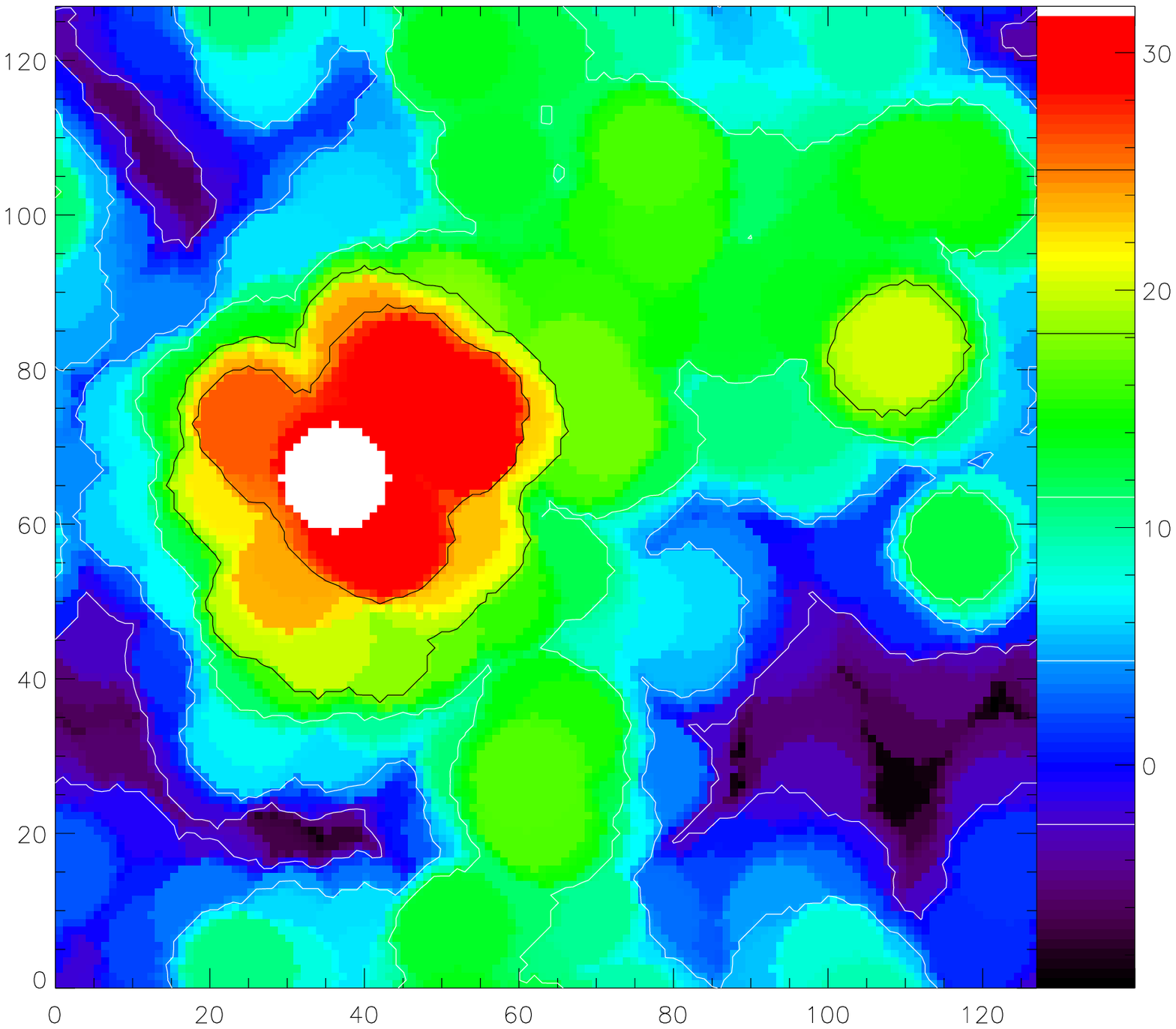}
   \includegraphics[width=3.cm,angle=90]{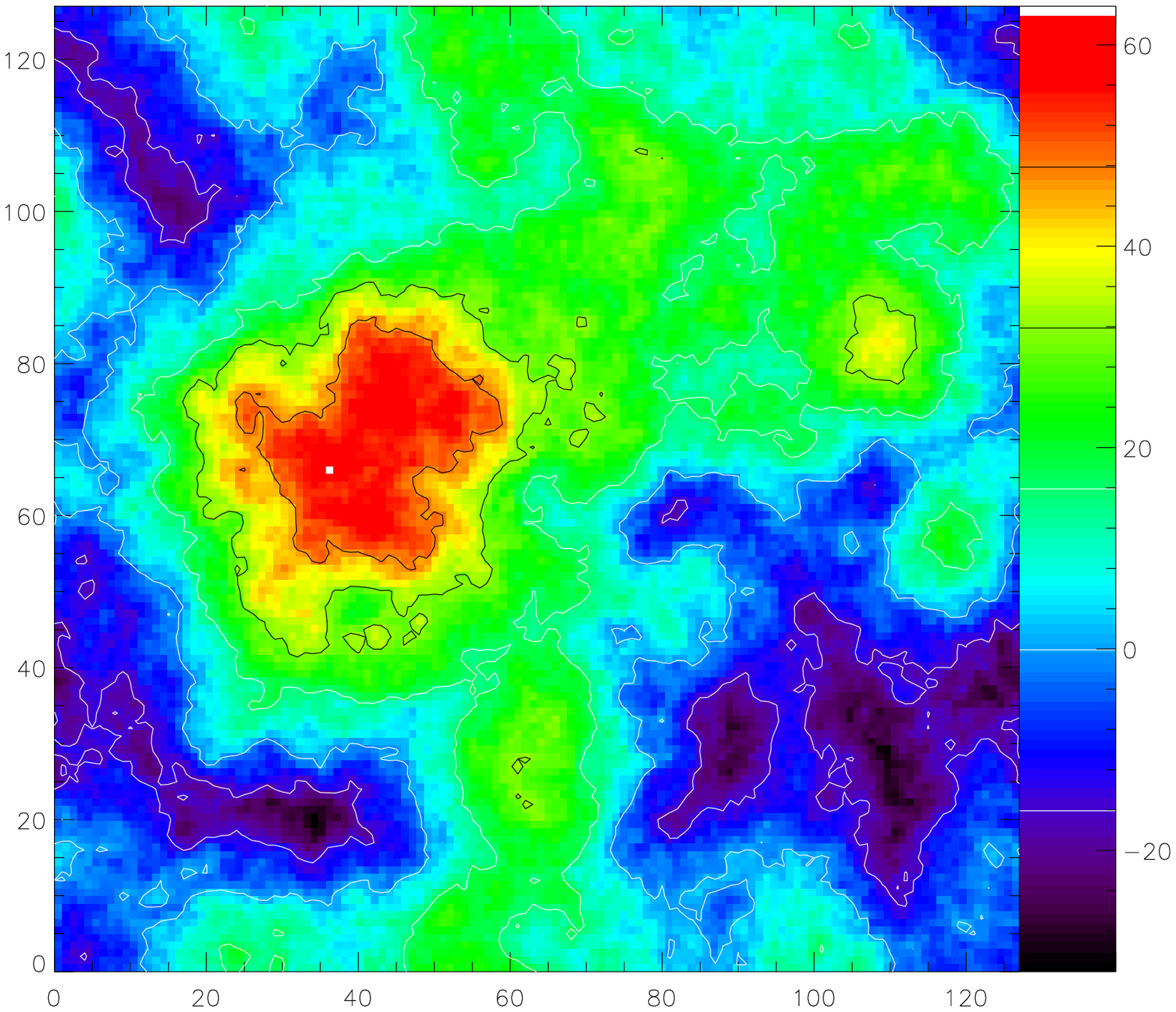}
   \includegraphics[width=3.cm,angle=90]{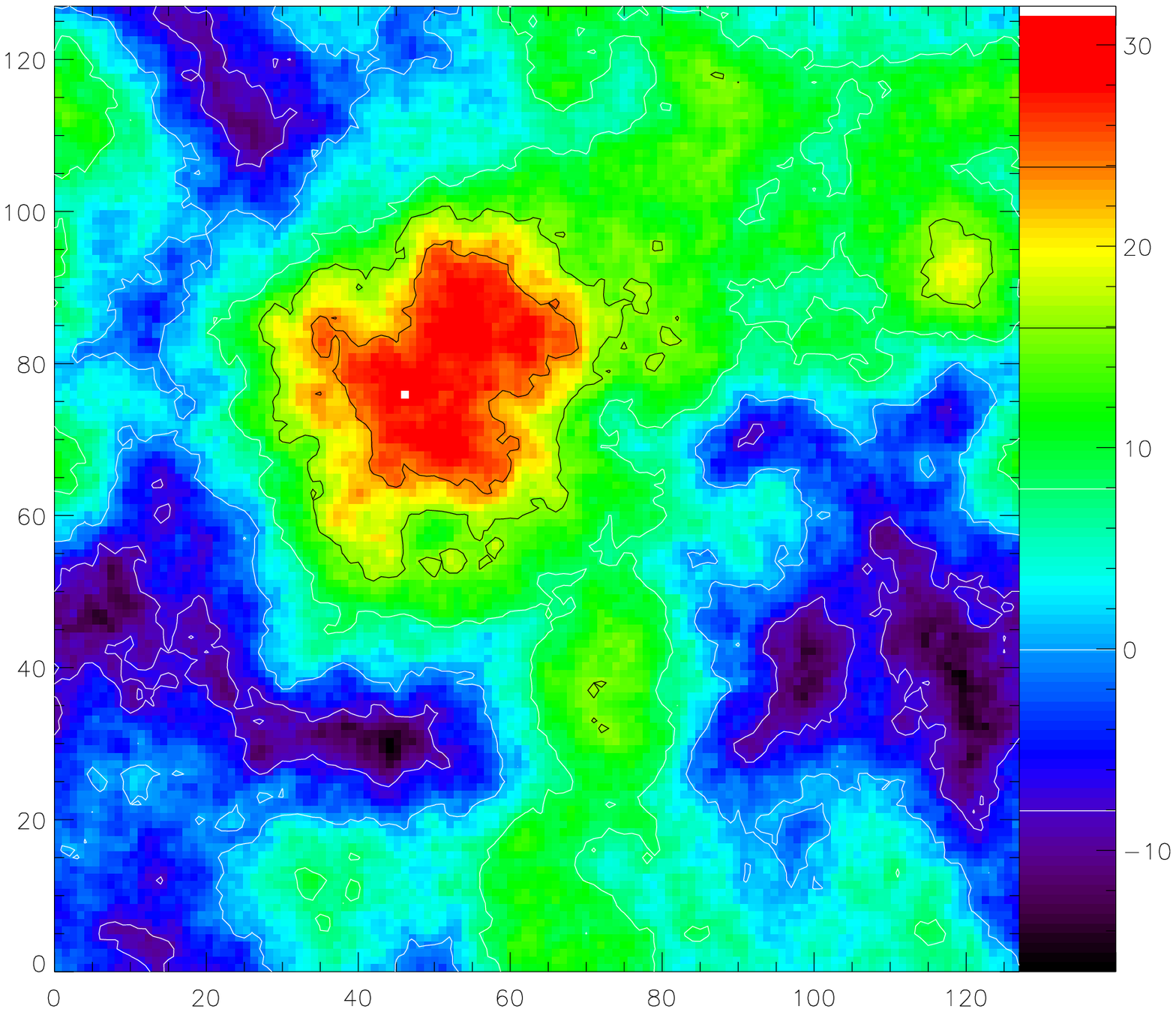}
      \caption{\emph{Top left.} A pure fBm map generated with $\mathrm{S/N}=\infty$ and spectral index 
      $\zeta=3$. \emph{Top right.} A pure fBm map filtered with maximum filter of size 15 pix.
       \emph{Bottom left.} efBm map which is generated by superimposing the pure fBm structure 
       (top left) and the filtered fBm structure (top right) and for brightness parameter $W=1$.
       \emph{Bottom right.} 
       efBm map shifted by $\tau_{i,x}=10$ pix and $\tau_{i,y}=10$ pix.
      }   
   \label{fig:fBm_maps}
\end{figure}
Pronounced scales in otherwise self-similar clouds may indicate special
physical processes acting on those scales, therefore giving access to
important characteristics of observed maps. Being able to 
find and compare those scales in different maps is therefore an essential
step in understanding interstellar turbulence. One process producing
such enhanced scales is the opacity of partially optically thick lines.
They provide a saturated picture for column densities above a specific 
threshold. To mimic this opacity saturation effect in simulations, we
enhance spatial structures in an fBm map $F(\vec{x})$ by smoothing it with a 
\emph{maximum filter} described by 
\begin{equation}
\mathcal{M}_{l_{\rm f}}(F(\vec{x})) = \max{\vec{x}':|\vec{x}'-\vec{x}|<l_{\rm f}/2} F(\vec{x}')\;,
\end{equation}
where $l_{\rm f}$ is the filter size. The maximum filter enhances the 
signal on the scale of the filter size and washes out structure on scales 
less than the filter size. {\changed In this way we introduce a prominent
scale with size $l_{\rm f}$ into the map.} This is demonstrated
in Fig.~\ref{fig:fBm_maps}. The upper left panel shows the original fBm structure
and the upper right panel the result of the convolution with a maximum filter
of size $l_{\rm f}=15$ pix. {\changed Structures below $l_{\rm f}$ pix are 
washed there while circles of 15 pix diameter
represent the dominant structures now.}
As the pure opacity effect is always superimposed by additional small scale
variations this would give an unrealistic picture. We
construct a more realistic picture by combining the original fBm, $F(\vec{x})$, with the
{\changed scale-enhanced fBm map,} $\mathcal{M}_{l_{\rm f}}(F(\vec{x}))$: 
\begin{equation}
  F_{\rm m}(\vec{x}) = W F(\vec{x}) + (2-W)\mathcal{M}_{l_{\rm f}}(F(\vec{x})),
  \label{eq:FmW}
\end{equation}
where we introduce the inverse brightness contrast parameter $W$, 
($0<W<2$), characterizing the contribution of the pure fBm in the total 
map. Then, $2-W$
is the relative brightness of the scale-enhanced map. For values of 
$W<1$ the enhanced structure is brighter than the pure
fBm structure while for $W>1$ it is fainter. This combination should
account for the characteristics of the globally self-similar 
structure and opacity effects. The resulting fBm with enhanced scale 
(in the following {\sl efBm}) map is shown in the lower left panel 
of Fig.~\ref{fig:fBm_maps} for $W=1$.

\subsubsection{Recovering the enhanced scale.}
\label{sect_scale_measure}

\begin{figure}
   \centering
   \includegraphics[width=6.2cm,angle=90]{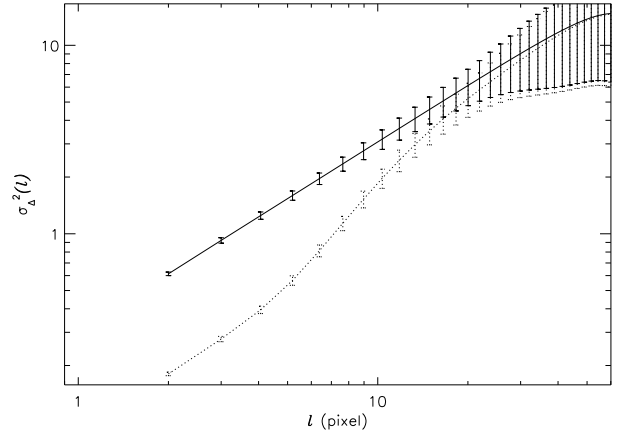}
      \caption{$\Delta$-variance spectra of the pure fBm structure and fBm structure 
	filtered at the scale of 15 pix by the maximum filter (full and dotted lines, respectively).  
      }   
   \label{fig:fBm_dvar}
\end{figure} 

In our numerical experiment we know the size of the maximum filter,
but in real observations, one first has to detect and quantify the
prominent structures, i.e. the manipulation of the data in our case.
\citet{MLO} and O08 have shown that the $\Delta$-variance
is an appropriate tool to {\changed detect these scales in
general astrophysical data sets with irregular boundaries and
variable data reliability across the maps. The $\Delta$-variance 
spectrum measures all prominent scales} that are strong 
enough to create a peak in the spectrum.
In a globally self-similar structure like our fBm (or efBm) maps,
there is, however, a monotonically increasing $\Delta$-variance spectrum.
Figure~\ref{fig:fBm_dvar} compares the $\Delta$-variance spectra
of the original fBm with the spectrum obtained for the efBm
map from Fig.~\ref{fig:fBm_maps}. We find no peak in
the $\Delta$-variance spectrum -- the underlying fBm structure
dominates -- but 
information about the filter size, $l_{\rm f}$, is apparently present
in the relative drop of structure variation at scales below the filter size.
The convolution of the map with 
the maximum filter of size $l_{\rm f}=15$ pix filters out
smaller scale structure. The efBm 
has less structure on scales $\lesssim 15$ pix, while at scales
larger than the size of the maximum filter the
$\Delta$-variance of the original fBm is basically recovered in the efBm.

Instead of searching for a peak in the $\Delta$-variance spectrum, we
therefore need a new approach to find enhanced scales by looking
at the steepening of the $\Delta$-variance spectrum that is visible 
in Fig.~\ref{fig:fBm_dvar}. As an auxiliary step, we first look at the 
pure maximum filter that is described by a filled circle with given diameter 
$l_{\rm f}$.
 
\paragraph{Circular structures with constant intensity} ~\\
\label{sec:csci}

\begin{figure}
   \centering
   \includegraphics[width=6.cm, angle=90]{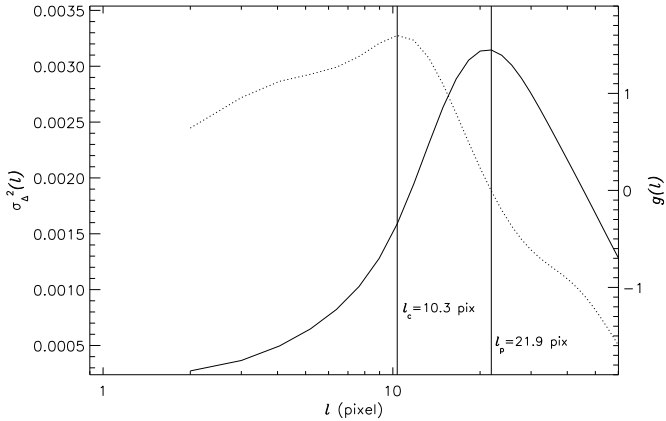}
      \caption{ $\Delta$-variance spectrum $\sigma_{\Delta}^2(l)$ of a filled circle of diameter 20 pix
      (full line) and derivative of the logarithmic $\Delta$-variance spectrum $g(l)$ of the filled circle
      (dotted line) for $\mathrm{S/N}=\infty$. 
      Vertical lines denote the measured prominent scale of the filled circle at $l_{\mathrm p}=21.9$ pix and 
      the critical scale $l_{\mathrm c}=10.3$ pix characterizing the peak of 
      the gradient spectrum (Eq.~(\ref{eq:dvar_lc})).}
      \label{fig:cm_deltavar}
\end{figure}

The $\Delta$-variance spectrum of a filled circle of diameter $d=20$ pix
is shown in Fig.~\ref{fig:cm_deltavar} as a full line. The maximum of the $\Delta$-variance falls at 
$l_{\mathrm p} \approx 20$ pix $\approx l_{\rm f}$, as predicted by O08. 
We verified this for filled circles with $l_{\rm f}=4,8,12,16,20,24,28$ pix. 

To measure the prominent scale even without clear $\Delta$-variance peak,
we examine the gradient of the logarithm of the $\Delta$-variance spectrum,
\begin{equation}
  g(l) =   \frac{\mathrm{d}\log \sigma^2_{\Delta}(l)}{\mathrm{d}\log l},
  \label{eq:gradient}
\end{equation}
in Fig.~\ref{fig:cm_deltavar} (dotted line). The steepening of the 
$\Delta$-variance spectrum below the prominent scale $l_{\rm f}$ is
reflected by a pronounced peak in the gradient spectrum. The root 
$g=0$ corresponds to the prominent scale $l_{\rm p}$.
We define the \emph{critical scale} as the maximum of the gradient
spectrum:
\begin{equation}
   l_{\rm c} =\argmax{l} \frac{\mathrm{d}\log \sigma^2_{\Delta}(l)}{\mathrm{d}\log l }.
   \label{eq:dvar_lc}
\end{equation}
For the filled circle with 20 pixels diameter the measured critical 
scale is $l_{\rm c} = 10.3$ pix, approximately half of the circle's diameter.

\begin{figure}
   \centering
   \includegraphics[width=6.cm, angle=90]{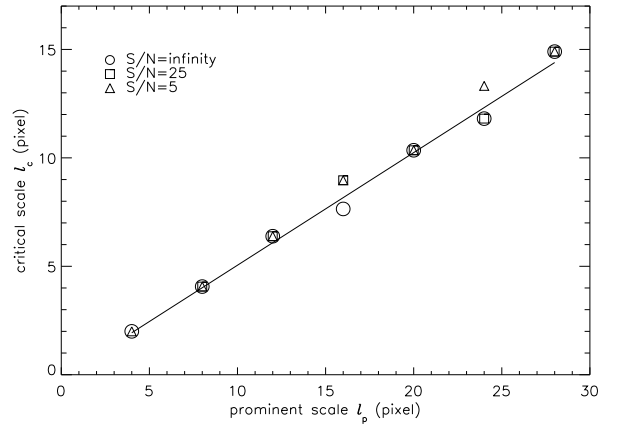} 
      \caption{Relation between critical scales and diameter of the circles for 
      $\mathrm{S/N} = \infty,25,$ and 5.
      The best-fit linear regression line 
	($l_{\mathrm c} \approx 0.52\,l_{\mathrm p}$) 
	for $\mathrm{S/N} = \infty$ is shown as solid line.}
   \label{fig:lc-lp}
\end{figure}

When computing the $\Delta$-variance gradient spectra for circles of
$l_{\rm f}=4,8,12,16,20,24,28$ pix (circles in Fig.~\ref{fig:lc-lp}),
we find a linear relation between measured $l_{\mathrm c}$ and $l_{\mathrm f}$ 
\begin{equation}
  l_{\mathrm c} = (0.52\pm0.02)\,l_{\mathrm p}-(0.14\pm0.36) \approx 0.52\,l_{\mathrm f},
  \label{eq:lc-lp_circle}
\end{equation}
shown as solid line in Fig.~\ref{fig:lc-lp}.

\paragraph{Noisy data} ~\\

To test the robustness of this approach against observational noise,
we add different levels of white noise to the circle maps and
compute their $\Delta$-variance gradient spectra. In noisy maps, the 
$\Delta$-variance spectrum at small scales is dominated by 
noise fluctuations (compare Figs.~\ref{fig:gm_delta2} and
\ref{fig:gm_delta1}) so that the critical scale $l_{\rm c}$ is not directly
measurable through Eq.~(\ref{eq:dvar_lc}). To still derive
 $l_{\rm c}$, one needs to subtract the noise contribution
from the $\Delta$-variance spectra of the contaminated circular structures:
\begin{equation}
 \sigma_{\Delta}^2(l)_{\rm structure} = \sigma_{\Delta}^2(l)_{\rm map} 
  - \sigma_{\Delta}^2(l)_{\rm noise}.
\label{eq:noisesubtract}
\end{equation}
The corresponding noise spectrum can be obtained by running the
$\Delta$-variance analysis on the noise map, obtained e.g. from
emission-free channels in an observed spectral line cube.
As we construct the noisy map in our numerical simulations by 
simply adding a noise map to the original map we do not need
to extract it separately here. 

Using this noise subtraction, we repeated the computation of 
the gradient peak scale $l_{\mathrm c}$ from maps with different 
circle sizes and noise levels $\mathrm{S/N} = 25$ (squares in Fig.~\ref{fig:lc-lp})
and $\mathrm{S/N} = 5$ (triangles). We find a very good match 
to the noise-free results with deviations 
by up to at most 2 pixels induced by the noise in the data. 
The regression coefficient 
($a=0.54 \pm 0.02$) for the noisy maps coincides with one for $\mathrm{S/N} = \infty$ 
($a=0.52 \pm 0.02$) within the error limits.

For observed maps the determination of the noise $\Delta$-variance
$\sigma_{\Delta}^2(l)_{\rm noise}$ may, however, be affected by 
uncertainties that will propagate into the determination of the
critical scale $l_{\rm c}$ if the latter falls into the noise-dominated
regime. Errors in the noise subtraction will affect the
$\Delta$-variance gradient spectrum at scales below the
$\Delta$-variance minimum, $l \le l_{\rm min}$ (see e.g.
Fig.~\ref{fig:fBm2mn}\,(b), $l_{\rm noise\,dominated} 
\lesssim l_{\rm min} = 5$ pix). This puts a lower limit on the 
scale $l > l_{\rm c} = l_{\rm min}$ that can be reliably measured.

\paragraph{efBm structures} ~\\
\hangindent=1.7cm

\begin{figure}
   \centering
   \includegraphics[width=7.2cm, angle=-90]{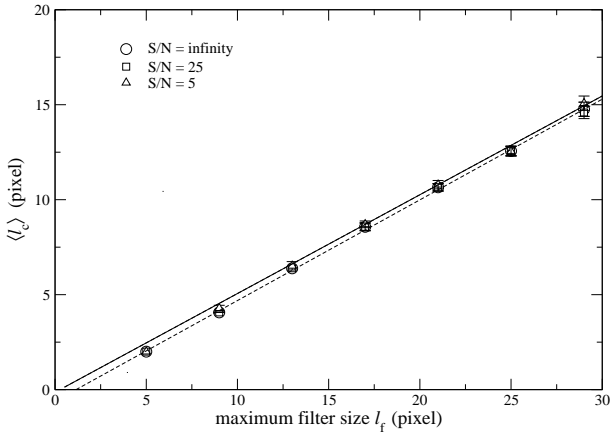}
   \caption{Mean critical scale $l_{\rm c}$ measured for eight fBm
   maps filtered with the maximum filter of size $l_{\rm f}$ for $\mathrm{S/N}=\infty$, 
   25 and 5 (circles, squares, and triangles, respectively). 1\,$\sigma$ error bars measured in a set of
   10 different realizations are superimposed. }
   \label{fig:fbm_lc-lf}
\end{figure}

In Figure~\ref{fig:fbm_lc-lf} we show the result of the equivalent experiments
for the fBm structures with enhanced scales and an inverse brightness contrast $W=1$.
The efBms are generated from a fBm map with spectral index $\zeta=3$ which is convolved with the 
maximum filter of the sizes $l_{\rm f}$ = 5, 9, 13, 17, 21, 25, 29 
pix.

For each efBm we use the gradient of the $\Delta$-variance spectrum to 
measure the critical scale (Eq.~(\ref{eq:dvar_lc})). Measured critical scales against 
maximum-filter sizes are plotted in Fig.~\ref{fig:fbm_lc-lf}. The mean critical 
scale and its uncertainty are estimated from 10 different random realizations of efBms.  
The circles with error bars represent the noise-free results. 
The data can be described by a linear relation,
\begin{equation}
   \langle l_{\rm c}\rangle = (0.53\pm0.01)\,l_{\rm f} - (0.61\pm0.1),
   \label{eq:lc-lf}
\end{equation}
represented by the dashed line in Fig.~\ref{fig:fbm_lc-lf}. 
Within 1 pix accuracy this is identical to the simple approximation without
offset from the origin $l_{\mathrm c} \approx 0.52\,l_{\mathrm f}$ (in agreement
with Eq.~(\ref{eq:lc-lp_circle})), obtained for the circular filter, shown
as solid line in Fig.~\ref{fig:fbm_lc-lf}.

The squares and triangles in Fig.~\ref{fig:fbm_lc-lf} show the measured
critical scales for the equivalent noisy maps with $\mathrm{S/N}=25$ and 
$\mathrm{S/N}=5$ after applying the noise correction from 
Eq.~(\ref{eq:noisesubtract}). One can see that the squares and triangles coincide well with 
circles for $l_{\rm f} > 1$ pix and they can be fit by the same linear relation given by 
Eq.~(\ref{eq:lc-lp_circle}) and (\ref{eq:lc-lf}) (see Fig.~\ref{fig:fbm_lc-lf}).

\begin{figure}
   \centering
   \includegraphics[width=6.2cm, angle=90]{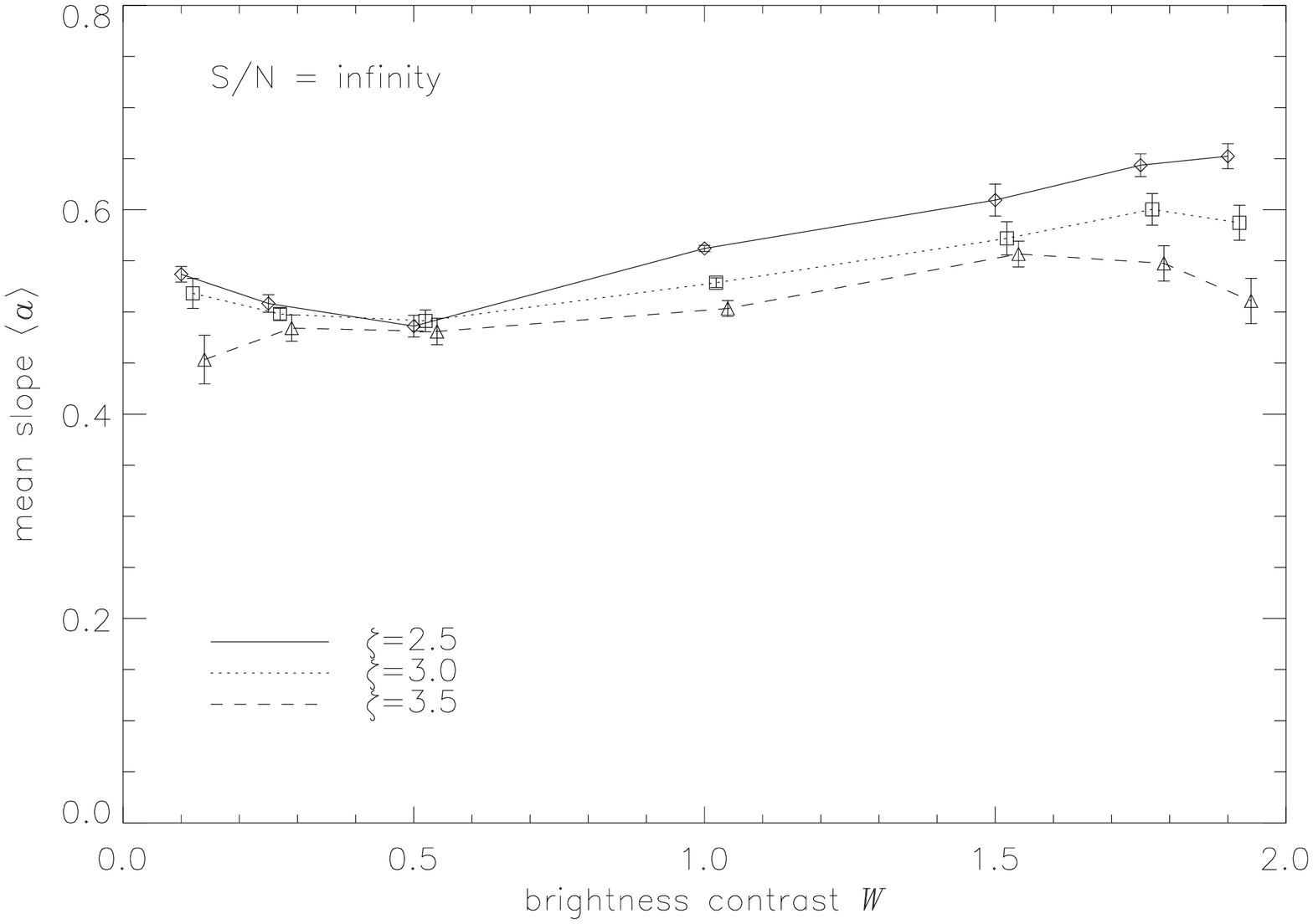}\\
   \includegraphics[width=6.2cm, angle=90]{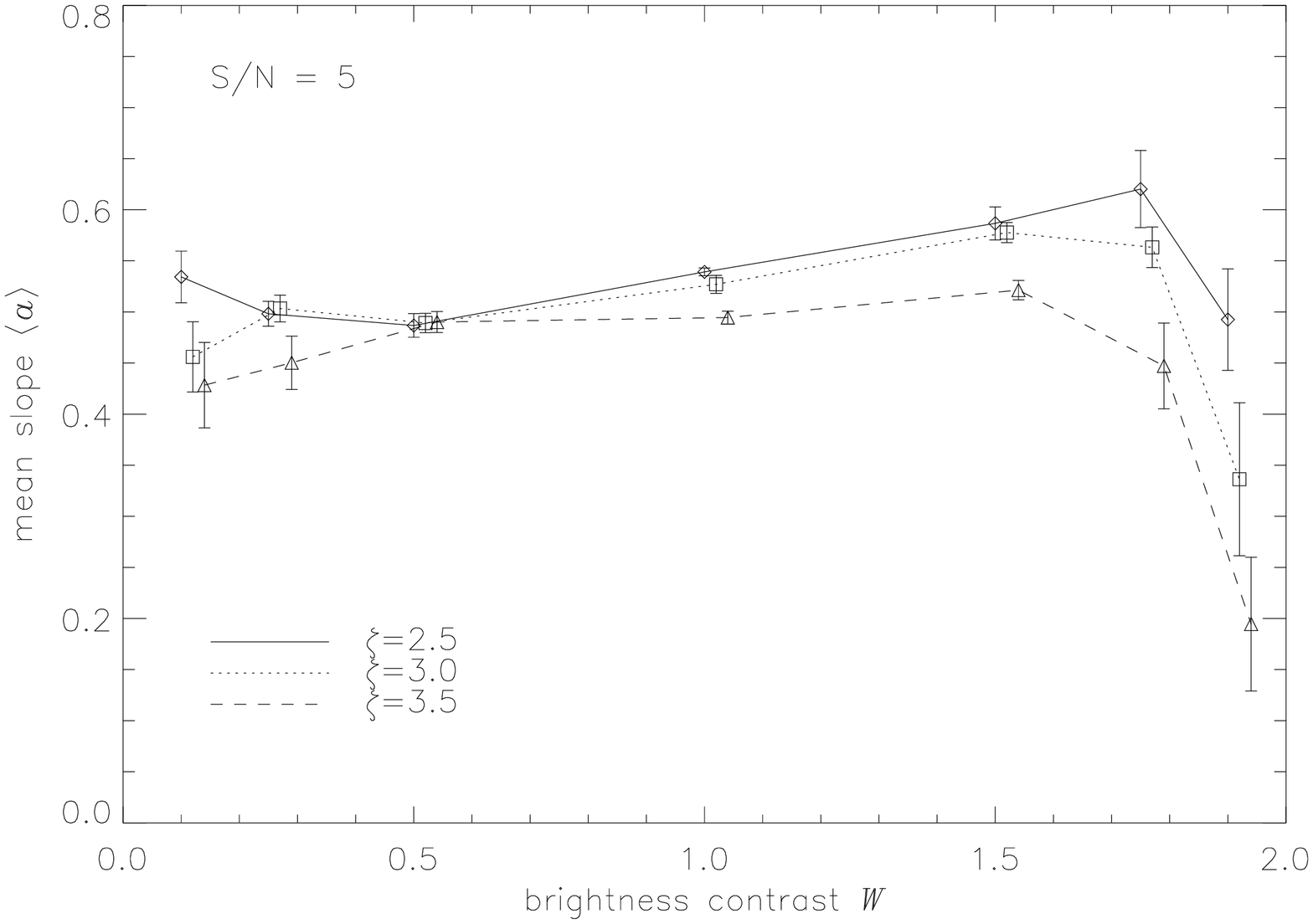}
   \caption{Measured slope $a$ against inverse brightness contrast $W$ for efBm maps
    having spectral indexes $\zeta=2.5,3,3.5$ (full, dotted, and dashed lines,
    respectively). The upper plot shows the result for
   $\mathrm{S/N}=\infty$, the lower plot for $\mathrm{S/N}=5$. The data points
   for $\zeta=3$ and 3.5 (squares and triangles) are horizontally displaced 
   by $W=0.02$ and $W=0.04$, respectively, from the $\zeta=2.5$ data points 
   (diamonds) for a better visibility. 
   The error bars represent 1\,$\sigma$ variations in the statistical sample.
   }
   \label{fig:a-W_StoN=5}
\end{figure}

To verify that the relation is robust against changes of the efBm spectral
index and the inverse brightness contrast $W$ we repeated the
experiment for a range of parameters. We vary the fBm spectral index in the
range of 2.5 to 3.5 typical for interstellar clouds \citep{Falgarone2007}
and brightness parameters $W = 0.1,0.25,0.5,1,1.5,1.75,1.9$. 
The previous experiment used $\zeta=3$ and $W=1$.
For each combination of $\zeta$ and $W$, we recover the critical scales 
$l_{\rm c}$ for $l_{\rm f}$ = 5, 9, 13, 17, 21, 25, 29 pix and compute the 
slope $a$ of the $l_{\rm c}-l_{\rm f}$ relation in the range $l_{\rm f} > l_{\rm min}$ pix.
To estimate the statistical uncertainty of $a$ we used 10 different random realizations
so that the error is composed from the ensemble variation and the imperfection
of the linear fit.
Figure~\ref{fig:a-W_StoN=5} shows the results for $\mathrm{S/N} = \infty$ and 
$\mathrm{S/N} = 5$.
In case of $\mathrm{S/N} = \infty$ (upper panel), the slope varies in the range 
between 0.47 and 0.63. It tends to increase with decreasing the spectral
index $\zeta$ and increasing inverse brightness contrast $W$.
Averaged over the full range of spectral indexes ($\zeta=2.5-3.5$) and 
inverse brightness contrast parameters ($W=0.1-1.9$) we obtain
a mean slope of $\langle a \rangle = 0.54 \pm 0.03$. When knowing $\zeta$ from the 
overall slope of the $\Delta$-variance spectrum, the gradient can be 
even better constrained. For the three spectral indexes we obtain
$\langle a \rangle_{\zeta=2.5} = 0.57 \pm 0.03$, 
$\langle a \rangle_{\zeta=3} = 0.54 \pm 0.02$, and 
$\langle a \rangle_{\zeta=3.5} = 0.51 \pm 0.01$.
These values can be used directly to measure the enhanced scale from the
critical scale in the $\Delta$-variance spectrum for large S/N ratios.

In case of the noisy maps (lower panel in Fig.~\ref{fig:a-W_StoN=5}), the variation of the 
slope is only slightly larger for inverse brightness contrasts $W \le1.5$
but strongly deviating from the noise-free behaviour for low contrasts
of the enhanced structure, i.e. values of $W\ge 1.75$. This is intuitively clear as we 
see less and less enhanced structure in a map if we reduce its
contribution by increasing $W$ to values close to two. If the enhanced 
structure is strongly diluted ($W>1.5$), the efBm maps appear more like 
pure fBm maps with only small variations of the gradient of the 
$\log \sigma^{2}_{\Delta}(l)$ so that the determination of the gradient
peak $l_{\rm c}$, becomes very sensitive to the noise contribution.
Additional noise makes a reliable detection of the enhanced scale increasingly
difficult. 
\begin{figure}
   \centering
   \includegraphics[width=6.2cm, angle=90]{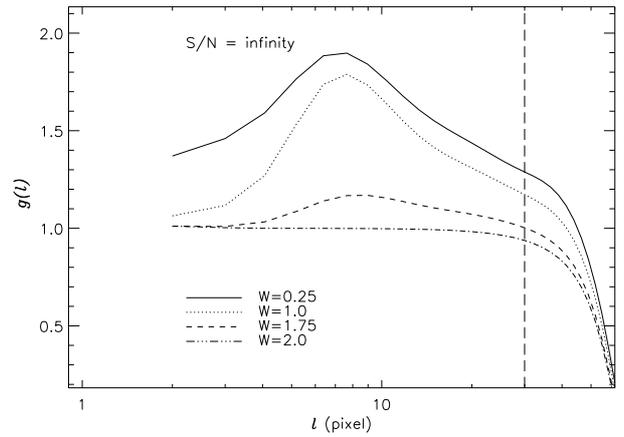}
   \caption{Gradient spectra of fBm maps generated with $\zeta=3$, $\mathrm{S/N}=\infty$, enhanced on scale of 
   $l_{\rm f}=15$ pix and for different inverse brightness contrast parameters $W=0.25,1,1.75$ and 2 (full, dotted, 
   dashed, and dot-dashed lines, respectively). The vertical dashed line indicates the upper 
   limit to the scale where the $\Delta$-variance slope can be measured reliably. Above that scale, 
   the spectrum is strongly affected by boundary effects of the assumed finite map.}
   \label{fig:gradient_spectra}
\end{figure}

The impact of the inverse brightness contrast on the $\Delta$-variance gradient
is visualized for the noise-free
case ($\mathrm{S/N} = \infty$) in Fig.~\ref{fig:gradient_spectra}.
It shows the gradient spectra of four efBm maps with $\zeta=3$,  
$l_{\rm f}=15$ pix, and inverse brightness contrasts $W=0.25, 1, 1.75, 2$.
The case of $W=2$ (dot-dashed line) represents
the pure fBm structure showing the constant gradient
$g(l) = \zeta -2 = 1$ for two-dimensional maps \citep{stutzki98} at scales
up to 30 pixels (dashed vertical line in Fig.~\ref{fig:gradient_spectra}). The drop
towards larger scales is due to the impact of the map boundary limiting
to size of large structures (O08). This is consistent with our results from
Sect. \ref{sect:limiting-cases} that only prominent scales up to 64 pixels,
i.e. half the map size can be detected. To omit the boundary effects we 
restrict the analysis to critical scales $l \le 30$ pix. When adding scale-enhanced structures,
$W<2$, we see a growing peak in the gradient spectrum at $l_{\rm c}=8$ pix.
It turns sharper if we lower $W$, i.e. the contrast of the peak gradient
$g_{\rm max}$ relative to the average gradient $\langle g \rangle$
grows when we increase the contribution of the scale-enhanced structures
by lowering the inverse brightness contrast parameter $W$.
The contrast of the peak value 
$g_{\rm max}$ relative to the average $\Delta$-variance gradient
$\langle g \rangle$ determines whether $l_{\rm c}$ can be reliably determined,
even if uncertainties are added to the spectrum in terms of noise.

\begin{figure}
   \centering
   \includegraphics[width=6.3cm, angle=90]{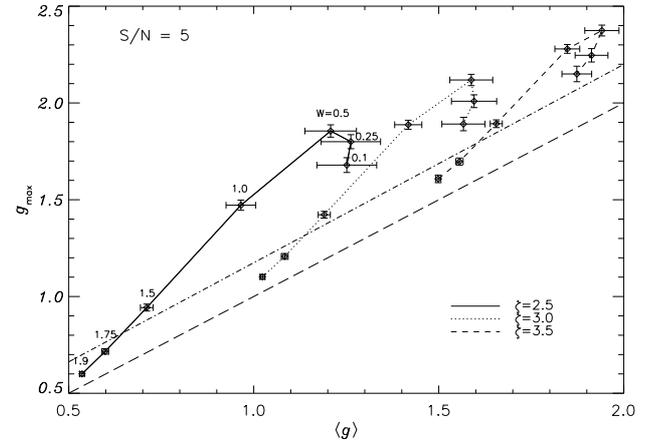}
   \caption{$\mathrm{S/N}=5$. Measured maximum gradient versus mean gradient 
   for efBm maps having spectral indexes $\zeta=2.5,3,3.5$ (full line, dotted and dashed lines, respectively)
   and different inverse brightness contrasts $W=0.1,0.25,0.5,1,1.5,1.75,1.9$.
   The error bars indicate the standard deviation of the ensemble of 10 different
   random realizations and seven different enhancement scales. 
   The long-dashed line is the identity representing straight spectra, i.e.
   $g_{\rm max}=\langle g \rangle$.
   The dash-dot line represents the lower limit of $g_{\rm max}$ as a function of $\langle g \rangle$ 
   below which the scale of the peak cannot be reliably determined (see Eq.~(\ref{eq:gmax-gmean})). 
   }
   \label{fig:gmax-gmean_StoN=5}
\end{figure}

In Fig.~\ref{fig:gmax-gmean_StoN=5} we show the result of the systematic study
of the relation between $g_{\rm max}$ and  $\langle g \rangle$ for different
values of the spectral index $\zeta$ and the inverse brightness contrast $W$
for $\mathrm{S/N} = 5$ where $\langle g \rangle$ is measured between $l_{\rm min}$
and 30 pixels. Averages and error bars are computed for efBms ensembles from
ten different random numbers and enhanced on scales of $l_{\rm f}=5, 9, 13, 17, 21, 25, 29$ pix.
The series of different inverse brightness contrasts $W$ for the same spectral index $\zeta$
are connected by solid ($\zeta=2.5$), dotted ($\zeta=3.0$), and dashed ($\zeta=3.5$)
lines. The long-dashed line indicates the identity, i.e. the behaviour of pure
fBms ($W=2$) not showing any maximum in the gradient (dot-dashed line in Fig.~\ref{fig:gradient_spectra}).
Weakly enhanced fBm maps ($W=1.9$) are positioned slightly above the 
identity. The deviation grows for stronger enhanced fBms, i.e. towards smaller inverse 
brightness contrasts $W<2$. 

As Fig.~\ref{fig:a-W_StoN=5} has shown that for noisy data, the scale of the
steepest gradient cannot be reliably determined if the efBm is only weakly enhanced,
so that the gradient peak is too shallow, we can translate Fig.~\ref{fig:gmax-gmean_StoN=5}
into an easily usable criterion for the contrast of $g_{\rm max}$ relative to $\langle g \rangle$
that still allows for a reliable measurement of $a$ and thus $l_{\rm f}$ through $l_{\rm c}$. For
$\mathrm{S/N}=5$, points with $W \ge 1.75$ had to be excluded. This corresponds to the
area below the dash-dotted line in Fig.~\ref{fig:gmax-gmean_StoN=5}. It can be 
described by $g_{\rm max}$ values that fall below
\begin{equation}
  g_{\rm max,\,limit} = (1.02 \pm 0.01)\langle g \rangle + 0.10 + 3\sigma_{\langle g_{\rm max} \rangle} 
  = 1.02\langle g \rangle + 0.15
  \label{eq:gmax-gmean}
\end{equation}
This relation is computed by connecting the data points with $W=1.75$ and adding the 
3$\sigma$ error margin of $\sigma_{\langle g_{\rm max} \rangle} = 0.017$ of the
individual data points. 
All efBm structures where the enhancement scale cannot be
measured reliably due to a too high inverse brightness contrast ($W \gtrsim 1.75$
in Fig.~\ref{fig:a-W_StoN=5}\,(b)) fall below this relation.
For $W \le 1.5$ critical scales can be reliably recovered for $\mathrm{S/N}=5$.
The mean slopes obtained for $\mathrm{S/N}=\infty$ and 
$\mathrm{S/N}=5$ ($0.54 \pm 0.03$ and $0.51 \pm 0.04$) agree within the error limits.

Figure~\ref{fig:gmax-gmean_StoN=5} can also be used to estimate spectral index and 
inverse brightness contrast of an efBm by measuring $g_{\rm max}$ and $\langle g \rangle$.
For high inverse brightness contrasts $W>0.5$ corresponding to about 
$g_{\rm max}-\langle g \rangle < 0.4$ spectral index and inverse brightness contrast can be 
uniquely determined, for lower values of $W$ there is an ambiguity in determination
of $\zeta$ and $W$ but it is still possible to determine $\zeta$ with an accuracy of 
about 0.5.

One can generalise the approach of measuring the enhancement 
scale $l_{\rm f}$ in efBm maps (Eq.~(\ref{eq:gmax-gmean})) {\changed for} S/N levels larger than 
$\mathrm{S/N}=5$. They show 
different thresholds of the inverse brightness contrast above which $l_{\rm f}$ can 
not be determined reliably anymore. When
lowering the noise, the threshold approaches $W=2$ approximately linearly with the noise
level, shifting down the dot-dashed line in Fig.~\ref{fig:gradient_spectra} towards
the identity line. This corresponds to a change of the constant in Eq.~(\ref{eq:gmax-gmean}) 
from 0.1 to 0 when S/N increases:
\begin{equation}
  g_{\rm max,\,limit} \simeq \langle g \rangle + 0.5(\mathrm{S/N})^{-1} +  
  3\sigma_{\langle g_{\rm max} \rangle}.  
  \label{eq:gmax-gmean_general}
\end{equation}

\paragraph{Global fit} ~\\

We can adopt the single value $\langle a \rangle = 0.52 \pm 0.04$ for estimating 
the enhancement scale,
\begin{equation}
  l_{\rm f} = l_{\rm c}/(0.52 \pm 0.04) = (1.92 \pm 0.15)\,l_{\rm c}.
  \label{eq:lf_StoN=inf_and_5}
\end{equation} 
Even without a priori knowledge of the
spectral index of an efBm cloud, one can measure its enhancement
scale with accuracy of about $\pm 8$\,\% if the prominent structure is
strong enough to meet the criterion of Eq.~(\ref{eq:gmax-gmean_general}).

However, Fig.~\ref{fig:gmax-gmean_StoN=5} can also be used to estimate 
spectral index, allowing for a more accurate determination of the
slope $a$. When including noise levels from  $\mathrm{S/N}=\infty$ to 
$\mathrm{S/N}=5$ in the fit, we measure the mean slopes 
$\langle a \rangle_{\zeta=2.5} = 0.55 \pm 0.04$, 
$\langle a \rangle_{\zeta=3} = 0.53 \pm 0.05$, and 
$\langle a \rangle_{\zeta=3.5} = 0.49 \pm 0.06$
allowing for a more accurate determination of the enhancement scale
$l_{\rm f}$ than through the general equation (\ref{eq:lf_StoN=inf_and_5}).
 
Altogether this gives a recipe for the measurement of a pronounced scale
in a structure that is globally dominated by self-similarity. After applying the
noise correction to the $\Delta$ variance spectrum, the $\Delta$-variance gradient
spectrum has to be evaluated to measure the mean gradient $\langle g \rangle$,
the peak gradient $g_{\rm max}$, and the location of the peak $l_{\rm c}$.
Depending on the noise in the data, the user has to evaluate whether the brightness 
contrast of the enhanced scale of a cloud is strong enough. 
If the gradient peak falls above Eq.~(\ref{eq:gmax-gmean_general}), i.e. 
$g_{\rm max} > g_{\rm max,\,limit}$, then it is sharp enough to allow for a 
reliable measurement of its position and one can use the critical scale 
$l_{\rm c} > l_{\rm min}$ to recover the enhancement scale using the relation 
$l_{\rm f} =1.92 l_{\rm c}$ (Eq.~(\ref{eq:lf_StoN=inf_and_5})).  

\subsubsection{Detecting and recovering displacements.} 

Next, we explore the power of the WWCC to quantify the displacement between 
different efBm structures through the CC coefficient spectrum and the spectrum
of displacement vectors.

\begin{figure}
   \centering
   \includegraphics[width=6.cm, angle=90]{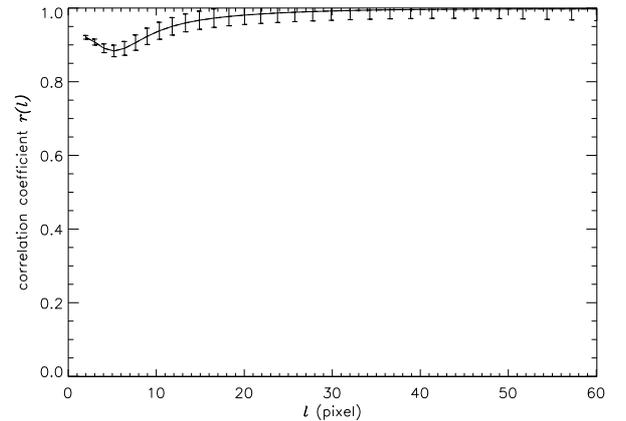}
      \caption{Spectrum of the CC coefficient for the original fBm map and efBm map from Fig.~\ref{fig:fBm_maps}
      (bottom left).
      }
      \label{fig:noshift}
\end{figure}

In Fig.~\ref{fig:noshift} we show the CC coefficient spectrum when comparing
the original fBm and the efBm map without displacement (Fig.~\ref{fig:fBm_maps},
top left and bottom left panels). The structures are highly 
correlated on small and large scales. The scale filtering with $l_{\rm f} = 15$~pix
produces a small minimum on a scale of about 5~pix due to the reduction of
structures below the filter size. Above the filter size
both maps turn very similar leading to a cross correlation coefficient
close to unity.

\begin{figure}
   \centering
   (a)\includegraphics[width=6.cm, angle=90]{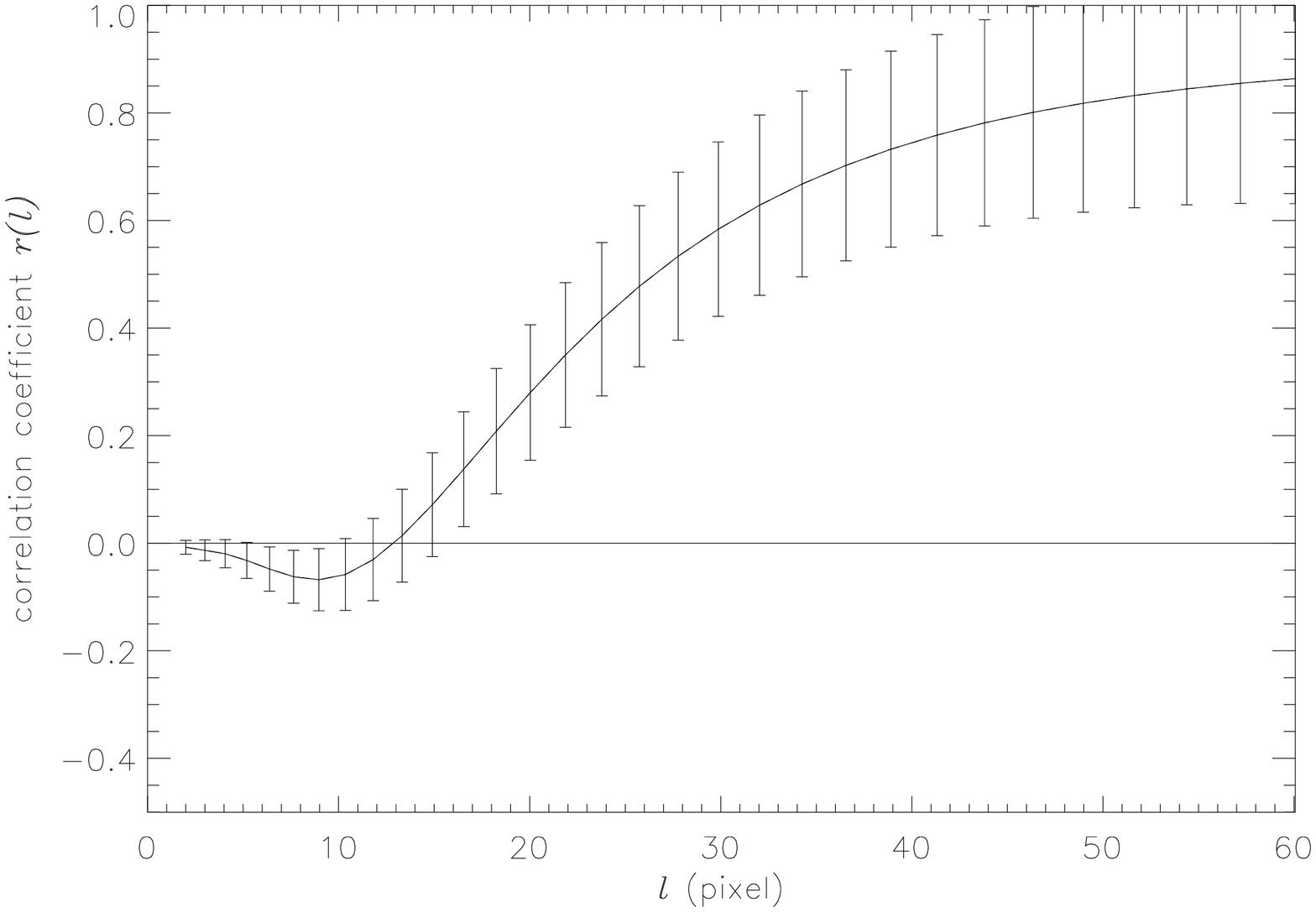}
   (b)\includegraphics[width=3.cm, angle=90, bb=26 23 268 780]{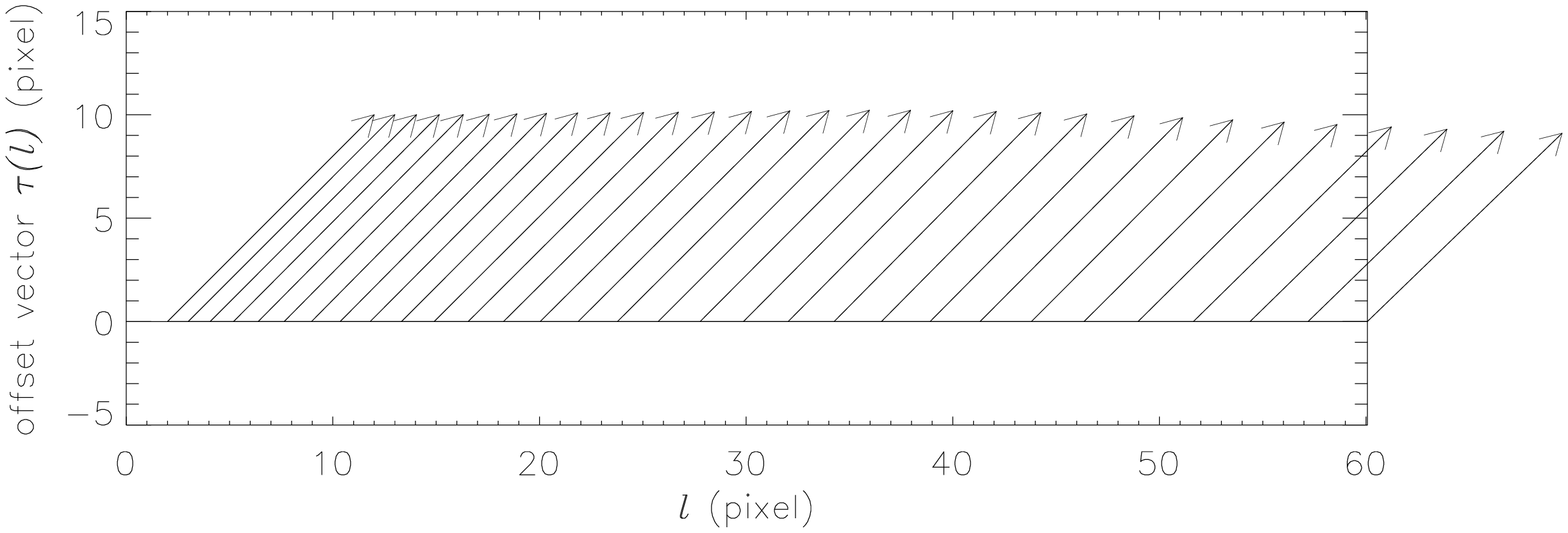}   
      \caption{Results of the WWCC for the original fBm map and the shifted efBm map from Fig.~\ref{fig:fBm_maps}
      (bottom right): (a) correlation coefficient and (d) displacement 
      vector as a function of scale. $1\sigma$ error bars are given.}
         \label{fig:fBmm}
\end{figure}

Figure~\ref{fig:fBmm}\,(a) shows the result of the WWCC when comparing
the fBm map and the efBm map displaced by $\tau_i=14.1$ pix ($\tau_{i,x}=10$ pix and 
$\tau_{i,y}=10$ pix, Fig.~\ref{fig:fBm_maps}, top left and bottom right panels). 
We find a behaviour very similar to that for the displaced Gaussians in
Fig.~\ref{fig:gm_cc0}.  The spectrum of the CC coefficients shows an
anti-correlation at small scales with a minimum at
$\sim 8$ pix, a steep increase from the minimum to about
30 pix, and a shallower increase at large scales.
The spectrum turns to positive CC coefficients at about 14~pix
approximately matching the offset amplitude. At larger scales, the
correlation becomes stronger as the wavelet filter only measures
structures with sizes above the enhancement and offset scale.
It is evident that the displacement weakens the correlation 
on all scales as compared with Fig.~\ref{fig:noshift}. 
The effect is strongest at small scales, i.e. below the offset
amplitude. The enhancement of the 15~pix scale in the efBm relative
to the original fBm does not affect the recovery of the displacement
vector (Fig.~\ref{fig:fBmm}\,(b)). It is accurately measured 
over all scales $l \lesssim 40$ pix. At larger scales, the finite
map size (see Sect.~\ref{sect_wwcc_gauss}) leads to a small underestimate of the 
amplitude of the displacement vector by $\sim 5\,\% \le 1$~pix.

\begin{figure}
   \centering
   (a)\,\,\,\includegraphics[width=3.1cm, angle=90]{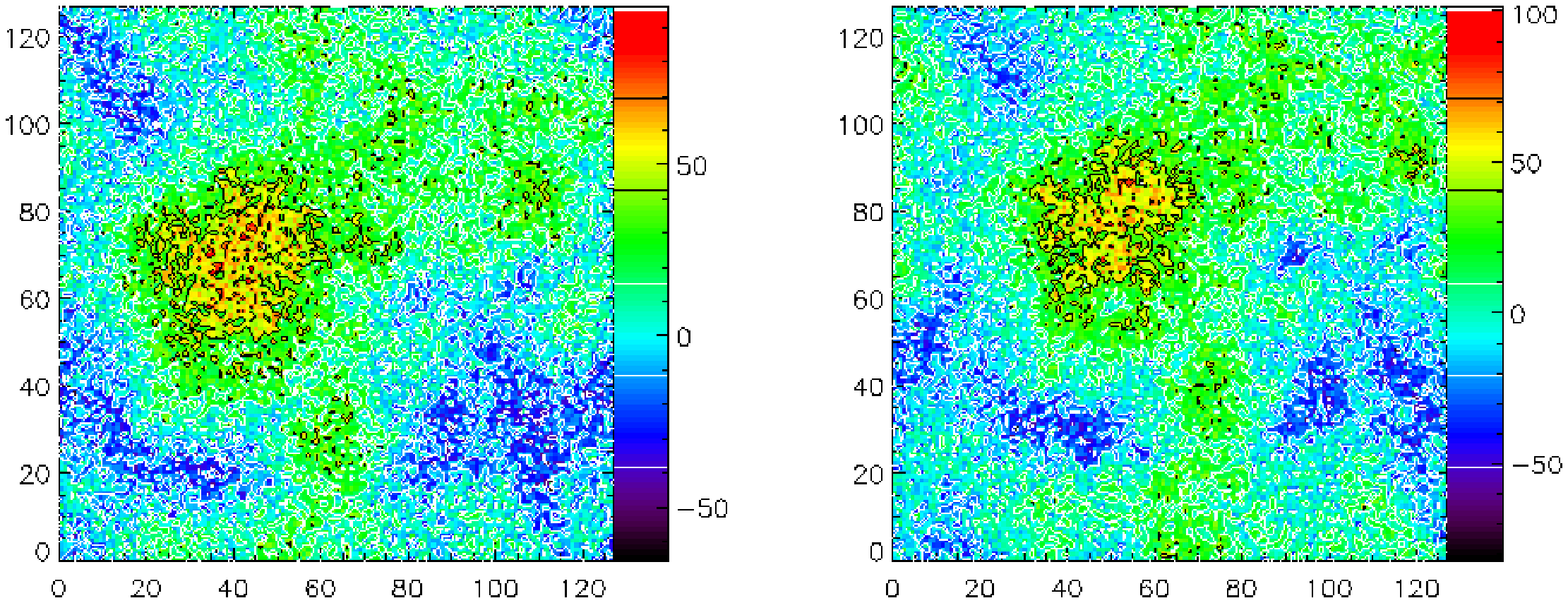}
   (b)\includegraphics[width=6.cm, angle=90]{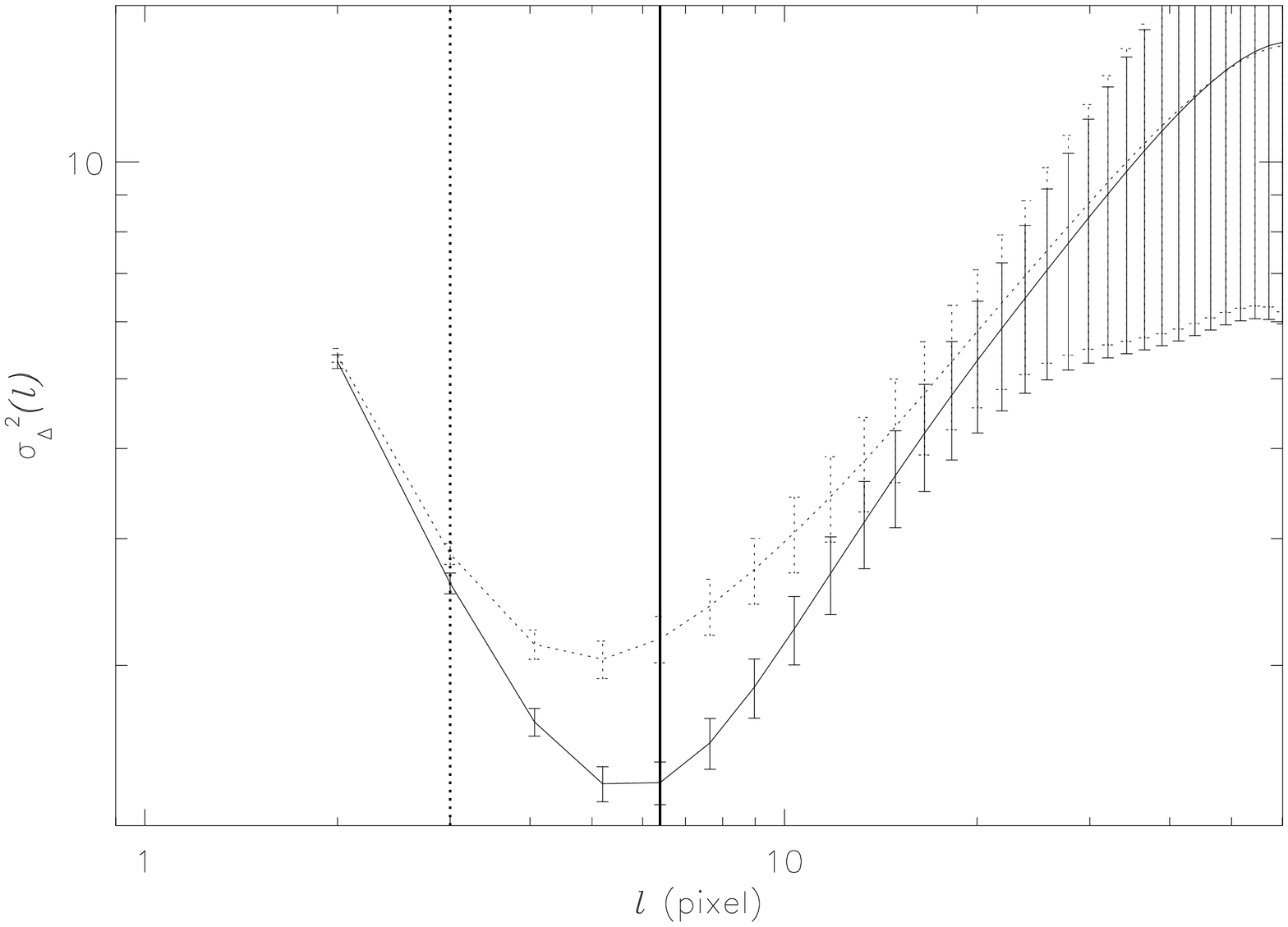}
   (c)\includegraphics[width=6.cm, angle=90]{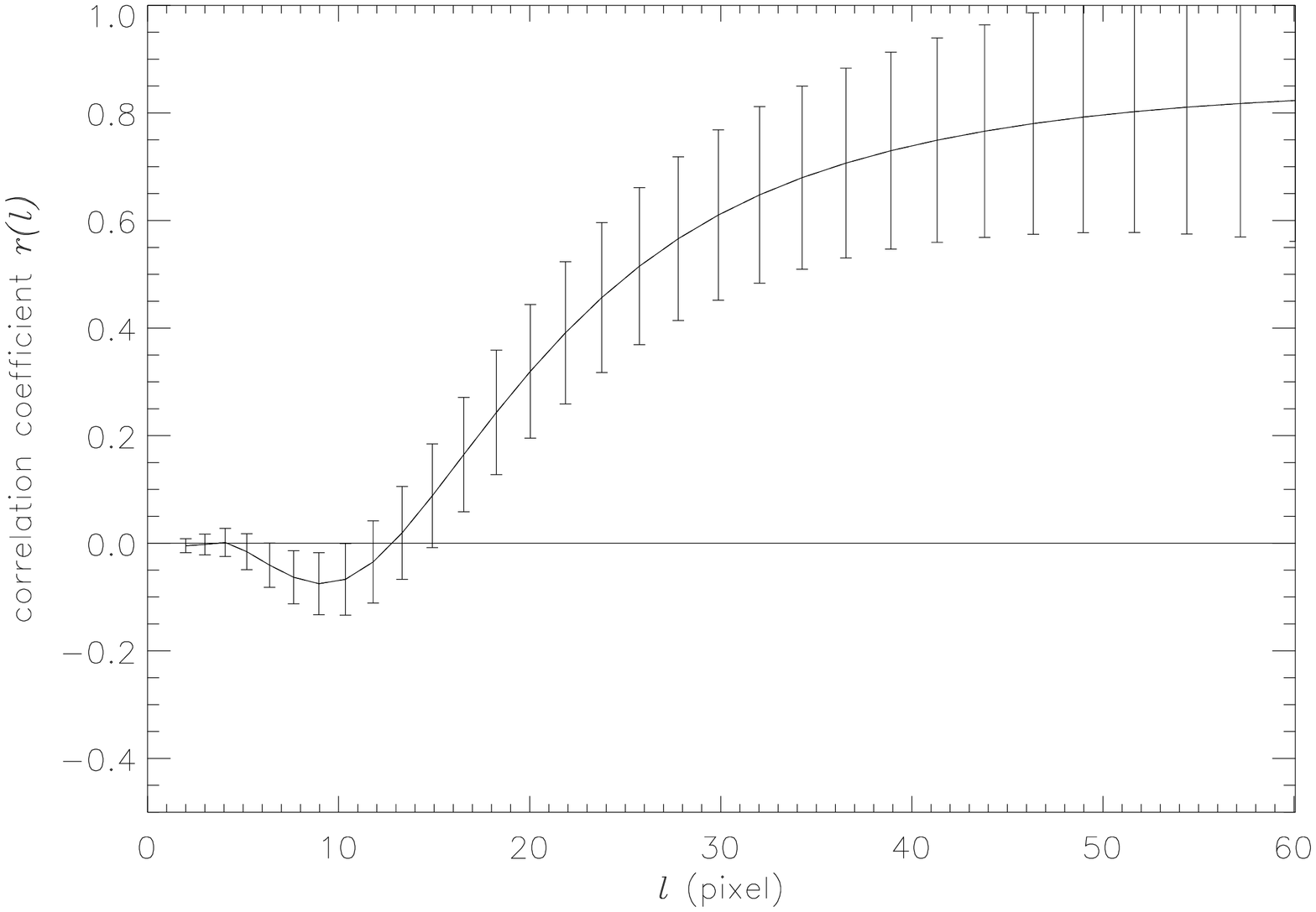}
   (d)\includegraphics[width=3.cm, angle=90, bb=26 23 268 780]{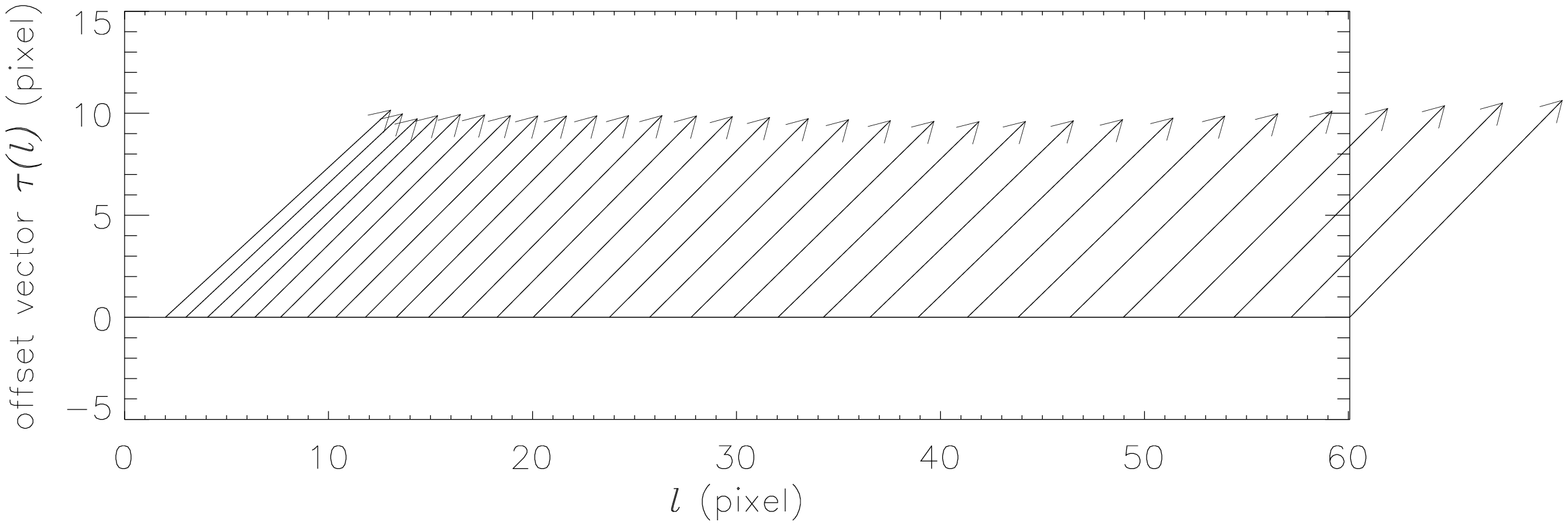}
      \caption{WWCC for two efBm maps ({\changed shown in (a)}, 
       maximum-filtered on scales of 15 pix and 5 pix) generated with a $\mathrm{S/N}=5$.
       The second efBm is shifted by $\tau_i=14.1$ pix ($\tau_{i,x}=10$ pix and $\tau_{i,y}=10$ pix).
       Part (b) shows the $\Delta$-variance spectra of {\changed the first map (full line) and the 
       second map (dotted line)} 
       with marks of critical scales ($l_{\rm c}=6.4$ pix and $l_{\rm c}=3$ pix, {\changed full and
       dotted vertical lines, respectively)} calculated from their noise-corrected spectra, 
      (c) the correlation coefficients and (d) the measured displacement vectors as a function of scale. 
       }
         \label{fig:fBmmn}
\end{figure}

To combine all of this, we finally consider a pair of efBm maps with 
additional noise and an offset by $\tau_i=14.1$ pix ($\tau_{i,x}=10$ pix and 
$\tau_{i,y}=10$ pix, Fig.~\ref{fig:fBmmn}). The first efBm map is generated using a maximum filter
of size $l_{\rm f1}=15$ pix,  the second efBm map with maximum filter of size 
$l_{\rm f2}=5$ pix. The noise in both maps gives a $\mathrm{S/N}=5$.
The maps are shown in Fig.~\ref{fig:fBmmn}\,(a), the $\Delta$-variance spectra
in \ref{fig:fBmmn}\,(b), the cross-correlation spectrum in \ref{fig:fBmmn}\,(c),
and the spectrum of detected displacement vectors in \ref{fig:fBmmn}\,(d). 
The effect of the noise is evident in the $\Delta$-variance spectra
as an increase of the variance towards small scales.  Below $l_{\rm min} \sim 6$ pix
the noise dominates, leading to a zero cross-correlation, but it has
little effect on the recovery of the displacement vectors. The change
of the vector amplitude remains within a 10~\% accuracy. Above the noise scale
the correlation curve behaves in the same way as in the case of 
$\mathrm{S/N}=\infty$ (see Fig.~\ref{fig:fBmm}\,(a)). The noise has no effect on $r(l)$ at scales 
larger than $\approx6$ pix. The amplitude of the displacement vector is 
also slightly underestimated on large scales as in the noise-free case.
The scale enhancement in both structures has basically no effect on the
outcome of the WWCC. It is only visible in the $\Delta$-variance
where the two critical scales can be measured as indicated in 
Fig.~\ref{fig:fBmmn}\,(b)\footnote{In the ideal case considered here, 
a recovery of the critical scale is still possible below the noise-limit
scale $l_{\rm min}$ but for observational data, this may not be reliable
any more (see Sect.~\ref{sect_scale_measure}).}.

\begin{figure}
   \centering
   \includegraphics[width=6.2cm,angle=90]{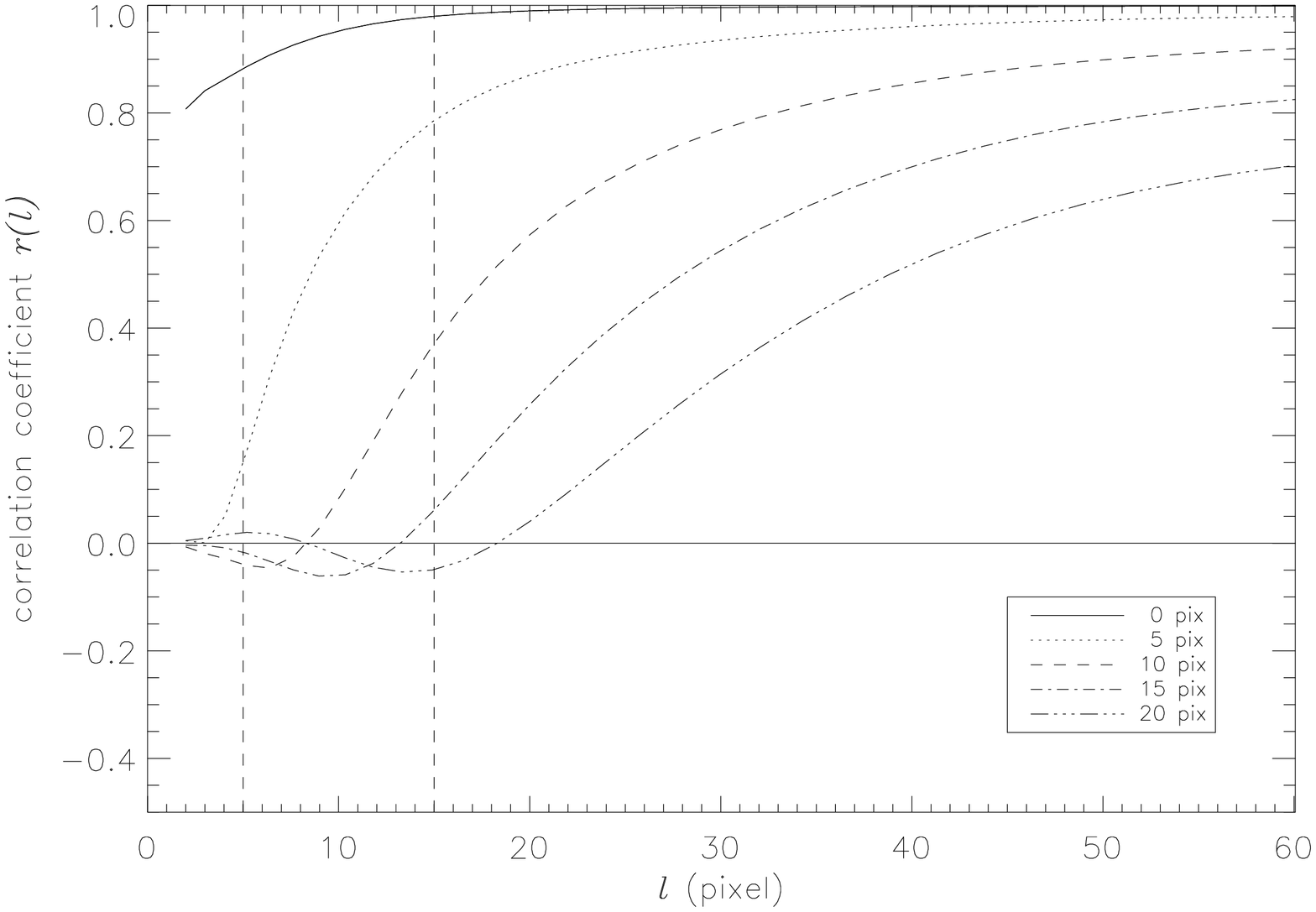}
      \caption{Correlation coefficient as a function of scale for efBm maps  
      filtered with maximum filter of size $l_{\rm f1}=5$ pix and $l_{\rm f2}=15$ pix and 
      offset by 0, 5, 10, 15, and 20 pix (full line, dotted, 
      dashed, dash-dotted, and dash-dot-dot-doted lines, respectively). The dashed vertical 
      lines denote the size of the filters. 
      }   
   \label{fig:cc-l_fbm}
\end{figure} 

To study the quantitative impact of the displacement of the efBm maps on the CC coefficient
more quantitatively, we perform the same experiment as for the Gaussians in
Fig.~\ref{fig:cc-l_gauss}, using two efBm maps filtered with maximum-filter 
sizes of $l_{\rm f1}=15$ pix and $l_{\rm f2}=5$ pix displacing them by $\tau_{i,\,y}=0, 5, 10,
15$, and 20 pix relative to each other. The result is shown in Fig.~\ref{fig:cc-l_fbm}.
The general behaviour is qualitatively similar to Fig.~\ref{fig:cc-l_gauss}. The shape of the
CC coefficient spectrum is regulated by the offset value and ratio between the smaller
to the larger filter size, $l_{\rm f2}/l_{\rm f1}$.
Without any offset between the maps, we find a strong correlation between structures
on all scales. It increases quickly to unity above the scale of the smallest filter 
size, $l_{\rm f2}=5$.  If the structures are offset by 5 pix 
(dotted line) there is no overlap any more below that scale so that the
cross correlation coefficient vanishes. The global shift leads to a lower
correlation on all scales, including the largest ones.
With increasing the offset (dashed, dashed-dotted, and dash-dot-dot-doted lines) 
the $r(l)$ curve shifts to larger scales, a scale range of negative
values (anticorrelation of filtered structures) appears,  
and the correlation becomes weaker on all scales. Obviously, the
WWCC detects the displacement in the same way for the simple Gaussian
structures in Fig. ~\ref{fig:cc-l_gauss} and for close-to self-similar
(e)fBm structures.

\begin{figure}
   \centering
   \includegraphics[width=7.2cm,angle=-90]{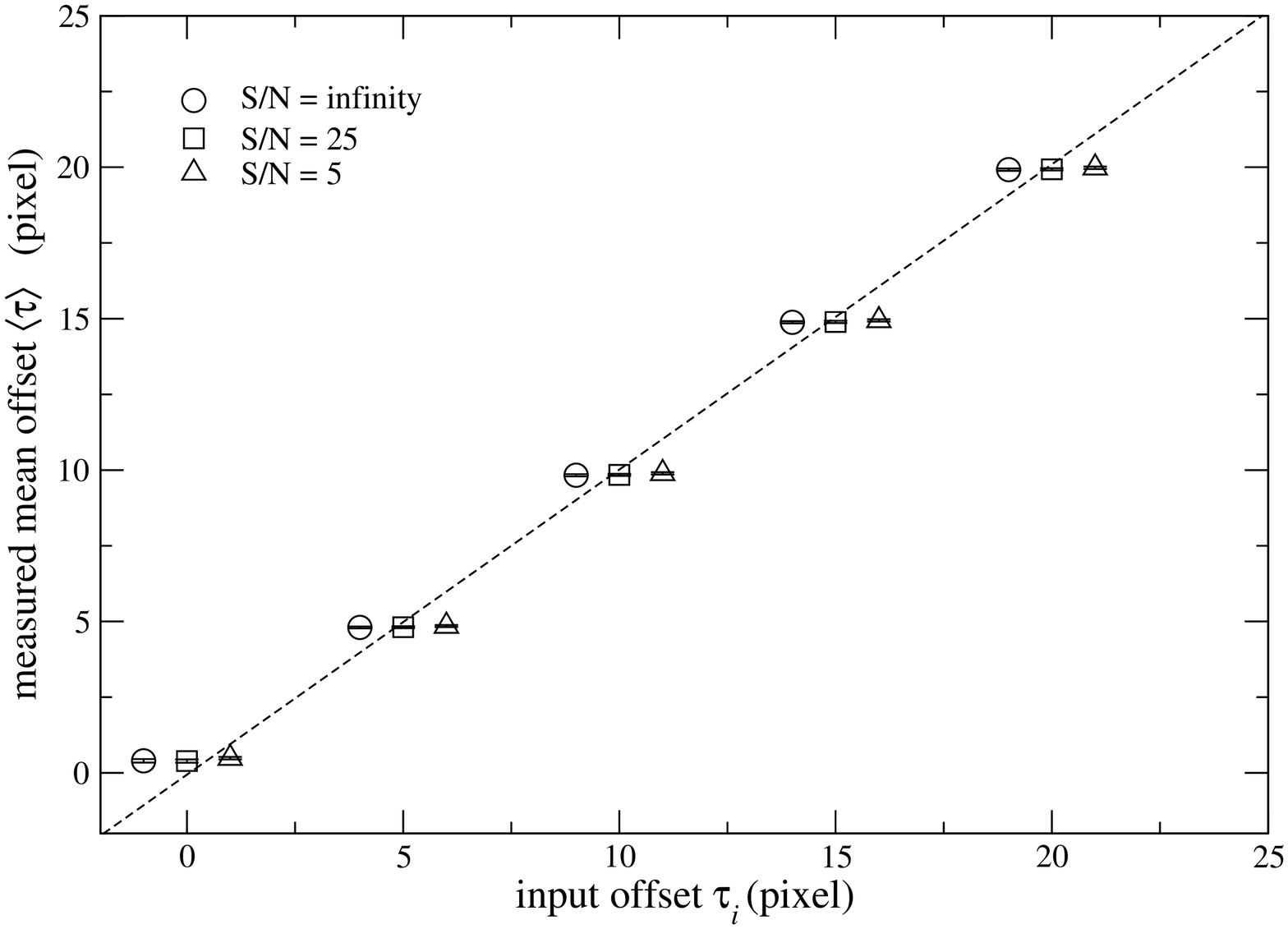}
   \caption{Measured mean 
   displacement ($\langle{\tau}\rangle$) vs. given displacement ($\tau_i$) for 
   $\mathrm{S/N}=\infty$, 25, and 5 (filled circles, squares, 
   and filled triangles, respectively). The circles and triangles are displaced by 
   1 pixel from the identity line (dashed line) for a better visibility. 
   The error bars represent $1\sigma$ variations. }
   \label{fig:fbm_shift-mshift}
\end{figure}

Finally, we test how well the displacement can 
be recovered for pairs of maps maximum-filtered on different scales. 
We filter the original fBm map with maximum filters of size 
$l_{\rm f}= 5, 9, 13, 17, 21, 25, 29$ pix to generate seven efBm maps with $\zeta=3.0$ and $W=1$.
Combining these seven maps (without repetition) gives 21 pairs of distinct maps. 
For a given pair of maps, one map is shifted by $\tau_i = 0, 5, 10, 15, 20$ pix, 
respectively and for each shift we perform the WWCC and recover the displacement 
($\tau$; Eq.~(\ref{eq:tau})) for ten random realizations of the fBm.
Figure~\ref{fig:fbm_shift-mshift} shows the mean recovered displacements
separately for $\mathrm{S/N} = \infty, 25$ and $5$.

Similar to the case of 
the Gaussian circular structures (Fig.~\ref{fig:goff-soff}), the mean displacements 
$\langle{\tau}\rangle$ are recovered with high accuracy for all noise levels
and all offsets less than 1/6 of the map size ($\tau_i \le 20$ pix) also for the 
efBm structures.
We also tested the identity between input and recovered displacement for 
efBm maps generated for all combinations of the 
spectral index $\zeta=2.5, 3$, and 3.5 and the inverse brightness contrast parameter $W=0.5, 1, 1.5$, 
(Eq.~(\ref{eq:FmW})) and found a perfect agreement 
for all combinations of $\zeta$ and $W$. The errors 
of the measured offset depends on the parameter $W$. For large $W$ the offset error 
slightly increases, however, it remains within the resolution limit $\pm 1$ pix.

\section{Application of WWCC to the molecular cloud G\,333}
\label{sec:application}

\begin{figure}
   \centering
   (a)\includegraphics[width=6.cm,angle=90]{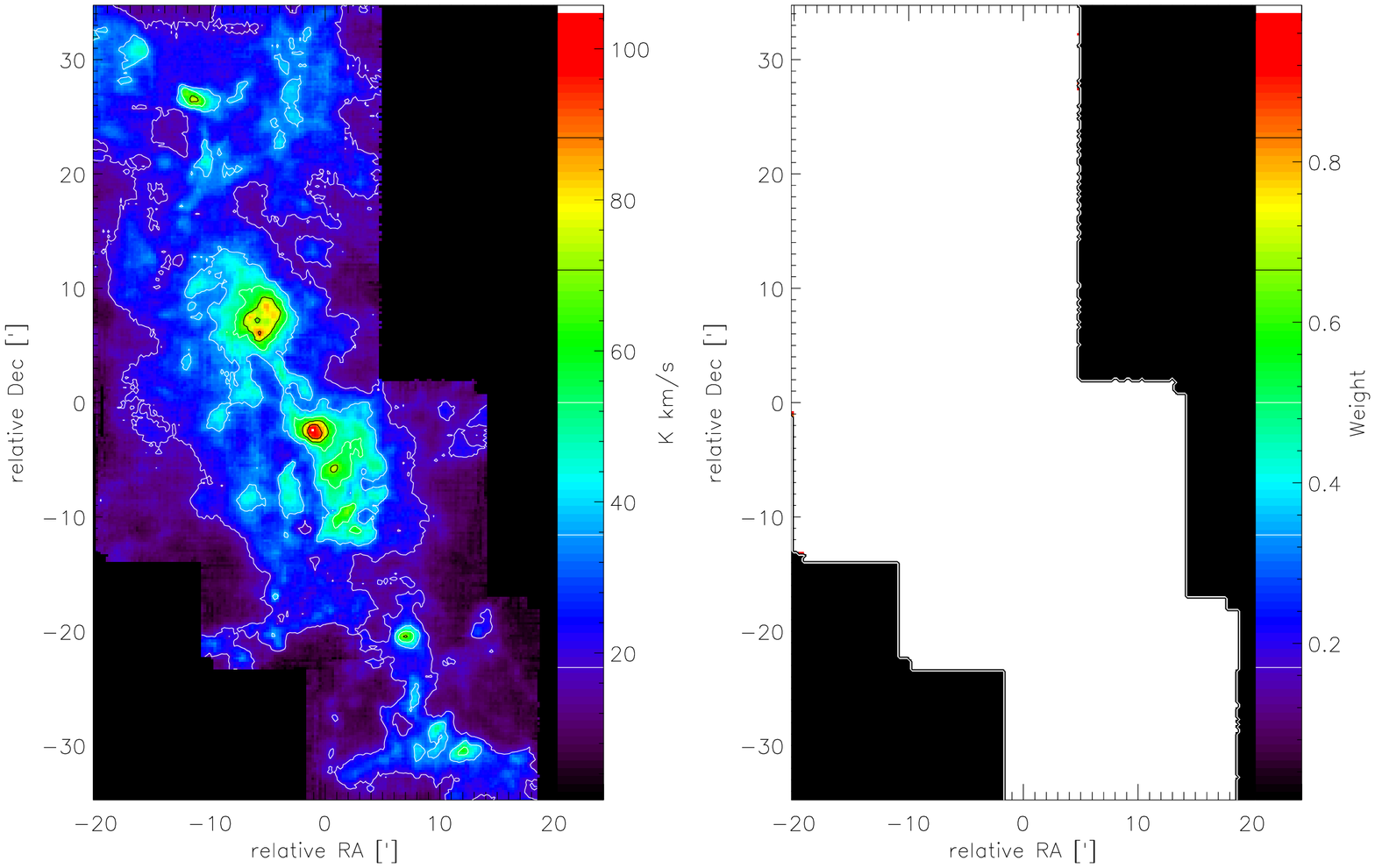}
   (b)\includegraphics[width=6.cm,angle=90]{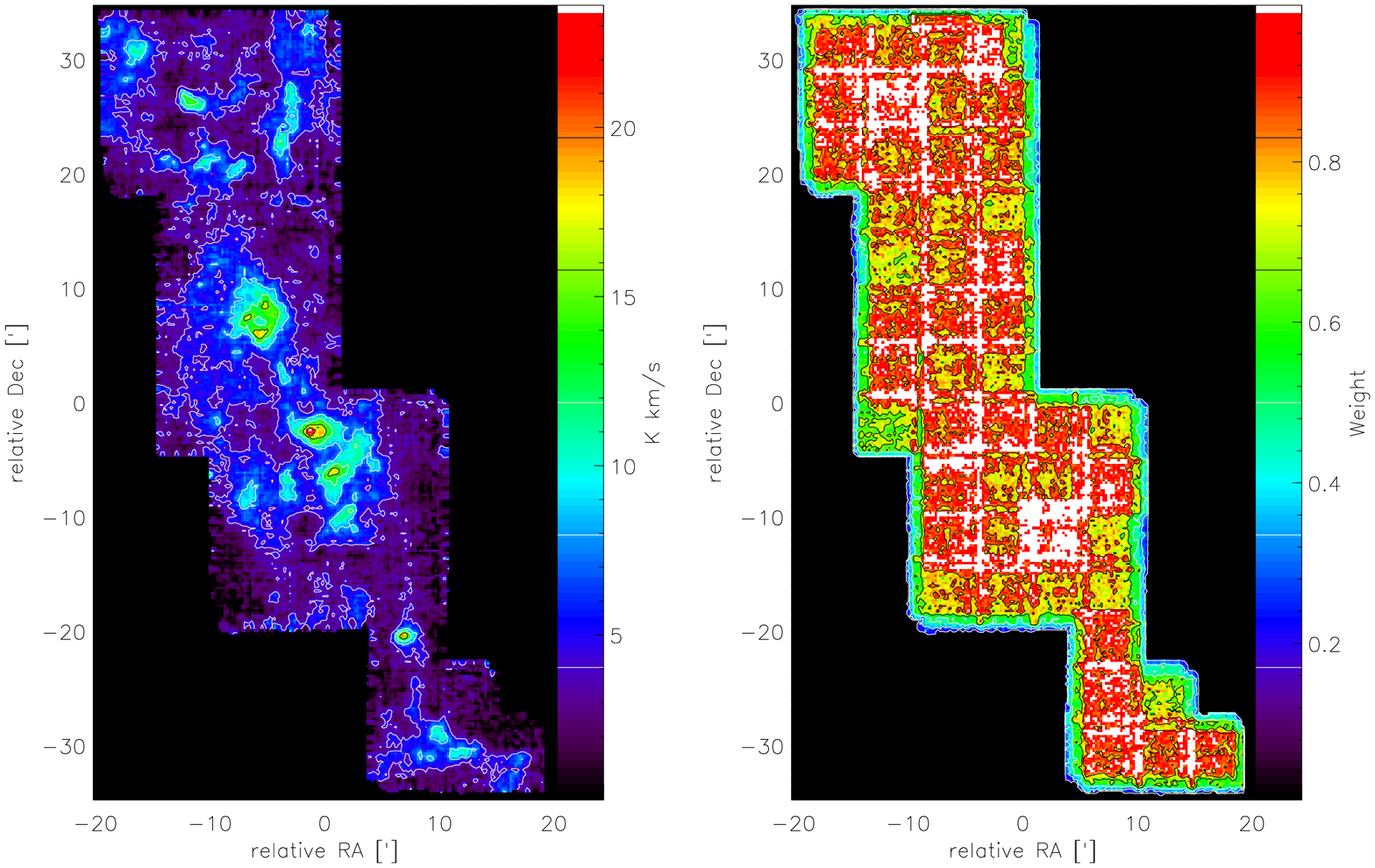}
   (c)\includegraphics[width=6.cm,angle=90]{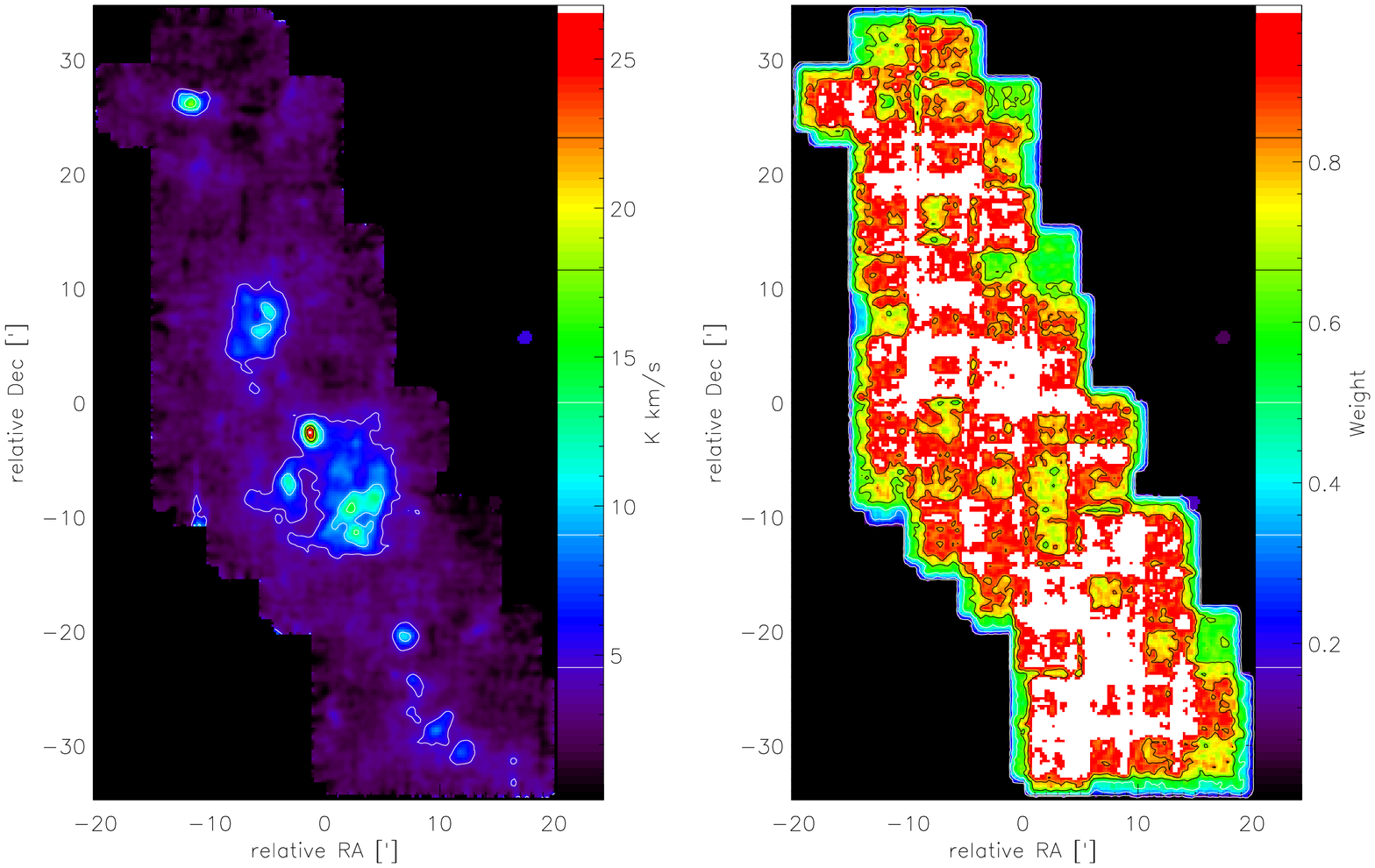}
   \caption{Integrated intensity maps of $^{13}$CO (a), C$^{18}$O (b) and HCN (c) emission lines of 
   the G\,333 taken by \cite{lo09} (left panels) and significance weights $w$ for the different 
   points in the maps (right panels). White color in the weight map indicates a significance
   of unity corresponding to a median signal-to-noise ratio of 25 or above. For the 
   $^{13}$CO map, only pixels close to the map boundary have a lower S/N, i.e. $w<1$.}
   \label{fig:G333}
\end{figure}

To test the application of the WWCC to observed maps we selected 
molecular line maps of the giant molecular cloud G\,333 
obtained with the Mopra telescope \citep{lo09}.
Over 20 molecular transitions were mapped and analysed through
the Principal Component Analysis (PCA) to investigate spatial correlations among different 
line transitions and classify molecules into high- and low-density tracers.
The PCA quantified the similarity of the different maps for eight
significant principal components. Each component can be visualized
and the components follow a hierarchy of sizes,
but the PCA is not able to clearly identify individual spatial scales
at which the similarity of different maps is broken or where gradients
between the mapped structures occur. To prove the value of the WWCC
in identifying these scales, we selected three of the maps from
\citet{lo09} that show a different behaviour in the PCA.
The maps of $^{13}$CO 1--0 and C$^{18}$O 1--0 show very similar eigenvalues
for the first five eigenvectors, indicating a very good correlation
at most scales. The map of HCN 1--0 in contrast only shows the same
eigenvalue for the first principal component, but a strongly
deviating behaviour for basically all other components. Here, we
will try to quantify the differences and commonalities seen by the 
PCA for the three maps in terms of the correlation and displacement 
between structures in the maps as a function of their spatial scales.

To stay comparable to the results of the PCA, we follow the 
same procedure in producing the analysed total intensity maps as performed by \cite{lo09}.
The maps were spatially smoothed with an Gaussian of FWHM=45$''$ = 0.75~arcmin\footnote{The
MOPRA beam at the frequency of the three lines is 36$''$. The
smoothing was performed by \cite{lo09} to allow for a uniform grid
for all lines from the OTF observations and a reduced noise.}
and the molecular lines of $^{13}$CO and C$^{18}$O were 
integrated over the velocity range from 
$-70$ to $-40$ km s$^{-1}$. A wider velocity range from 
$-80$ to $-30$ km s$^{-1}$ was chosen in case of HCN because of visible line wings.
The maps are shown in Fig.~\ref{fig:G333}. {\changed At
the distance of 3.6~kpc of G\,333 \citep{Lockman1979}, the
resolution of 0.75~arcmin corresponds to 0.79~pc.} 

In the next step, we need to generate significance maps (weights)
for the intensity maps to account for the uncertainty of the
information in the individual pixels due to observational noise,
stemming from variable sensitivities and finite integration times.
\citet{ossenkopf08b} proposed to use a saturating linear function of
the inverse noise level as weighting function, 
$w = {\rm min}(1,\alpha\times$S/N) with $\alpha$ being an
appropriate scaling factor. The choice of the saturation level,
i.e. the S/N ratio above which we assume that
the distortion of the structural information by noise is negligible,
is somewhat arbitrary. Based on our numerical experiments in the
previous sections and the inspection of obviously ``noisy structures''
in the maps, we use a value of S/N=25 here, where we define the
signal level as the median of the intensity over the map.

We estimate the noise at each pixel from the emission-free
channels in the velocity ranges from $-80$ to $-75$ km s$^{-1}$ and $-27$ to $-15$ km s$^{-1}$ for
$^{13}$CO and C$^{18}$O. For HCN, we use the velocity ranges $-80$ to $-70$ km s$^{-1}$ 
and $-30$ to $-20$ km s$^{-1}$. 
The significance maps are given next to the integrated intensity maps in Fig.~\ref{fig:G333}.
Due to the high signal-to-noise ratio of the $^{13}$CO observations, 
the S/N=25 limit means that we consider almost all the measured information
in the $^{13}$CO map as reliable, with the exception of a few
pixels at the map boundaries that have a lower integration time
coverage. For the C$^{18}$O and HCN maps, larger regions with
low coverage producing some noisy patterns are weighted 
down by their lower significance in the statistical analysis of the
$\Delta$-variance and the WWCC.
The significance maps basically represent the coverage of the map in the mosaicing observation
with multiple coverages of the individual subfields visible as
square areas in the significance maps. Saturated (white) strips and areas
in the maps of weights indicate a very good coverage, leading to
a low noise so that all details of the lines can be seen.
The combination of the mosaicing observations 
leads to a high significance (low noise) in overlapping areas and 
a lower significance (higher noise) at the map boundaries.
To test the sensitivity of our results on the choice of the saturation
limit we varied the limit by a factor two in both directions but found
no measurable impact on the $\Delta$-variance spectra and only a small
impact on the WWCC on the scale of a few pixels, well within
the error bars. 

 \begin{figure}
    \centering
    \includegraphics[width=6.4cm,angle=90]{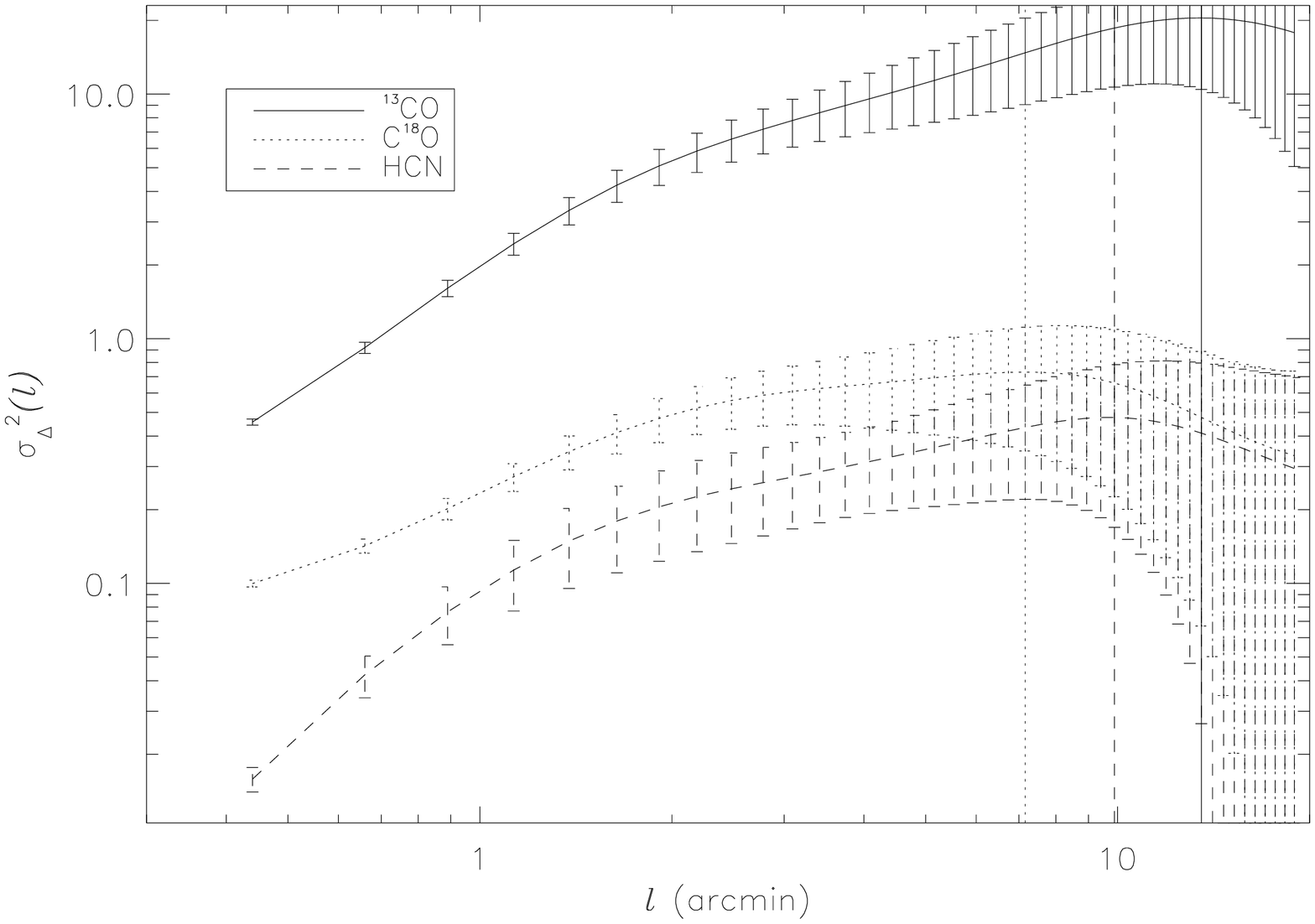}
    \caption{$\Delta$-variance spectra of $^{13}$CO, C$^{18}$O and HCN emission 
    maps (full, dotted and dashed lines, respectively). Prominent scales of the
    $^{13}$CO, C$^{18}$O and HCN maps are marked by vertical full, dotted, and 
    dashed lines, respectively. }
    \label{fig:G333_dvar}
\end{figure}

Figure~\ref{fig:G333_dvar} shows the $\Delta$-variance spectra for the integrated
intensity maps of $^{13}$CO, C$^{18}$O, and HCN. The C$^{18}$O and HCN clouds 
have a dominant scale size of 7.2 and 10 arcmin {\changed (7.5 and 10.5~pc)}, while the $^{13}$CO 
emission has a larger dominant structure with $l_{\rm f} \sim 12$ arcmin {\changed (12.5~pc)}. This
matches approximately to the size of the largest structures with more than half
of the peak intensity level in the individual maps.

\begin{figure}
    \centering
    \includegraphics[width=6.4cm,angle=90]{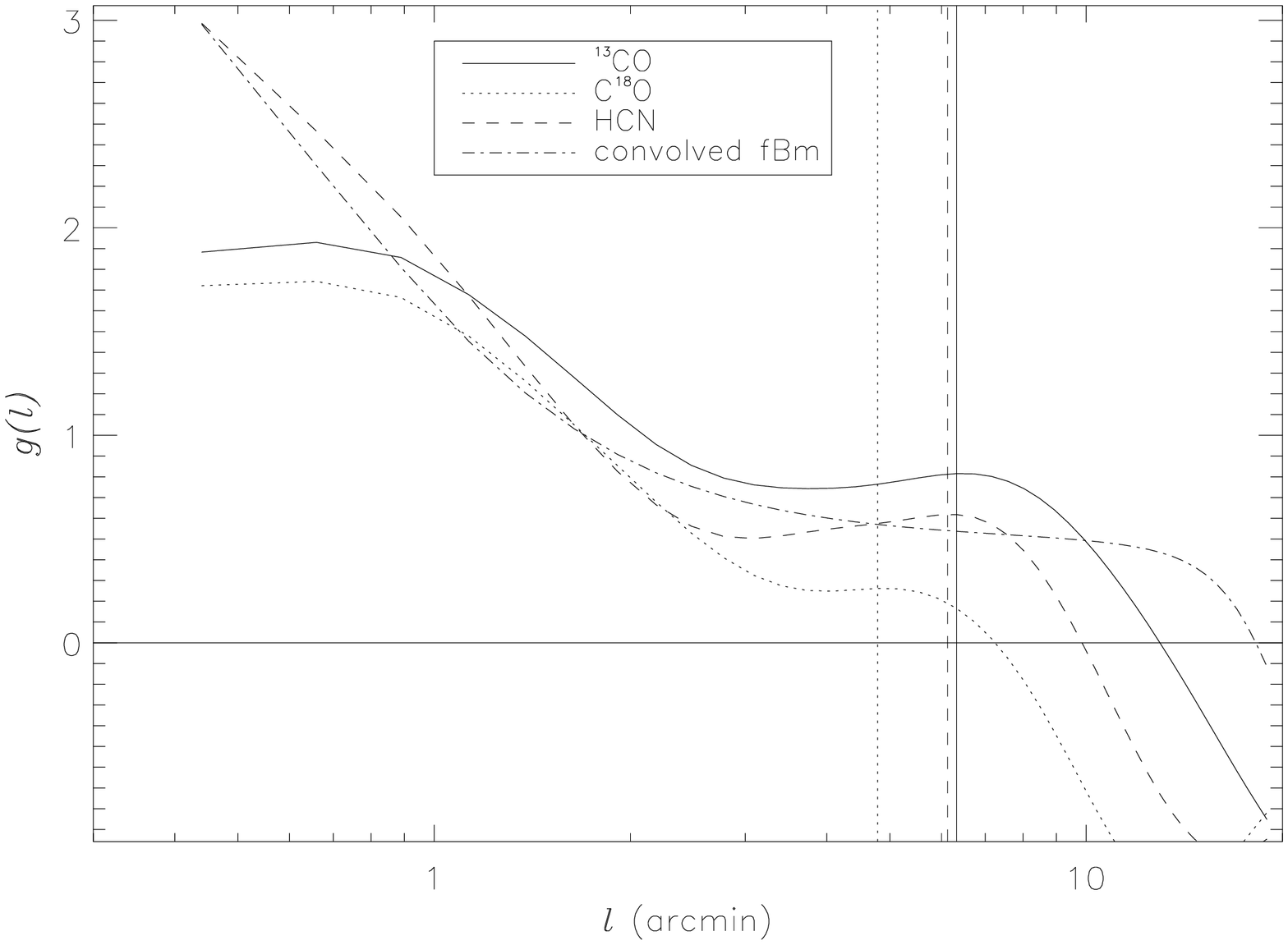}
    \caption{Gradients of the noise-corrected $\Delta$-variance spectra ($g(l)$) of $^{13}$CO, 
    C$^{18}$O and HCN emission maps (full, dotted and dashed lines, respectively). Critical scales 
    of the $\Delta$-variance gradients of $^{13}$CO, C$^{18}$O and HCN are marked by vertical full, 
    dotted, and dashed lines, respectively. The dashed-dotted line represents 
    the gradient spectrum of a pure fBm map ($\zeta=2.5$) convolved with a Gaussian filter of 
    FWHM=0.75 arcmin.}
    \label{fig:G333_dvar-gradient}
\end{figure}
 
Based on the results from Sect.~\ref{sect_scale_measure}, we can also estimate 
the enhanced scale of these maps using the critical scale $l_{\rm c}$ (Eq.~(\ref{eq:lf_StoN=inf_and_5}))
computed from noise-corrected maps (see Appendix A for an example of the noise-correction).
The gradient spectra of the noise-corrected $\Delta$-variance are shown 
in Fig.~\ref{fig:G333_dvar-gradient} in comparison to the spectrum obtained for
a pure fBm with $\zeta=2.5$ convolved with a Gaussian beam of 0.75~arcmin.

In the overall shape, we find a clear similarity between the spectra from the
simulated map and the observed maps. On small scales, there is a peak due
to the blurring of smaller structures through the telescope beam with a filter
size of 0.75~arcmin as discussed already by \citet{bensch01}. This dominates
the gradient spectra below 2~arcmin. A pure Gaussian beam should provide a
monotonous increase towards small scales as seen for the HCN data. The
drop of the gradient spectra for the $^{13}$CO and C$^{18}$O at the two smallest
lags may be either due to an imperfect noise correction or to a relative surplus
of structures at or close to the resolution limit. Therefore, we cannot reliably
quantify the structural properties of the maps at scales below 1~arcmin.

Above the scales affected by the beam, we also find a significant difference
between the convolved fBm and the observed maps. Instead of the flat spectrum
with a drop at scales approaching the map size for the fBm, the observed maps
show a clear peak around 5-6 arcmin {\changed (5.5~pc)} and a quick drop of the gradient towards
larger scales, providing the roots that match the prominent scales of 7-12~arcmin {\changed (7-12.5~pc)} 
discussed above. 
The secondary peaks in the gradient spectrum happen on critical scales $\approx 6.3$~arcmin
for $^{13}$CO and HCN and $\approx 4.8$ arcmin for C$^{18}$O. Using  
Eq.~(\ref{eq:lf_StoN=inf_and_5}) this translates into enhancement scales of
$\approx 12$ arcmin ($^{13}$CO and HCN, {\changed 12.5~pc}) and 9 arcmin 
(C$^{18}$O, {\changed 9.4~pc}), differing 
from the measured pronounced scales of 12, 10, and 7~arcmin by at
most two arcminutes, proving both methods to be approximately consistent.  

We can compare the scales found by the $\Delta$-variance analysis with the 
results from the PCA analysis by \citet[][]{lo09}. The {\changed large-scale
distribution of the emission in all lines is traced by the first 
{\changed principal component} (their Fig.~14), showing 
typical structure sizes of 5--10$'$, but the PCA only measures to what
degree the three maps match to that component. It does not quantify the
match or mismatch in terms of a size scale and it does not measure
the difference in the dominant structure size for the three maps.} 
The clump
analysis also performed by  \citet[][]{lo09} only finds clumps that are much
smaller, with sizes $< 2'$.

Measurements of $\langle g \rangle$  and $g_{\rm max}$ of a cloud allow to estimate 
the spectral index by interpolating between $\zeta=2.5,3,$ and 3.5 
(Fig.~\ref{fig:gmax-gmean_StoN=5}). In the inertial range, we expect to
find $\langle g \rangle = \zeta -2$. To evaluate $\langle g \rangle$
we are restricted to the range not affected by beam smearing, i.e. to $l > 3$~arcmin.
We use the plateau before the second peaks (Fig.~\ref{fig:G333_dvar-gradient}) 
to estimate the spectral indices of the cloud emission structure as
$\zeta_{\rm C^{13}O} = 2.8$, $\zeta_{\rm ^{18}CO} = 2.3$, 
$\zeta_{\rm HCN} = 2.5$.

  \begin{figure}
    \centering
    \includegraphics[width=6.4cm,angle=90]{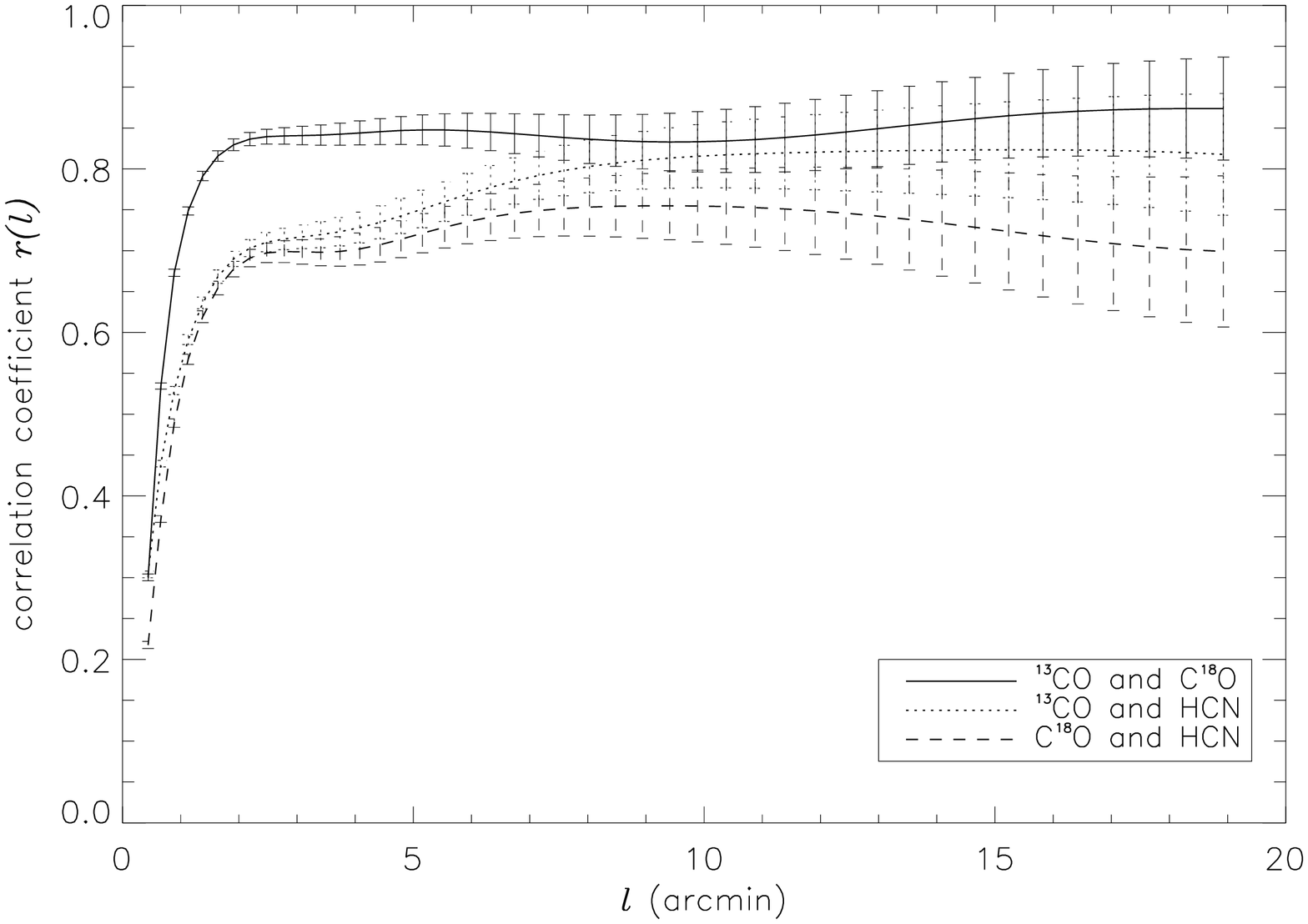}
    \caption{Dependence of the CC coefficient on spatial scale for three pairs of emission maps, 
    $^{13}$CO and C$^{18}$O, $^{13}$CO and HCN, and C$^{18}$O and HCN (full, dotted and dashed lines, 
    respectively).}
    \label{fig:G333_ccf}
 \end{figure}
 
 \begin{figure}
   \centering
   (a)\includegraphics[width=1.6cm, angle=90, bb=27 127 146 750]{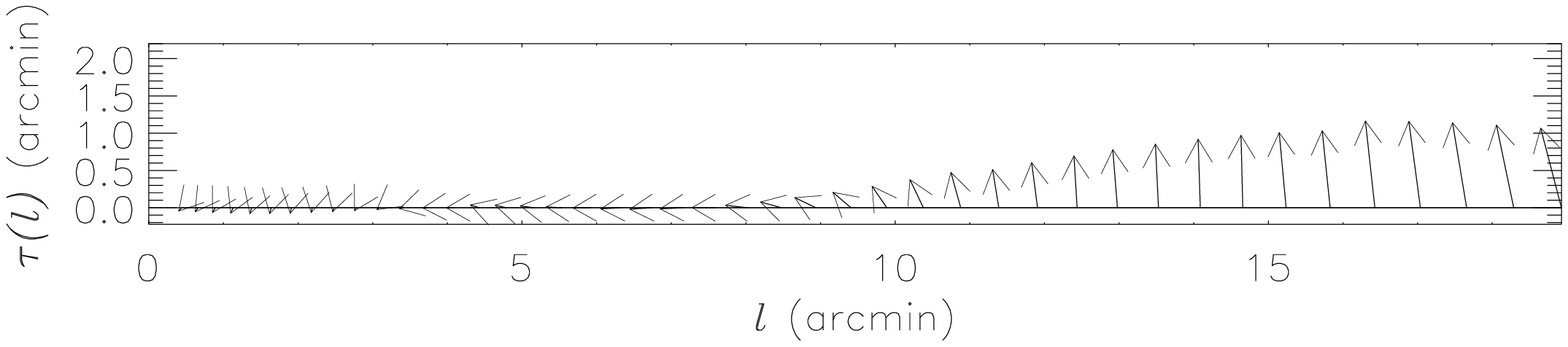}
   (b)\includegraphics[width=2.03cm, angle=90, bb=27 63 179 755]{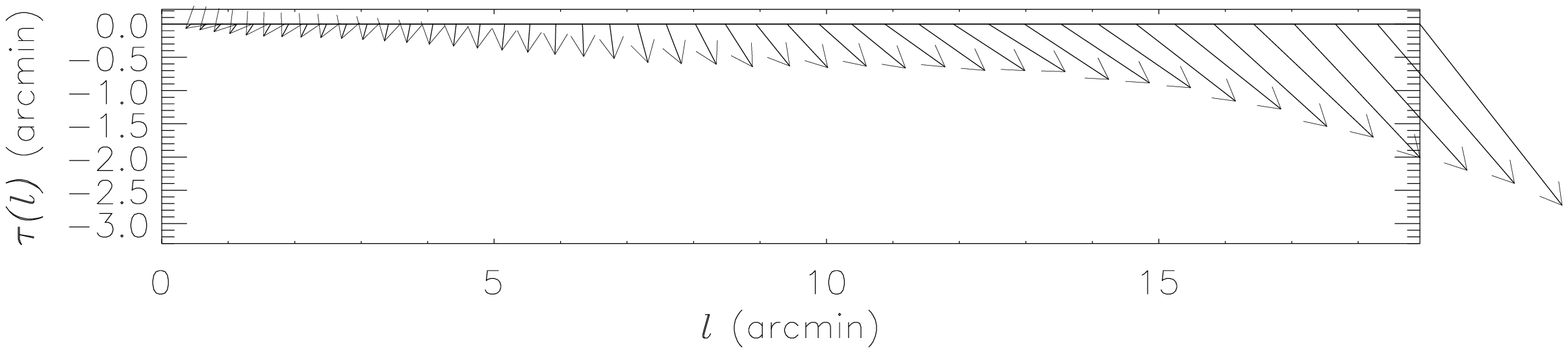}
   (c)\includegraphics[width=2.03cm, angle=90, bb=27 63 179 755]{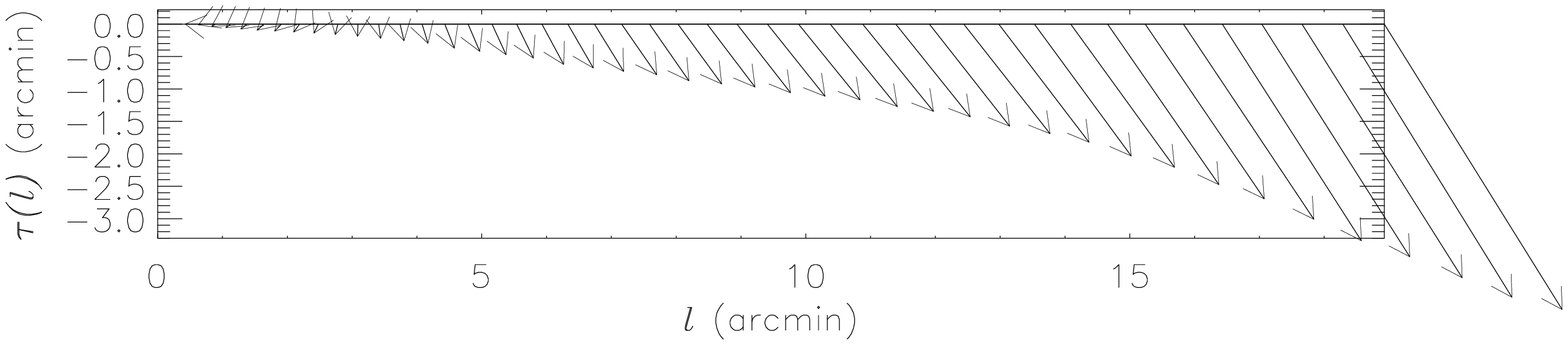}
   \caption{Dependence of the displacement vector on spatial scale for three pairs of emission maps, 
    $^{13}$CO and C$^{18}$O (a), $^{13}$CO and HCN (b), and C$^{18}$O and HCN (c). }
   \label{fig:G333_off}
\end{figure}

In Figs.~\ref{fig:G333_ccf} and \ref{fig:G333_off} we show the results of
the application of the WWCC to the three pairs of maps $^{13}$CO 
and C$^{18}$O, $^{13}$CO and HCN, and C$^{18}$O and HCN, in terms of 
the cross-correlation coefficient spectra and the spectra of displacement vectors.
For the pair of $^{13}$CO and C$^{18}$O maps the cross-correlation coefficient 
shows a strong match ($r(l) \approx 0.85-0.89$) for all scales $> 2$ arcmin
(Fig.~\ref{fig:G333_ccf}).  In spite of the apparent differences in the
two maps, this confirms the strong similarity of the emission in both
molecules, tracing the same conditions of interstellar gas. The difference
at small scales is partially due to the noise and partially due to
optical depth effects in $^{13}$CO at the densest spots in the map.
The fact that the correlation does not reach unity at larger scales must be
basically due to the insufficient excitation of C$^{18}$O in low density gas
leading to the apparently smaller size of the emitting area and to a lesser degree a result of 
relatively small systematic displacement. The WWCC
result, however proves that this effect is scale-independent (above 2~arcmin),
i.e. we find the same amount of thin gas, dark in C$^{18}$O but bright in
$^{13}$CO on all spatial scales in the molecular cloud. This is fully in
line with the fractal description of molecular clouds.

When comparing the CO isotopes with HCN, however, we find a weaker
correlation on all scales and in particular a significant drop of the
CC coefficient at the scale below $\approx 7$~arcmin {\changed (7~pc)}. The weakest correlation
is found between C$^{18}$O and HCN ($r(l>2 \,\,{\rm arcmin}) \approx 0.7$). 
The increase of the CC coefficient at 8~arcmin scale is also visible here 
but less pronounced. This weak correlation contradicts the frequently
made assumption that both species are high-density tracers, {\changed being 
sensitive to the same dense gas concentrated in small clumps and
cores. As both molecules have very different critical densities
our weak correlation at small scales probably reflects different
excitation conditions there, i.e. a thermalization of the CO isotopes
over larger scales (lower densities) compared to HCN. As all high-density
clumps are embedded in lower density envelopes, the correlation
is still well above 0.5, but significantly different from unity.}
Possible physical reasons for kink of the CC curve at the scale of 
$\approx 7$ arcmin will be discussed in Sect.~\ref{sec:interpretation}.

As the PCA is based on the computation of the cross-correlation matrix
(Eq.~(\ref{eq:ccf})) we can directly compare our scale dependent
cross-correlation coefficients (Eq.~(\ref{eq:wwccf}), Fig.~\ref{fig:G333_ccf})
with the global cross correlation coefficients already computed by \citet{lo09}.
They measured coefficients of 0.88, 0.79, and 0.75 for $^{13}$CO and C$^{18}$O,
$^{13}$CO and HCN, and C$^{18}$O and HCN maps, respectively, when comparing
all pixels of their integrated intensity maps. {\changed \citet{lo09} did
not specify the uncertainty of the coefficients, so that we repeated their
computation of the correlation coefficients and obtained errors of about 0.01.} 
When we compare these {\changed global CC coefficients} with the average of the spectrum of the CC 
coefficient $r(l)$ over the full range of spatial scales,
the scale-averaged CC coefficients are slightly lower ({\changed $0.82\pm0.04$, $0.75\pm0.05$, and $0.69\pm0.05$} for 
$^{13}$CO and C$^{18}$O, $^{13}$CO and HCN, and C$^{18}$O and HCN maps) than the
global ones computed without weighting function. The difference is {\changed marginally significant
and can be} due to two effects: the inclusion of the weighting function tends to lower
the CC coefficients on large scales by 0.02-0.05, and the small CC coefficients 
on the noise-dominated scales $\lesssim 2$ arcmin further reduce the average
cross correlation coefficient. When restricting ourselves to scales without
significant noise distribution, $l > 2$ arcmin, we obtain
numbers closer to the direct correlation coefficients measured from PCA
(0.85, 0.78 and 0.73). This indicates that the global cross-correlation coefficient
traces differences in the maps on scales independent from noise quite
well, but it does, of course, not deal with a variable reliability of the
data and it cannot distinguish between differences on 
different noise-free scales, such as the varying correlation with HCN below and
above 7$'$. The WWCC method is capable of recovering the CC coefficients,
also measured by the PCA, and, in addition,   
it allows the analysis of the CC coefficients on scale-by-scale basis. 

Figure~\ref{fig:G333_off} shows the distribution of measured displacement vectors 
as a function of scale for the three pairs of maps\footnote{Note that the sum of 
the recovered displacement vectors on large scales does not exactly add up to zero
because of the finite map size. The boundary treatment tends to lead to a small
underestimate of the displacement amplitudes on large scales, 
as was shown for the displaced Gaussian circular structures (see Fig.~\ref{fig:gm_dis}).}. 
There is no displacement between the structures of $^{13}$CO and C$^{18}$O clouds 
(Fig.~\ref{fig:G333_off}\,(a)) on small scales ($\lesssim 10$ arcmin), while the structure is 
offset at larger scales with an amplitude of $\sim 1$ arcmin, where all
structures in $^{13}$CO are shifted to the south-west relative to C$^{18}$O. 
If we consider the pairs with HCN, however, we find a much larger offset
at large scales (Fig.~\ref{fig:G333_off} (b) and (c)). There is a gradual increase of the offset 
up to $\approx 3.5$ arcmin (HCN relative to $^{13}$CO) and up to 
$5$ arcmin (HCN relative to C$^{18}$O), where HCN is offset relative to the CO isotopes 
in south-west direction along the filament.

\subsection{Identification of correlated structures}
\label{ssec:identification}

\begin{figure}
   \includegraphics[width=6.cm,angle=90]{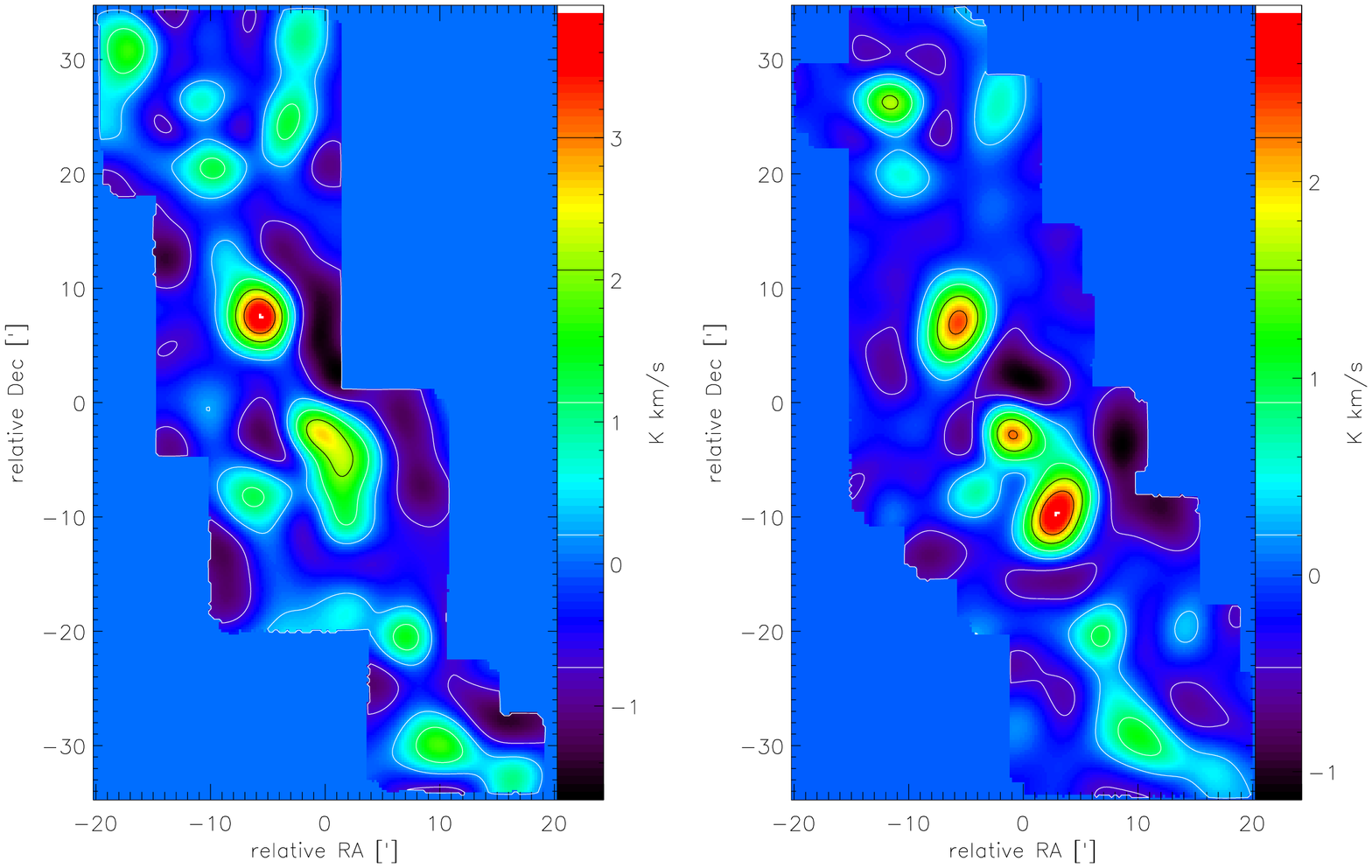}
   \includegraphics[width=5.7cm,angle=90]{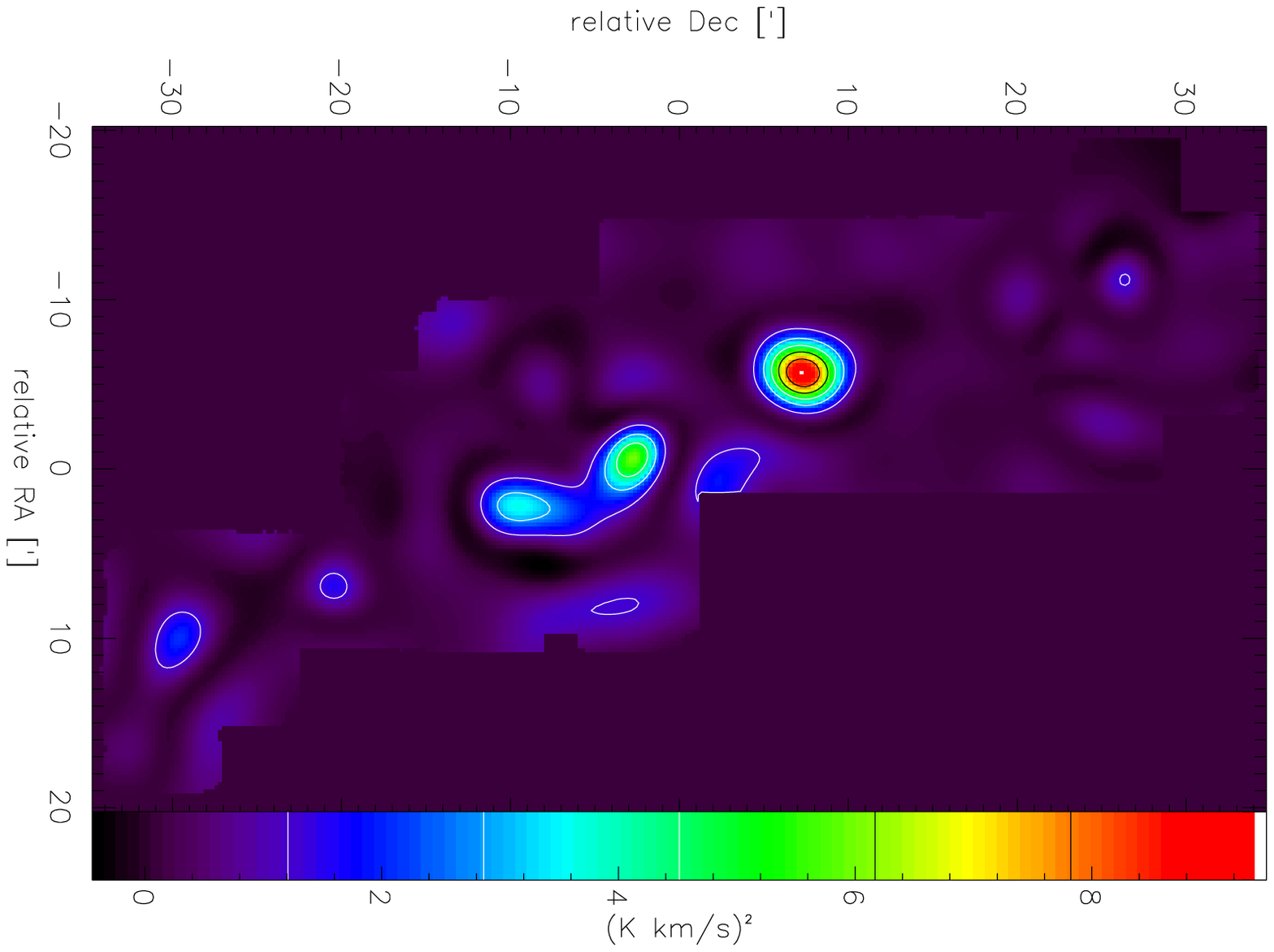}
       \caption{\emph{Top panels.} Integrated intensity maps of C$^{18}$O and HCN emission lines 
       wavelet-filtered on scale of $l=5$ arcmin (left and right panels, respectively). 
       \emph{Bottom left.} The product map generated by multiplying the two wavelet-filtered maps (top panels).  
      }
   \label{fig:g333_5pix}
\end{figure}

\begin{figure}
   \centering
   \includegraphics[width=6.cm,angle=90]{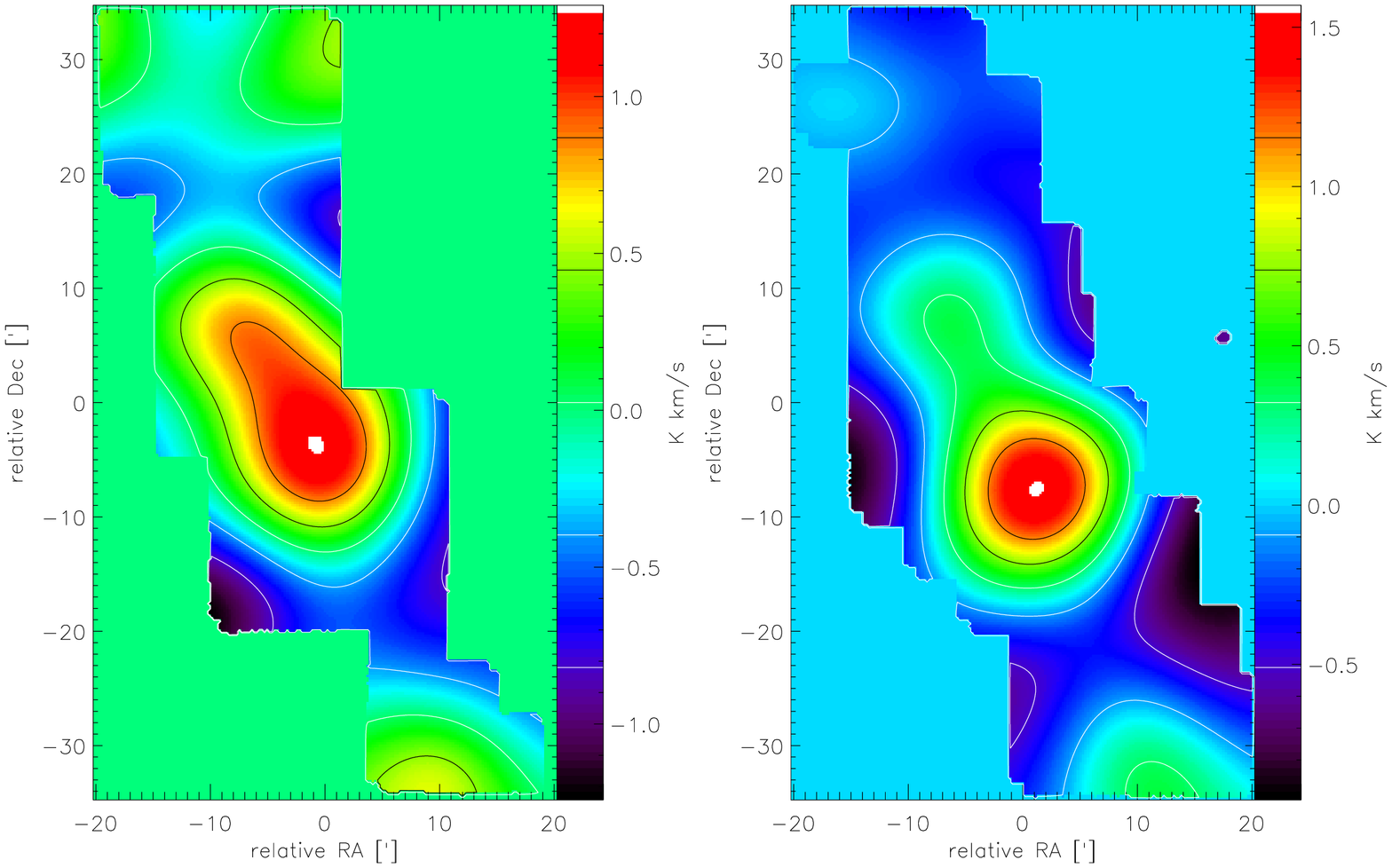} \\
   \includegraphics[width=6.cm,angle=90]{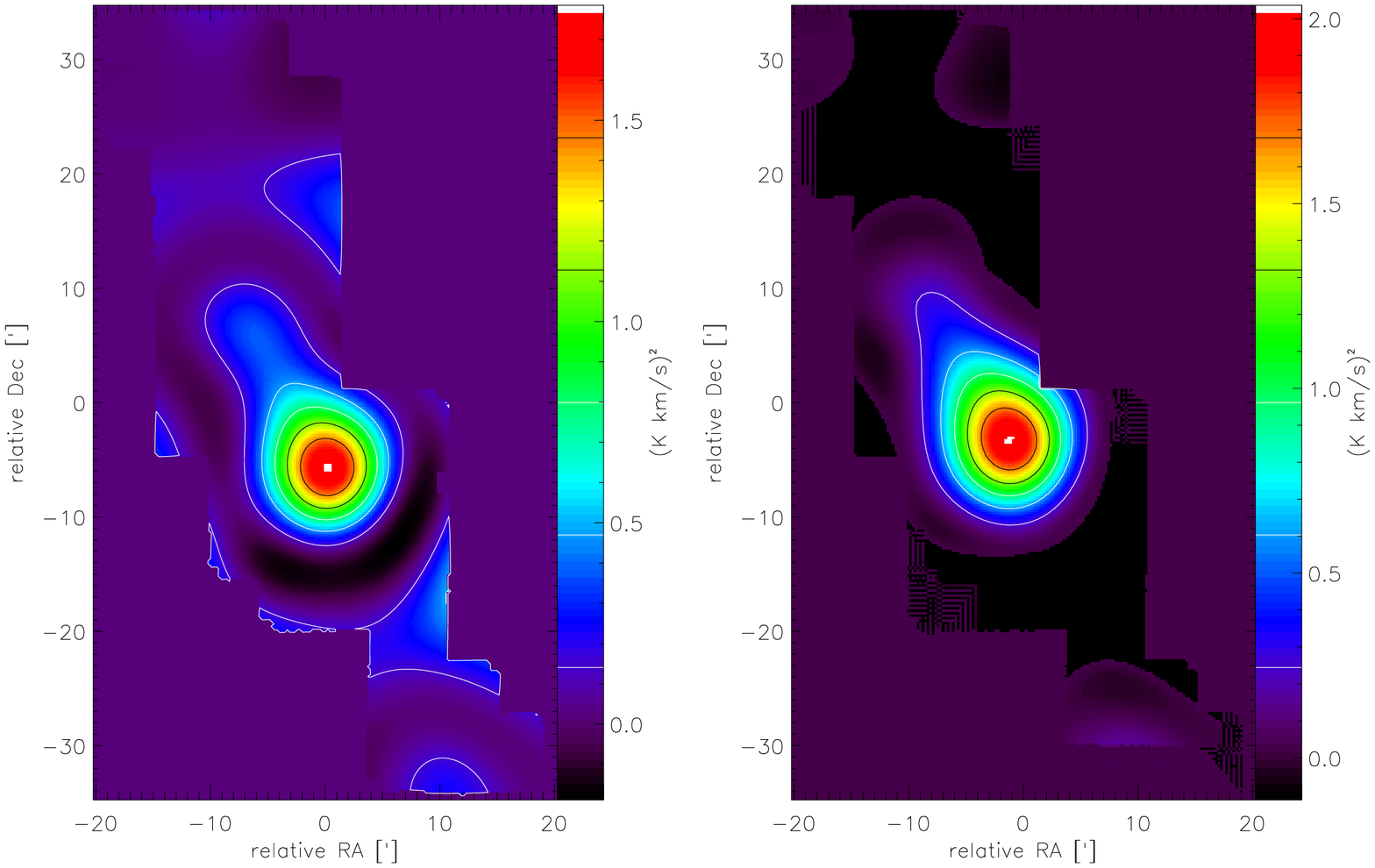}
      \caption{\emph{Top panels.} Integrated intensity maps of C$^{18}$O and HCN emission lines 
       wavelet-filtered on scale of $l=18$ arcmin (left and right panels, respectively). 
       \emph{Bottom left.} The product map generated by multiplying the two wavelet-filtered maps (top panels).  
       \emph{Bottom right.} The shifted product map of the top left and top right maps where the latter one 
       is shifted by measured offset in the North-East direction ($\tau_x=-2.8$ arcmin and $\tau_y=4.5$ arcmin). 
       Note the increase of the signal for the shifted combination compared to the
       direct product.
      }   
   \label{fig:g333_18pix}
\end{figure}

To understand the nature of the structures that are actually responsible for the
measured shift and correlation we go one step back and look at the individual
filtered maps and compare them as for the fBm maps in Sect.~\ref{sect_fbm_displaced}.
Based on the outcome of the WWCC, we examine here structures with a scale of 5~arcmin, i.e.
below the kink in the cross-correlation spectrum at 7~arcmin, and structures at
a scale of 18~arcmin, i.e. showing the maximum offset between the maps,
for the pair of C$^{18}$O and HCN maps.

Figure~\ref{fig:g333_5pix} (top panels) shows the maps wavelet-filtered on 
the scales of 5 arcmin. The correlation coefficient of the maps at this
scale is about $r(l=5^{'})\approx 0.7$ and the offset between them is small
$\tau(l=5^{'}) \approx 0.5^{'}$. In the figure, we can identify two relatively
bright cores in C$^{18}$O in the central region of the cloud, 
one of them also matching a peak in HCN, while the second one breaks up 
into two knots in HCN.
The peak in the filtered HCN map is further offset to the south-west, not 
having any direct counterpart in C$^{18}$O. 
Multiplication of these maps ($P(\vec{x},l=5^{'})$; see Eq.~(\ref{ssec:fbm1.1}),
Fig.~\ref{fig:g333_5pix}, bottom left) reveals three relatively bright correlated 
features at the locations of the three knots seen in HCN. In the northern core
(relative R.A. $\approx -2^{'}$, Dec $\approx -2^{'}$) we find only a small 
shift of the HCN peak relative to the C$^{18}$O peak. In the southern core
we find one feature (R.A. $\approx -7^{'}$, Dec $\approx 7^{'}$) that is
prominent both in C$^{18}$O and HCN. The third peak (relative R.A. $\approx 0.3^{'}$ and 
Dec $\approx -10^{'}$) stems from the global peak in the filtered HCN map
that has no bright counterpart in the C$^{18}$O map, but falls on a relatively 
faint elongated region there. The matching peaks dominate the cross correlation
while the southern knot drives the measured offset. The WWCC result
at small scales combines the contribution from all three features.
The two matching knots, and in particular the northern core, brightest in C$^{18}$O,
are very similar in both maps, probably caused by warm and dense core on the
scale of $\la 8$~arcmin, have similar emission profiles in both transitions.
They create a cross-correlation coefficient of about 0.7 and do not contribute
to any measured offset. In the third feature, HCN is clearly offset from
the C$^{18}$O emission. It hardly contributes to the cross-correlation function,
but it produces the measured small global offset.

In Fig.~\ref{fig:g333_18pix} we show the same analysis for the C$^{18}$O and 
HCN maps wavelet-filtered on the scale of 18~arcmin. The WWCC showed a 
cross-correlation coefficient $r(l=18^{'}) \approx 0.7$ level and a displacement
of the HCN structure by $\tau_x=2.8$ arcmin and $\tau_y=-4.5$ arcmin in south-west
direction relative to C$^{18}$O (Figs.~\ref{fig:G333_ccf} (dashed line) and 
\ref{fig:G333_off}\,(c)). The filtered maps are dominated by a
cometary structure with a relatively similar shape, but a clear mutual offset.
The simple product map (Fig.~\ref{fig:g333_18pix}, bottom left panel)
reflects this shape but the displacement between the maps adds a negative arc-like 
distribution in south-west direction and leads to a low amplitude of the
product maps of at maximum 1.8. 
When accounting for the offset in shifting the HCN map by the measured displacement
vector from the WWCC before computing the product map (bottom right panel), the 
negative arc-like structure disappears, the signal becomes more concentrated 
in the center of the cometary structure, and the amplitude of the product map
grows to 2.1. The weakening of 
the correlation coefficient on large scales (Fig.~\ref{fig:G333_ccf}, dashed line) 
can be attributed to the offset between C$^{18}$O and HCN structures which 
progressively increases towards large scales (Fig.~\ref{fig:G333_off}\,(c)). 
It becomes clear that at large scales, the WWCC is dominated by the southern core 
also seen at small scales. This is systematically offset between HCN 
and C$^{18}$O along the filament and shows a cometary-shape structure at that scale.

\subsection{Interpretation}
\label{sec:interpretation}

The $\Delta$-variance spectra show the largest difference between the
$^{13}$CO map and the C$^{18}$O map while the HCN map exhibits an
intermediate behaviour (Fig.~\ref{fig:G333_dvar}). The C$^{18}$O map 
has a shallow spectrum with the
smallest prominent size, i.e. a relative surplus of small-scale structures
relative to the other two maps. This can be seen by eye in Fig.~\ref{fig:G333}.
All emission structures and in particular the peaks are smaller than
the corresponding structures in the other two tracers. This should be mainly 
an optical depth effect. Following \citet{lo09} most clumps in the source
are optically thick in $^{13}$CO so that one only sees an outer ``$^{13}$CO-photosphere''
while C$^{18}$O traces the full centrally-peaked column density structure.
HCN 1--0 has an intermediate optical depth. Consequently, the optical depth
effects produce exactly the structure blurring discussed in Sect.~\ref{sec:fbm2}.
As a result the $\Delta$-variance spectrum becomes steeper and the pronounced
structures appear larger than in the underlying column density profile. The
true structure is probably best traced by the C$^{18}$O map.

In contrast to this effect, we find a much weaker spatial correlation between
the HCN map and the other two maps compared to the mutual correlation between
the two CO isotopologue maps. $^{13}$CO and C$^{18}$O emission line maps are
strongly correlated above the noise scale. This is explained by the fixed
relative abundance and similar excitation conditions, with a critical density 
of only $2\times 10^3$~cm$^{-3}$, that lead to a match of the locations of all
local features. In that sense, the opacity blurring is only a second-order
effect. The small offset of the largest large structures between $^{13}$CO and C$^{18}$O
measured in the WWCC indicates that the optical depth is somewhat higher
in the south of the C$^{18}$O peaks compared to their northern environment,
so that the corresponding larger $^{13}$CO emission pattern is slightly shifted to
the south of the C$^{18}$O peaks. This means that the column density profile
around the cores should have shallower decrease to the south compared to the north.

The globally lower correlation coefficient between HCN and the CO isotopologues
indicates a systematically different structure on all scales. We find in particular
a clear ``kink'' towards smaller coefficients at scales below 7~arcmin.
At those scales we see in Fig.~\ref{fig:g333_5pix} that the emission of the
two main cores in the cloud is very different for HCN and C$^{18}$O. In the
northern core (-6$'$,+6$'$) the HCN emission is systematically shifted to the south relative to
C$^{18}$O and in the southern core (0,-10$'$) HCN even peaks at a location south of the
C$^{18}$O that is not prominent in C$^{18}$O at all. This systematic difference
could indicate {\changed a chemical transition or a density gradient at the
measured scale of about 7~pc.} The HCN emission peaks in the south of the column density
peaks traced by C$^{18}$O could stem from an increased HCN abundance, e.g. due
to UV illumination \citep[see e.g.][]{fuente05}, or from an increased volume
density. As HCN 1--0 has a critical density of $10^6$~cm$^{-3}$, it traces
much denser gas than the CO isotopologues. {\changed The measured scale is
significantly larger than the size of individual clumps in G~333, measured
by \citet{wong08} and \citet{lo09} to fall between about 0.4 and 2~pc,
i.e. it must reflect a global process that affects the whole cloud. As such
a process continued sequential star-formation was proposed by \citet{nguyen15}.
It would be in line with the compression of gas by newly formed stars.
\citet{wong08} have observed this process on a scale of 9~arcmin around
G~333.6-0.2, the northern region in our map, but that region contributes only
to a small degree to the statistics in our whole map.
The inclusion of more species that are separately sensitive either to different
densities or to chemical changes produced by UV radiation from young stars 
should allow to distinguish between both effects.}

The systematic offset of the HCN structures relative to structures of $^{13}$CO 
and C$^{18}$O at large scales (Fig.~\ref{fig:G333_off} (b) and (c)) can be understood
from the emission filtered patterns on large scales shown in Fig.~\ref{fig:g333_18pix}.
The structures correlated on scales of 18~arcmin have a cometary-like structure oriented 
in the same north-east direction. The HCN peak at that scale sits at the head
of the structure. The measured displacement vector follows the cometary structure
due to the shift of the individual HCN peaks towards the south-east of the corresponding
C$^{18}$O peaks. The combination of this structure and the derived displacement vector
indicate a global anisotropy either in the density structure or in the chemical
structure. A density structure that is consistent with the observations asks for
a statistically higher density in the south-west compared to the north-east. This
could be a global feature or a structure where every core shows a low density tail 
towards the north-east. This would also explain the small shift between $^{13}$CO
and C$^{18}$O due to the higher optical depths associated with higher densities in the
south-east. A global anisotropy in the abundance structure asks for a preferred
direction of the external UV illumination of the cloud. This is not impossible,
but there are no independent observational evidences for such a field from the
surrounding stars.

Altogether we see that the combination of the $\Delta$-variance analysis and the 
WWCC allows {\changed in principle} to distinguish the structural effects of optical
depth, excitation, and variable abundance as a function of spatial scale. {\changed
For a real discrimination between the three effects, however, more data need to be 
included in the analysis, covering well-defined ranges of critical densities, 
chemical formation conditions, and optical depths.}

\section{Summary}
\label{sec:discussion}

We developed the weighted wavelet-based cross correlation (WWCC)
to study the correlation and displacement between structural
changes in molecular clouds as a function of scale.
By inheriting the properties of the $\Delta$-variance analysis
from \citet{ossenkopf08a} the WWCC method can use a weighting function 
for each data point in the map to quantify their significance or
variable signal-to-noise ratios. The weighting function also allows us to deal with
irregular map boundaries and it permits the WWCC analysis through Fast Fourier 
transform which considerably speeds up the computation. 

The method can be applied to observations of interstellar clouds
taken in different tracers or at different velocity ranges.  
The {\changed WWCC measures} the correlated structural
changes between different maps as a function of their spatial scale.
{\changed Characteristic scales where the behaviour of different 
tracers starts to match or deviate from each other can be interpreted e.g. as scales
of chemical or phase transitions, or} driving and dissipation scales 
in interstellar turbulence.

We performed a series of numerical tests of the WWCC for simulated 
maps with well-defined spatial properties. It reveals that
\begin{itemize}
  \item For structures with a well-defined size, the spectrum of cross-correlation (CC) coefficients
  strongly depends on the displacement between structures and ratio of 
  their prominent scales. The CC spectrum turns negative on small scales if 
  the overlapping area between the two structures becomes small compared to
  the {\changedn not overlapping} area. The degree and scale range of the anticorrelation
  depends mainly on the displacement and much less on the size ratio.
  Beyond the scale of the mutual displacement, the correlation monotonically
  increases. For structures without dominant scales, the CC spectrum
  also traces the mutual displacement: it reaches a minimum close to the scale
  of the offset between the structures. 
  In case of noisy data, the amplitude of the CC spectrum is lowered
  at small scales, both for correlated and anti-correlated structures.
  
  \item  The displacement vectors can be reliably recovered at scales 
  dominated by actual structural information, not observational noise.
  For overall self-similar
  structures, independent of an enhancement of particular scales,
  there is always enough structural information on all scales that
  the displacement vector can be robustly recovered even for relatively
  low signal to noise levels ($\mathrm{S/N} \ge 5$). In case of
  individual structures with a fixed size, the lower limit for the
  reliable measurement of the displacement can be estimated from the
  minimum of the $\Delta$-variance spectrum, $l_{\rm low} = 1.5 l_{\rm min}+1.5$~pix
  and the upper scale is estimated from $l_{\rm upp} = -0.38 \,l_{\rm p} + 65$~pix,
  where $l_{\rm p}$ is the prominent scale of the largest structure.

  \item \citet{MLO} showed that the $\Delta$-variance can be used
  to measure the scale of individual prominent structures in maps.
  We extended the $\Delta$-variance analysis to allow for such an
  identification of enhanced structure scales also in maps
  that have structure on all scales, but with some scales being
  slightly more prominent than the rest.
  By using the logarithmic gradient of the $\Delta$-variance
  spectrum we can use the scale of the steepest gradient $l_{\rm c}$ to
  measure the size of the enhanced structure $l_{\rm f} = 0.52 l_{\rm c}$
  as long as the critical scale falls above the noise limit $l_{\rm min}$.
  
\end{itemize}

As the results of the WWCC are sometimes non-intuitive, it can be useful
to inspect the individual wavelet-filtered maps to understand the origin of a varying cross-correlation and measured
displacement vectors in terms of the individual features in the maps
that drive the statistical parameters. The product of the maps filtered (and shifted)
on a specific scale of interest allows to identify the locations of correlated structures and 
recognize their shapes.

We analysed molecular line maps ($^{13}$CO, C$^{18}$O, and HCN) of the G\,333 
giant molecular cloud. The $\Delta$-variance spectrum and its gradients (Eq.~(\ref{eq:lf_StoN=inf_and_5})) allow to measure prominent scales in the maps at 12~arcmin ($^{13}$CO), at 7--9~arcmin (C$^{18}$O), 
and at 10--12~arcmin (HCN). For the {changedn turbulent, self-similar structure, 
covering all spatial scales} we can estimate
the {\changedn spectral indices of the power spectrum} 
from the $\Delta$-variance gradient spectrum as $\zeta=2.8$ ($^{13}$CO),
$\zeta=2.3$ (C$^{18}$O), and $\zeta=2.5$ (HCN). This indicates a gradual difference
between the three maps with most small-scale structure and the smallest prominent
features in the C$^{18}$O data and the dominance of larger structures and a
relatively larger prominent structure in $^{13}$CO. The HCN map shows an
intermediate characteristics. {\changedn All indices are lower than expected
from simple Kolmogorov turbulence, but in the typical range observed in
molecular line observations of other interstellar clouds \citep{Falgarone2007}.}

Application of the WWCC to these molecular line maps shows that:
\begin{itemize}
  \item $^{13}$CO and C$^{18}$O emission line maps are strongly correlated above the noise scale
  in {\changedn spite of} their overall very different scaling behaviour. This is {\changedn
  expected from a constant relative abundance, similar excitation conditions, and the higher
  optical depth of $^{13}$CO. A fixed abundance ratio and identical excitation conditions
  for both species provide a match of all local features of the map, but a perfect match
  of all structures is prevented by the larger $^{13}$CO optical depth. It leads to
  a suppression of the high density peaks relative to ther lower-density environments in
  the intensity map so that all emission patches appear somewhat larger in $^{13}$CO compared 
  to C$^{18}$.} The measured $1'$ offset of the large 
  structures between $^{13}$CO and C$^{18}$O in southern direction {\changedn indicates that 
  this optical depth effect is slightly anisotropic, due to a shallower density gradient to 
  the south.}
  
  \item {\changedn Both the HCN--$^{13}$CO and HCN-C$^{18}$O correlations decrease on 
scales less than 7 arcmin (or 7 pc), where the maps trace different structures. 
On larger scales, all species trace the overall structure of the molecular cloud
resulting in tighter correlations. A chemical differentiation of HCN relative to the CO isotopologues or a very different volume density structure could be the origin of those differences on scales smaller than 7 pc.} 
  The offset of the HCN structure relative to structures seen in $^{13}$CO and C$^{18}$O 
  {\changedn measures the relative shift of the HCN emission to the south along the filament.
  It may be produced by a higher HCN abundance in the south, due to an overall
  anisotropic radiation field, or a somewhat higher density in that direction.}
\end{itemize}

\begin{acknowledgements}
We thank Robert Simon and Peter Frick for valuable comments {\changed and
anonymous referee for careful reading and useful suggestions}. 
This project was financed through DFG project number Os 177/2-1.
\end{acknowledgements}

\begin{appendix}
\section{Recovering the $\Delta$-variance spectrum of an observed map}
\label{subs:rec_dvar}

\begin{figure}
   \centering
   \includegraphics[width=6cm,angle=90]{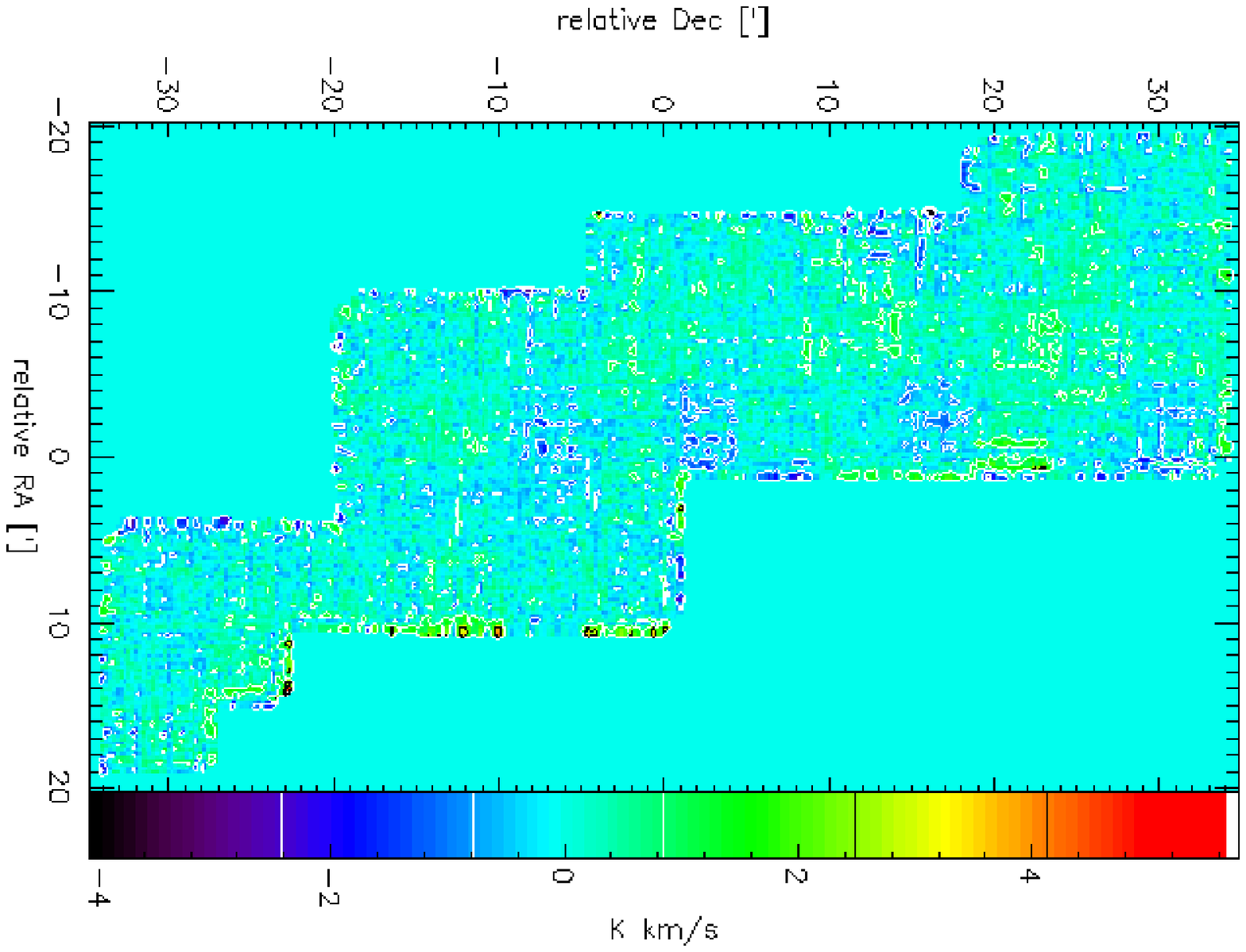}
   \caption{The noise extracted from the C$^{18}$O emission line map of the GMC G\,333.}
   \label{fig:G333_noise_C18O}
\end{figure}

To recover the enhancement scale in an efBm-like molecular cloud, we need
to correct the measured $\Delta$-variance spectrum for the noise 
contribution of the observations. Here, we demonstrate the procedure
for the G333 C$^{18}$O 1-0 map introduced in Sect.~\ref{sec:application}
(Fig. \ref{fig:G333} (b)). 
In a first step we measure the noise distribution across 
the map by integrating off-line channels free of emission. 
To avoid the need for a radiometric correction, we integrate the noise over 
the same width of a velocity range that is used to integrate the line emission   
($\Delta v_{\rm noise} = \Delta v_{\rm line}=30$ km s$^{-1}$). 
In the C$^{18}$O data cube this is achieved by integrating the noise over the velocity ranges  
from $-80$ to $-65$ km s$^{-1}$ and from $-35$ to $-20$ km s$^{-1}$.
The resulting noise map is shown in Fig.~\ref{fig:G333_noise_C18O}. 
In case of $^{13}$CO and HCN maps, the observed spectra do not contain enough
channels free of emission to select the same velocity width for integrating 
the noise and line emission. To correct for the smaller integration range of
the noise channels, $\Delta v_{\rm noise} < \Delta v_{\rm line}$ we have to
scale the integrated noise intensity radiometrically by 
$\sqrt{\Delta v_{\rm line}/\Delta v_{\rm noise}}$. 

\begin{figure}
   \centering
   \includegraphics[width=6cm,angle=90]{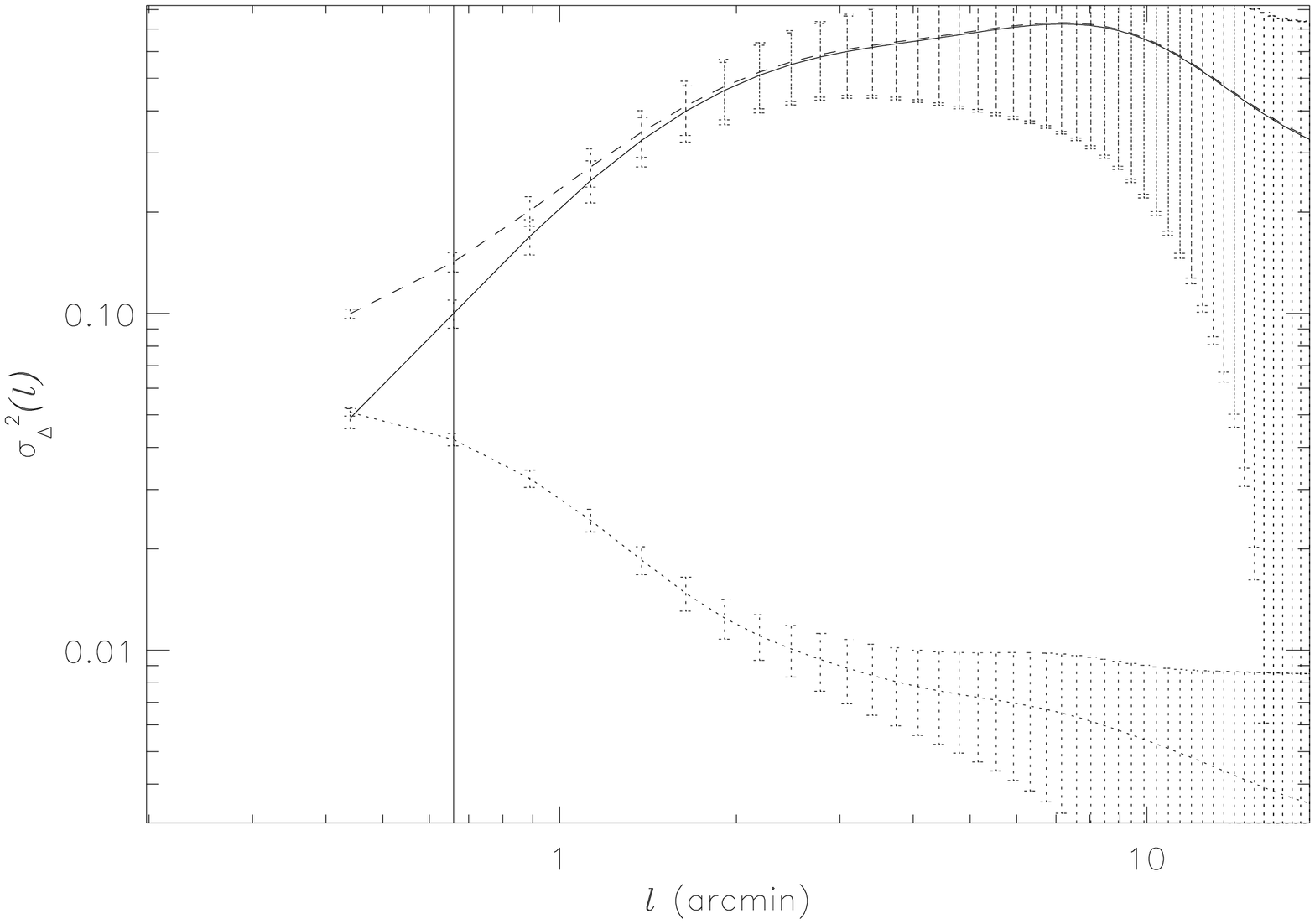}
   \caption{$\Delta$-variance of the observed C$^{18}$O emission line map (dashed line) 
   and the noise emission across the C$^{18}$O map in Fig.~\ref{fig:G333_noise_C18O} (dotted line). 
   Their difference represents a pure $\Delta$-variance free of the noise effect (full line). 
   The vertical line denotes the critical scale $l_{\rm c} = 0.66$ arcmin at which the pure 
   $\Delta$-variance curve is the steepest. $1\sigma$ error bars are presented.}
   \label{fig:G333_dvars_C18O}
\end{figure}

In the next step, 
we compute the $\Delta$-variance spectra of the observed C$^{18}$O map and 
the noise map (respective dashed and dotted lines in Fig.~\ref{fig:G333_dvars_C18O}). 
A pure $\Delta$-variance spectrum corrected for the noise (full line) is then 
recovered by subtracting the $\Delta$-variances of the observed C$^{18}$O map
and the noise. The critical scale can be measured from the gradients in the
corrected $\Delta$-variance spectrum (Eq.~(\ref{eq:dvar_lc})).

\end{appendix}


\end{document}